%% file: spinning-OPE.tex
\documentclass[11pt,reqno,preprint]{article}
\usepackage{jheppub}
\usepackage{epsfig}
\usepackage{amssymb}
\usepackage[utf8]{inputenc}
\usepackage{amsmath}
\usepackage{mathrsfs}
\usepackage{hyperref}
\usepackage{multirow}
\usepackage{subcaption}

\def\be{\begin{equation}}
\def\ee{\end{equation}}
\def\ba{\begin{eqnarray}}
\def\ea{\end{eqnarray}}

\def\a{\alpha}

\def\l{\langle}
\def\r{\rangle}

\def\.{{\cdot}}

\def\pa{a}
\def\pb{b}
\def\pp#1{{#1}}

\newcommand{\bea}{\setlength\arraycolsep{2pt} \begin{eqnarray}}
\newcommand{\eea}{\end{eqnarray}}
\newcommand{\nn}{\nonumber}

\def\OO{\mathcal{O}}

\newcommand{\lsim}{\mathrel{\hbox{\rlap{\lower.55ex \hbox{$\sim$}} \kern-.3em \raise.4ex \hbox{$<$}}}}
\newcommand{\gsim}{\mathrel{\hbox{\rlap{\lower.55ex \hbox{$\sim$}} \kern-.3em \raise.4ex \hbox{$>$}}}}

\def\ft#1#2{{\textstyle{\frac{\scriptstyle #1}{\scriptstyle #2} } }}
\def\fft#1#2{{\frac{#1}{#2}}}

\def\0{{\sst{(0)}}}
\def\1{{\sst{(1)}}}
\def\2{{\sst{(2)}}}
\def\3{{\sst{(3)}}}
\def\4{{\sst{(4)}}}
\def\5{{\sst{(5)}}}
\def\6{{\sst{(6)}}}
\def\7{{\sst{(7)}}}
\def\8{{\sst{(8)}}}
\def\9{{\sst{(9)}}}

\def\sst#1{{\scriptscriptstyle #1}}

\title{Helicity basis for three-dimensional conformal field theory}
\author{Simon Caron-Huot, Yue-Zhou Li}
\affiliation{Department of Physics, McGill University, 3600 Rue University, Montr\'eal, QC Canada H3A 2T8}
\emailAdd{schuot@physics.mcgill.ca, liyuezhou@physics.mcgill.ca} \abstract{
Three-point correlators of spinning operators admit multiple tensor structures compatible with conformal symmetry.
For conserved currents in three dimensions, we point out that helicity commutes with conformal transformations
and we use this to construct three-point structures which diagonalize helicity.
In this helicity basis, OPE data is found to be diagonal for mean-field correlators of conserved currents and stress tensor. Furthermore, we use Lorentzian inversion formula to obtain anomalous dimensions for conserved currents
at bulk tree-level order in holographic theories, which we compare
with corresponding flat-space gluon scattering amplitudes.
}
\begin{document}

\maketitle

\def\x#1{x_{#1}^2}
\def\xsq#1{x_{#1}^4}
\def\y#1{y_{#1}^2}
\def\zb{\overline{z}}
\def\a{\alpha}
\def\ab{\overline{\alpha}}
\def\j{J}
\def\zbar{\bar{z}}
\def\eps{\epsilon}
\def\GN{G}
\def\mubar{\bar{\mu}}
\def\gammaE{\gamma_{\rm E}}

\section{Introduction}
\label{intro}

It is an old proposition to use self-consistency conditions, such as unitarity, analyticity and crossing symmetry,
to ``bootstrap'' physical observables like the S-matrix of Lorentz invariant quantum field theories.
Nonperturbatively, this philosophy has been successfully applied in recent years to conformal field theories (CFT).
This has allowed to nonperturbatively explore the space of conformal theories,
and to extract precision spectra for a number of specific theories (for a review see \cite{Poland:2018epd}).

A surprising feature of the bootstrap is that a small number of correlators often suffice to obtain interesting constraints. Many studies therefore focus on four-point correlators of scalar operators.
Spinning correlators are technically more complicated but much progress has been made
and numerical studies involving them are now possible
\cite{Albayrak:2019gnz,Iliesiu:2017nrv,Dymarsky:2017yzx,Karateev:2019pvw,Erramilli:2020rlr}.
As nontrivial representations of rotation groups, spinning operators are bound to involve fancier structures.
Three-point functions, for example, can be constructed using the embedding formalism \cite{Costa:2011mg,Costa:2011dw}, and four-point conformal blocks, key ingredient to the bootstrap, may then be obtained by acting with corresponding spinning-up or weight-shifting operators on scalar seeds \cite{Costa:2011dw,Karateev:2017jgd}. This heavy machinery comes at a cost. This is especially visible in analytic work, which has so far
specialized to limits such as free theories, the Regge limit,
or conformal collider kinematics (see for example \cite{Afkhami-Jeddi:2016ntf,Costa:2017twz,Karateev:2018oml,Sleight:2018epi,Kulaxizi:2017ixa,Kologlu:2019bco,Kologlu:2019mfz}).

There are several motivations to pursue analytic work with spinning correlators.
A main one is the analogy with perturbative S-matrices, where massless spinning particles obey stringent self-consistency
conditions. These include
Weinberg's derivation of perturbative general relativity from soft limits \cite{Weinberg:1965nx},
or to give just one more modern example, on-shell recursion relations for gluon amplitudes \cite{Britto:2004ap,Britto:2005fq}.
For strongly coupled conformal theories with a holographic AdS dual that includes weakly coupled gravity,
stress-tensor correlators are thus expected to strongly constrain not only gravity, but its coupling to matter.
Indeed any CFT has a stress tensor, which, like gravity, couples to every degree of freedom.

A useful starting point for analytic approaches is good control of mean-field theory, around which one can start
various approximations, be these in large spins, large $N$, small $\epsilon$, or other quantities
\cite{Komargodski:2012ek,Fitzpatrick:2012yx,Kaviraj:2015cxa,Alday:2015eya,Kaviraj:2015xsa,Alday:2015ewa,Alday:2016njk}.
When the mentioned
technology is applied to spinning correlators, the OPE data become matrices in the space
of tensor structures. But even making seemingly natural choices,
one finds dense, non-diagonal matrices already in mean field theory (MFT) \cite{Karateev:2018oml}!
It is difficult to bring oneself to study corrections to such a zeroth approximation.




A possible way forward is the fascinating observation that the number of spinning structures in CFT$_d$ is identical to the number of structures for scattering amplitudes in QFT$_{d+1}$ \cite{Kravchuk:2016qvl}.  While physically natural from the viewpoint  of the bulk-point
or flat space limits of correlators, it is still unclear whether this counting extends to a useful map beyond that limit.
Indeed, the non-diagonal nature of MFT correlators stands in sharp contrast with the QFT side,
where diagonalizing trivial scattering $S=1$ was never a big challenge!
We should then ask: can one find a basis of CFT three-point structures in which MFT correlators are diagonal?

In this paper we address this question in the special case of CFT${}_3$, exploiting the fact that in QFT${}_4$ massless particles come with two helicity states $\pm$.
We point out that the ``helicity'' of a conserved current is a meaningful (crossing-symmetric) concept
also in CFT${}_3$, which formally implies that a helicity basis of three-point structures will
automatically diagonalize crossing symmetry.
We will confirm this by computing explicit OPE data in MFT,
as well as the first correction to CFT${}_3$ current correlators dual
to tree-level gluon scattering in AdS${}_4$.

This paper is organized as follows. In section \ref{gene}, we construct the helicity basis for three-point functions and explain
that it diagonalizes a well-defined operator $h$.  We also introduce the group-theoretic concepts to be used in later sections,
including three-point pairings, shadow transforms, Euclidean and Lorentzian inversion formula.
In section \ref{GFF}, we use both inversion formulas to independently obtain mean-field
OPE data for conserved currents of various spins.
In section \ref{application}, we apply our scheme to study YM$_4$/CFT$_3$, using the Lorentzian inversion formula to extract the analytic-in-spin part of the leading-order double-twist anomalous dimensions of currents. In section \ref{fromamp}, we explicitly check that the anomalous dimensions of
the double-twist states $[VV]_{n,\j}$ at large-$n$ agree with flat-space partial waves for tree-level gluon scattering.

This paper contains a number of technical appendices.
In appendix \ref{Witten}, we relate CFT${}_3$ three-point functions conserved currents to the bulk YM${}_4$ couplings, using the AdS embedding formalism.  In appendix \ref{app:spinors}, we explain how to simplify certain calculations by representing polarization vectors as spinors and give formulas for Fourier transforms.
In appendix \ref{more conformal blocks}, we review the series expansion of scalar conformal blocks.
Moreover, we show how to compute OPE data for correlators that are powers of cross-ratios multiplied with Gegenbauer polynomials, which may have applications to other problems; we also record simplified expansions for certain scalar, currents and stress-tensors exchanges. Finally, flat-space gluon amplitudes, including Yang-Mills and higher-derivative couplings, are reviewed in appendix \ref{gluonampdo}.

\section{Generalities}
\label{gene}

\begin{figure}[t]
\centering \hspace{0mm}\def\svgwidth{100mm}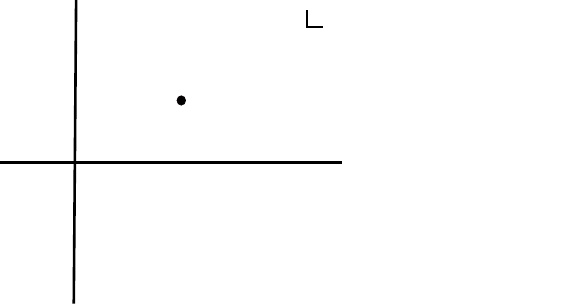
\caption{Conformal frame used for three-point functions: $\langle\mathcal{O}_1(0)\mathcal{O}_2(x)\mathcal{O}_3(\infty)\rangle$.}\label{fig:frame}
\end{figure}

The structure of conformal correlators for spinning external operators is by now well understood.
Here we aim to concisely summarize key results so as to state our new three-point structures
as early as possible (eq.~\eqref{Tpm} below).
We eschew the use of embedding space and cross-ratios. Rather, we use
conformal symmetry to place local operators at standard locations such as
$(0,x,\infty)$ as shown in figure \ref{fig:frame}, or $(0,x,y^{-1},\infty)$ for four-points.

In this frame, three point functions for scalar operators are
determined by dimensional analysis up to a normalization:
\be
T_{123}(x)=\langle \OO_1(0)\OO_2(x)\OO_3(\infty)\rangle = \frac{1}{|x|^{\Delta_1+\Delta_2-\Delta_3}}, \quad |x| \equiv \sqrt{x_\mu x^\mu}\,. \label{plTs}
\ee
We define $\OO_i(\infty)$ by taking the limit $x^{-1}\to 0$ in an inverted frame (with an inversion tensor inserted for spinning operators),
so it behaves as an operator of dimension $-\Delta_i$ (see eq.~(114) of \cite{Simmons-Duffin:2016gjk}).
We will often Fourier transform with respect to the second position $x$:
\be \label{T in p}
T_{123}(p)=
\langle \OO_1(0)\OO_2(p)\OO_3(\infty)\rangle = \int d^dx e^{-ip{\cdot}x} \langle \OO_1(0)\OO_2(x)\OO_3(\infty)\rangle.
\ee
This was used in \cite{Karateev:2018oml} to simplify calculations of shadow transforms
and to compute conformal pairings, which all become simple algebraic operations.

\begin{figure}[t]
\centering \hspace{0mm}\def\svgwidth{100mm}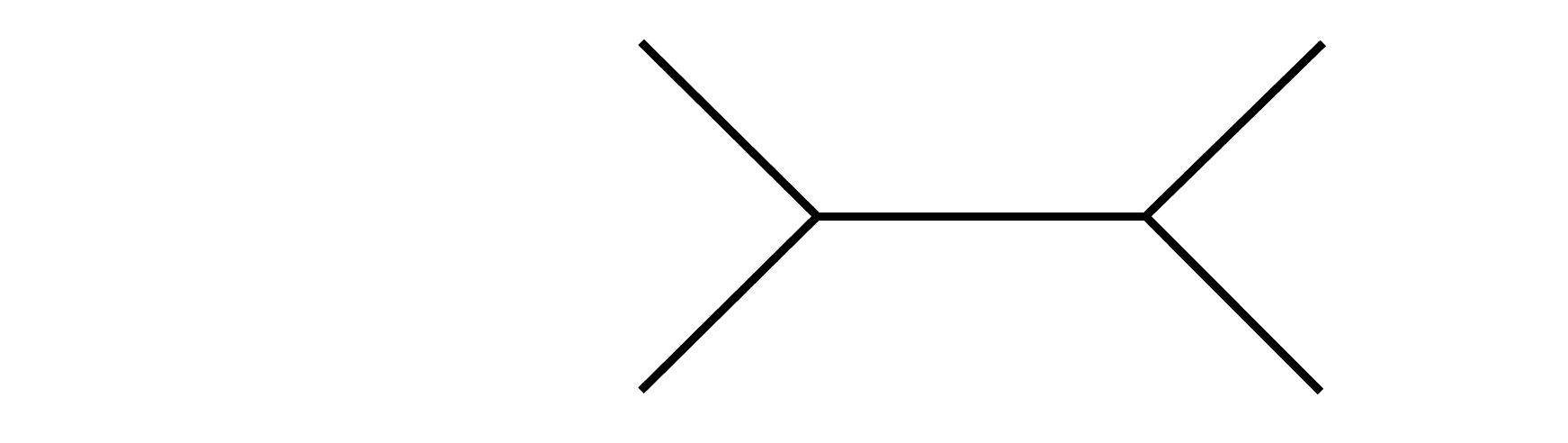
 \caption{Four-point function factorized into three-point functions.}\label{fig:fourptexp}
\end{figure}

It is important to note that we do not Fourier transform all operators, as is sometimes considered in the literature, e.g. in \cite{Bzowski:2013sza}.
The only Fourier integrals we will compute involve powers of a single variable as in (\ref{plTs}) which are rather straightforward.
Physically, singling out one operator is natural in conformal bootstrap applications, as we typically treat external and internal states asymmetrically.
We think of the third operator as the exchanged one $\mathcal{O}$ in the conformal block decomposition of a four-point correlator,
as shown in figure \ref{fig:fourptexp}.

\subsection{Three-point functions: helicity basis}
\label{helicity}

Multiple index contractions generally exist between spinning operators,
and three-point structures are correspondingly no longer unique.
They are straightforward to classify in the above frame \cite{Costa:2014rya}.
For pedagogical reasons, let us focus on the case where all operators are symmetric traceless tensors,
$\OO^{\mu_1{\ldots}\mu_\j}$, where $\j$ is the spin of the operator. In $d=3$, this covers all bosonic operators.
We work in index-free notation \cite{Costa:2011mg} and dot into the $J$'th power of a null polarization vector $\epsilon^\mu$.
Our two-point functions follow the standard normalization:
\be
\langle \mathcal{O}(0)\mathcal{O}(\infty)\rangle=(\eps_1\.\eps_2)^{\j}\,.
\ee

Any index contraction between the $\epsilon_i^\mu$ and $x^\mu$ defines an allowed three-point function.
For example, for two operators of spin-1 and a third of spin $\j_3$ $\langle V_1V_2\mathcal{O}_3\rangle$,
a basis of five independent (parity-even) monomials is easily enumerated:
\be
B_V= \left\{\eps_1{\cdot}\eps_2, \quad
\frac{\eps_1{\cdot}x\ \eps_2{\cdot}x}{x^2}, \quad
 \frac{\eps_1{\cdot}x\ \eps_2{\cdot}\eps_3}{\eps_3{\cdot}x}, \quad
 \frac{\eps_1{\cdot}\eps_3\ \eps_2{\cdot}x}{\eps_3{\cdot}x}, \quad
   \frac{\eps_1{\cdot}\eps_3\ \eps_2{\cdot}\eps_3}{(\eps_3{\cdot}x)^2}x^2
\right\}
\times \frac{(\eps_3{\cdot}x)^{\j_3}}{|x|^{\Delta_1+\Delta_2-\Delta_3+\j_3}}\,.  \label{xspace structure}
\ee
Each monomial has homogeneity $(1,1,\j_3)$ with respect to the three $\eps_i$.
It will be useful to treat structures analytically in the third spin $\j_3$.
The fact that $1/(\eps_3{\cdot}x)$ appears in the denominator implies that certain
structures cease to exist at low spin.  It will be possible to use a common labelling scheme for all values of $\j_3$,
but we will have to remember that certain structures do not contribute at low $\j_3$.

Although our frame choice breaks permutation symmetry it is trivial to restore it.
For example to exchange 1 and 2, we simply take translation by an amount $-x$ and substitute $x^\mu\mapsto -x^\mu$.\footnote{
This is really a substitution, not a symmetry transformation. It can be done whether or not the theory is parity symmetric.}
Less trivially, to interchange operators 1 and 3, we use the inversion $x^\mu\mapsto x^\mu/x^2\equiv x^{-1}$,
acting with the inversion tensor on $\eps_2$:
\be
T_{123}(\infty,x,0) = x^{-2\Delta_2}T_{123}(0,x^{-1},\infty)\Big|_{\eps_2^\mu\mapsto I^{\mu\nu}(x) \eps^\nu_2},
\quad I^{\mu\nu}(x) = \delta^{\mu\nu}-2\frac{x^\mu x^\nu}{x^2}.\label{exinforder}
\ee
There is no need to include inversions acting on $\eps_{1}, \eps_{3}$ because inversion is included in the definition of $\OO(\infty)$.
The structures in eq.~(\ref{xspace structure}) become
\be
 \left\{\eps_1{\cdot}\tilde{\eps}_2, \quad
 -\frac{\eps_1{\cdot}x\ \eps_2{\cdot}x}{x^2}, \quad
 \frac{\eps_1{\cdot}x\ \tilde{\eps}_2{\cdot}\eps_3}{\eps_3{\cdot}x}, \quad
 -\frac{\eps_1{\cdot}\eps_3\ \eps_2{\cdot}x}{\eps_3{\cdot}x}, \quad
  \frac{\eps_1{\cdot}\eps_3\ \tilde{\eps}_2{\cdot}\eps_3}{(\eps_3{\cdot}x)^2}x^2
\right\}
\times \frac{(\eps_3{\cdot}x)^{\j_3}}{|x|^{\Delta_3+\Delta_2-\Delta_1+\j_3}}  \label{xspace structure 1}
\ee
where $\tilde{\eps}^\mu_2 = \eps^\mu_2-2x^\mu \eps_2{\cdot}x/x^2$.

Let us now improve this in steps.  Instead of just ``listing all monomials'', a good idea is to use the SO$(d{-}1)$ symmetry
which preserve the point $x$.  An SO$(d)$ traceless symmetric tensor of rank $\j$ can be written as a
direct sum of multiple SO$(d{-}1)$ tensors, with rank $0\leq \j'\leq \j$ indices, roughly,
how many indices are perpendicular to $x$.
Three-point structures are then in one-to-one correspondence with SO($d{-}1$) singlets in the tensor products of the
three representations from the three legs.
Such a scheme was used for example in ref.~\cite{Costa:2014rya}. While effective for generic operators,
this is \emph{not} the scheme we shall use, since we are interested in conserved currents.
In $x$-space, conservation is a cumbersome differential constraint.

The next improvement is to use instead SO$(d{-}1)$ tensors in \emph{momentum} space, separating indices that are parallel or perpendicular to $p$ in the frame in eq.~\eqref{T in p}.  For conserved currents one simply has to drop all the structures that are not fully perpendicular to $p$.
For example, for two conserved currents in $d$ dimensions
(which have scaling dimension $\Delta_1=\Delta_2=d-1$) there are just two allowed structures, proportional to:
\be\begin{aligned}
 &
\Big\{
\fft{\left[p^2 (\eps_1\.\eps_3)-(p\.\eps_1)(p\.\eps_3)\right]\left[p^2 (\eps_2\.\eps_3)-(p\.\eps_2)(p\.\eps_3)\right]}
{(p\cdot\epsilon_3)^2}
- \frac{p^2 (\eps_1\.\eps_2)-(p\.\eps_1)(p\.\eps_2)}{d-1}\,,\\
&\quad
p^2 (\eps_1\.\eps_2)-(p\.\eps_1)(p\.\eps_2)\Big\}\, \times (p\cdot\epsilon_3)^{J_3}|p|^{d-4-\Delta_3-J_3}\,.
\label{VVmom}
\end{aligned}\ee
These two structures are transverse with respect to $\eps_{1}$ and $\eps_2$
and are respectively SO$(d-1)$ traceless symmetric tensors of rank 2 and 0 with respect to $\eps_3$.
The first structure is analytic for spin $J\geq 2$, and the second for $J\geq 0$.
In this example ``transverse'' simply means invariant under $\epsilon_i\mapsto \epsilon_i + p_i$.
For higher-rank conserved currents, the correct statement will
involve an operator $\mathcal{D}$ designed to preserve the constraint $\epsilon_i^2=0$ \cite{Dobrev:1975ru}:
\be
p^\mu \mathcal{D}_\mu^{\epsilon_1}T=p^\mu \mathcal{D}_\mu^{\epsilon_2}T=0,\qquad
\mathcal{D}_\mu^\epsilon\equiv \left(\fft{d}{2}-1+\epsilon\cdot \fft{\partial}{\partial\epsilon}\right)\fft{\partial}{\partial\epsilon^\mu}-
\frac{\epsilon_\mu}{2} \fft{\partial}{\partial\epsilon}\cdot\fft{\partial}{\partial\epsilon}\,.\label{Tod}
\ee
Such a scheme could be used to label three-point structures in any dimension
$d$, including operators $\OO_3$  in mixed representations of SO($d$).\footnote{There are momentum-space constructions for spinning operators in the literature, where all three positions are Fourier transformed, see, e.g., \cite{Bzowski:2013sza,Bzowski:2017poo,Bzowski:2018fql,Jain:2021wyn} and references therein, which enjoy potential applications to inflationary cosmology \cite{Maldacena:2011nz,Baumann:2020dch}.}
We now specialize to $d=3$, where further simplifications occur.

In $d=3$, ${\rm SO}(d{-}1)$ irreps (transverse to $p$)
are one-dimensional and labelled by helicity $\pm J$.
For two conserved currents of any spin there are thus only four structures.
A projector onto the positive-helicity component of $\eps_2$ can be written by combining parity-even and odd structures:
\be
 \eps_{2\mu} \Pi_{\pm p}^{\mu\nu} \eps_{3\nu}
 \equiv \frac12\left(\eps_2\.\eps_3 - \frac{(p\.\eps_2)(p\.\eps_3)}{p^2} \pm\frac{i}{|p|}(\eps_2,p,\eps_3)\right).
\ee
Here $(a,b,c)=\epsilon_{\mu\nu\sigma}a^\mu b^\nu c^\sigma$ denotes contraction
with $\epsilon_{123}=+1$ the antisymmetric tensor in Euclidean signature.
The projector satisfies $\Pi^2_{\pm p}=\Pi_{\pm p}$ and $p{\cdot}\Pi_{\pm p}=0$.
For $p$ along the z axis, it can be written as $\tfrac12(1,i,0)^\mu (1,-i,0)^\nu$.

Given two conserved currents of spin $J_1$ and $J_2$ in $d=3$, we thus define a complete basis of four possible three-point couplings, including a convenient factor, as:
\be \label{Tpm}
\boxed{ T^{\pm,\pm}_{123} \equiv
\frac{(4\pi)^{\fft{3}{2}}(-i\sqrt{2})^{\j_1+\j_2+\j_3}}{2^{\tau_1+\tau_2-\Delta_3}}\times (\eps_1 \Pi_{\mp p} \eps_3)^{J_1}(\eps_2 \Pi_{\pm p} \eps_3)^{J_2}\times
  (p\cdot\epsilon_3)^{\j_3-\j_1-\j_2} |p|^{\beta_{12;3}-3}\,,}
\ee
where $\beta_{12;3} = (\Delta_1+\j_1)+(\Delta_2+\j_2)-(\Delta_3+\j_3)$ and $\tau_i=\Delta_i-\j_i$ is the twist.
The two superscripts represent the helicity of each operator.
Note the reversal of the momentum in the first projector, since the first operator has momentum $-p$, so that helicity retains its physical interpretation as spin along momentum axis.
The transversality condition \eqref{Tod} is readily verified for any $\j_i$.

Eq.~\eqref{Tpm} defines the helicity basis we will use throughout.
The opposite-helicity structures $T^{+-}_{123}$ and $T^{-+}_{123}$ are only allowed for local operators
(polynomial in $\eps_3$) when $J_3\geq J_1+J_2$. On the other hand,
since SO(2) representations are one-dimensional, the projectors satisfy the identity:
\be
 (\eps_1 \Pi_- \eps_3)(\eps_2 \Pi_+ \eps_3) = (\eps_1 \Pi_- \eps_2)(\eps_3 \Pi_- \eps_3) =  -\frac{(p\.\eps_3)^2}{2p^2} (\eps_1 \Pi_- \eps_2)\,,
\ee
which extends the range of same-helicity structures
$T^{++}_{123}$ and $T^{--}_{123}$ to: $J_3\geq |J_1-J_2|$\,.
These ranges coincide with the usual selection rules for the total angular momentum of two massless particles in flat four-dimensional space.

Although eq.~\eqref{Tpm} is primarily meant to be used for conserved currents, where $\Delta_i=1+\j_i$ for $i=1,2$,
we kept $\Delta_i$ free since the structures make sense for any $\Delta_i$. In particular,
we will use the same expressions below for shadow-transformed operators.
For spin-0 states, we keep the same formula but drop superscripts.

Once the helicity basis is defined in momentum space, it is often necessary to
transform it to coordinate space. The Fourier-transform of a power-law is straightforward
\be \label{Fourier}
\int \fft{d^dp}{(2\pi)^d}e^{ip\cdot x}p^{2k}=\fft{4^k}{x^{2k+d}}\fft{\Gamma(\fft{d}{2}+k)}{\pi^{\fft{d}{2}}\Gamma(-k)}\,.
\ee
Our strategy is to perform Fourier-transform for pure power-laws at first, and then replace
\be
p\cdot\epsilon\rightarrow -i\epsilon\cdot\partial\,.
\ee
Doing so, one finds that the parity-even and odd components produce disparate gamma-functions that don't nicely combine.
Many calculations are thus simplified by switching to an Even/Odd basis of parity eigenstates.
Each parity sector contains two elements, representing states with opposite or same helicity:
\be\begin{aligned}\label{even odd}
\Big\{ T^{E,\rm opp}_{123},T^{E,\rm same}_{123}
\Big\} &\equiv \fft{\Gamma\big(\fft{3-\tau_1-\tau_2+\Delta_3+\j_3}{2}\big)}
{\Gamma\big(\fft{\tau_1+\tau_2-\tau_3}{2}\big)} \times
\Big\{\frac{T^{+-}_{123}+T^{-+}_{123}}{\sqrt{2}},\, \frac{T^{++}_{123}+T^{--}_{123}}{\sqrt{2}}\Big\}, \\
\Big\{ T^{O,\rm opp}_{123},T^{O,\rm same}_{123} \Big\} &\equiv
\fft{\Gamma\big(\fft{2-\tau_1-\tau_2+\Delta_3+\j_3}{2}\big)}
{\Gamma\big(\fft{1+\tau_1+\tau_2-\tau_3}{2}\big)} \times \Big\{ \frac{T^{+-}_{123}-T^{-+}_{123}}{\sqrt{2}},\, \frac{T^{++}_{123}-T^{--}_{123}}{\sqrt{2}}\Big\}\,,
\end{aligned}\ee
where we introduced gamma-factor normalizations for future convenience.
These ensure that the transform produces polynomials in $\Delta_3$ and $\j_3$ of the lowest possible degree,
as the denominator cancels spurious double-twist poles from the Fourier transform.

Fourier transforms may now be straightforwardly computed, by expanding
the even/odd structures into dot products of $p$ with polarizations,
up to a possible single odd factor $(p,\eps_i,\eps_j)$.

As a trivial example, in the scalar case $\j_1=\j_2=0$, there is just a single structure
\be
  T^{E}_{00\mathcal{O}} = 2^{\frac{\j_3}{2}} |x|^{\Delta_3-\j_3-\Delta_1-\Delta_2} (x\.\eps_3)^{\j_3}\,. \label{T00}
\ee
As a more illustrative example, for two spin-1 currents $\langle V_1 V_2 \mathcal{O}\rangle$
the two even structures turn out to be proportional to eq.~\eqref{VVmom} (in the same order).
As it should, the transform takes the form of a matrix acting on the basic structures $B_V$ in eq.~\eqref{xspace structure}:
\be
\begin{small}
\left(\begin{array}{c} T^{E,\rm opp}_{11\mathcal{O}} \\T^{E,\rm same}_{11\mathcal{O}}\end{array}\right)
=2^{\frac{\j_3+1}{2}}n\left(
\begin{array}{ccccc}
 2 (n{-}\tilde{\j}_3)  &2(n{-}1)& (3\tilde{J}_3{-}4 n{+}1) &
 (3\tilde{J}_3{-}4 n{+}1) &  \ft{\tilde{\j}_3^2{-}(8 n{+}1)\tilde{\j}_3{+}8n^2}{2 n} \\
 2({-}n{-}J_3) & 2(n{-}1) & J_3 & J_3 &\ft{(J_3{-}1) J_3}{2 n}
\end{array}
\right)\.B_V\,,\label{VVstcor}
\end{small}
\ee
where $n$ is defined through $\tau_3=\tau_1+\tau_2+2n$, and $\tilde{J}$ denotes the ``spin shadow'': $\tilde{J}=-1-J$ in $d=3$ \cite{Kravchuk:2018htv}.
The parity-odd structures can be similarly represented in terms of four odd monomials:
\be
B_V^\prime = \left\{
  \frac{ (\eps_1,x,\eps_3)\eps_2\.x}{x^2}, \quad
  \frac{(\eps_1,x,\eps_3)\eps_2\.\eps_3}{x\.\eps_3}, \quad
    \frac{(\eps_2,x,\eps_3)\eps_1\.x}{x^2}, \quad
  \frac{(\eps_2,x,\eps_3)\eps_1\.\eps_3}{x\.\eps_3}\right\}\frac{(\eps_3{\cdot}x)^{\j_3-1}}{|x|^{\Delta_1+\Delta_2-\Delta_3+\j_3-1}}\,,
  \label{xspace structure odd}
\ee
in which
\be
\left(\begin{array}{c} T^{O,\rm opp}_{11\mathcal{O}} \\T^{O,\rm same}_{11\mathcal{O}}\end{array}\right)
=2^{\frac{\j_3+1}{2}}\left(
\begin{array}{cccc} (1 {-} 2 n)& (1 {+} \j_3{+} 2 n)& (1{-}2 n) & (1{+}\j_3{+}2 n)\\
(1 {-} 2 n)& (-1 {+} \j_3{+} 2 n)& (-1{+}2 n) & (1{-}\j_3{-}2 n)\end{array}\right)\. B_V^\prime\,.
\label{VVstcor odd}
\ee
Notice that so far $n$ is simply a notation for the twist, but when $n$ takes on (half-)integer values it will represent so-called double-twist operators.
Parity-even double twists have integer $n$ while parity-odd ones have half-integer $n$.

A technical complication when dealing with higher-rank tensors and odd structures
is the presence of Gram determinant relations (antisymmetrizing any four vectors gives zero).
In our calculations below,
we circumvent this either by evaluating expressions on a symbolic three-dimensional parametrization,
or by using the spinor formulation in appendix \ref{app:spinors}.

The opposite-helicity structure in eq.~\eqref{VVstcor} is physically allowed for $\j_3\geq 2$,
but there is an important discrete exception:
when $\mathcal{O}_3$ is a conserved current ($\j_3=1$ and $\Delta_3=2$). Then the
complicated polynomial
in the fifth column vanishes, shielding the problematic denominator in eq.~\eqref{xspace structure}.
The three structures: $T^{E,\rm opp},T^{E,\rm same},T^{O,\rm same}$ then define valid (and independent)
couplings between three currents.  We verify in appendix \ref{Witten} that these map, \emph{respectively}, to bulk Yang-Mills couplings ${\rm Tr}F^2$, and to parity even/odd parts of ${\rm Tr}F^3$!

\subsection{Helicity is conformally invariant}
\label{ssec:helicity is conf}

The reader may worry that our definition of helicity structures in eq.~\eqref{Tpm} is tied to the specific frame $(0,x,\infty)$.
However, it turns out to be independent of this!
Here we construct a conformal integral transform, whose eigenvalue is helicity.
Its existence will automatically imply that crossing is diagonal in the helicity basis.

It is intuitively clear from holography that helicity should be frame-independent,
since momentum-space currents with definite helicity source AdS${}_4$ gauge fields that are either self-dual or anti-self-dual
near the boundary \cite{Maldacena:2011nz,Raju:2012zs}.
Helicity structures for correlators of three higher-spin currents in momentum space,
and their relation with bulk AdS couplings, were discussed in \cite{Skvortsov:2018uru}.
(For a spinor-helicity formalism in AdS${}_4$, see also \cite{Nagaraj:2018nxq}.)
Since the self-dual decomposition is invariant under conformal isometries,
we expect it to be independent of frame and agree between all channels.

In momentum space, the operation which measures helicity is simply
\be \label{h in p}
 h J^\mu(p) \equiv -i\frac{\epsilon^{\mu\nu\sigma}p_\sigma}{|p|} J^\nu(p)\,.
\ee
Fourier transforming this defines an integral transform:
\be
 h J^\mu(x) = \int d^3y H^{\mu\nu}(x-y) J_\nu(y)\,,
 \qquad H^{\mu\nu}(x-y) \equiv \frac{\epsilon^{\mu\nu\sigma}}{2\pi^2} \frac{\partial}{\partial y^\sigma}\frac{1}{(x-y)^2}\,.
\ee
We now show that $h$ commutes with  conformal transformations.
Normally, this would require the kernel $H$
to transform like a two-point function between a current and its shadow, $\langle J^\mu(x) \tilde{J}^\nu(y)\rangle$.
For a generic operator, this is impossible: conformal two-point functions between operators of different dimension must vanish!
(This follows easily from scale invariance in the frame $(x,y)=(0,\infty)$.)
The loophole here is that since $J^\nu(y)$ is conserved, the shadow $\tilde J^\nu$
is defined \emph{modulo} a derivative: the kernel $H$ only needs to be conformally invariant
modulo a total derivative  $\partial_y^\nu X^\mu(x,y)$.

Let's thus check invariance under inversion $x^\mu\mapsto x^\mu/x^2$.
Applying the standard transformation laws, a short calculation gives:
\be\begin{small}
 \frac{\mathcal{I}^{\mu\mu'}(x) \mathcal{I}^{\nu\nu'}(y)}{x^4y^2} H^{\mu\nu}(x^{-1}-y^{-1}) =
\frac{1}{\pi^2} \left[\frac{\eps^{\mu\nu\sigma}(y-x)_\sigma}{(x-y)^4} +\frac{\eps^{\mu\nu\sigma} x_\sigma}{(x-y)^2x^2}
+2 \frac{(x-y)^\nu \epsilon^{\mu \rho\sigma}y_\rho x_\sigma }{x^2(x-y)^4}
\right]
\end{small}\ee
We have used the Schouten identity to eliminate terms with $x^\mu$ or $y^\mu$.
With a bit of inspection, we find that the sum of $H$ and its transform is indeed a total derivative:
\be
 H^{\mu\nu}(x-y) + \frac{\mathcal{I}^{\mu\mu'}(x) \mathcal{I}^{\nu\nu'}(y)}{x^4y^2} H^{\mu\nu}(x^{-1}-y^{-1})
 = \frac{\partial}{\partial y^\nu} \frac{\epsilon^{\mu\rho\sigma} y_\rho x_\sigma}{\pi^2(x-y)^2} \,. \label{total der}
\ee
This shows formally that $h$ is invariant under inversion (up to an overall sign change):
\be
(h J)^{-1} =  -h(J^{-1})
\ee
where $(J^{-1})^\mu(x)=\mathcal{I}^{\mu\mu'}J^{\mu'}(x^{-1})/x^4$ denotes the inversion map.
The sign change was expected since $h$ is parity-odd. One could equivalently say that $h$ is invariant
under the combination of inversion and parity.

To illustrate the action of $h$, let us briefly consider two-point functions.
A special feature of $d=3$ CFTs is that \emph{two} structures are allowed by conformal invariance \cite{Closset:2012vp}:
\be
 \langle J_\mu(x) J_\nu(0) \rangle = \left( \delta_{\mu\nu} \partial^2 - \partial_\mu\partial_\nu\right) \frac{\tau}{32\pi^2x^2}
+ \frac{i\kappa}{2\pi}\epsilon_{\mu\nu\rho}\partial^\nu\delta^3(x)\,,
\ee
where the coefficient $\kappa$ of the contact term is defined modulo an integer.
It is easy to see (for example using momentum space expressions from ref.~\cite{Closset:2012vp}) that acting with $h$ on $J_\mu(x)$ yields the same with
$\tau$ and $8\kappa/\pi$ interchanged. This confirms that $h$ takes conformal two-point functions to conformal two-point functions.
Of course, just like the shadow transform, $hJ$ is generally not a local operator.

For higher-spin conserved currents, a similar transform can be defined
\be
h T^{\mu_1\cdots \mu_J}(x) = \int d^3y
\left( H^{\mu_1\nu_1}(x-y) 
T^{\nu_1 \mu_2\ldots \mu_J}(y) + \mbox{$(J-1)$ permutations of $\mu_1$}\right)
\label{higher spin}
\ee
Generally, $h^2=J^2$, and one can easily verify that the structures in eq.~\eqref{Tpm}
are eigenstates:
\be
 h_1 T^{\pm_1,\pm_2}_{123} = (\pm_1 J_1) T^{\pm_1,\pm_2}_{123}\, \qquad
 h_2 T^{\pm_1,\pm_2}_{123} = (\pm_2 J_2) T^{\pm_1,\pm_2}_{123}\,. \label{hT}
\ee
Although we did not construct a total derivative akin to eq.~\eqref{total der} in the higher-spin case,
we believe $h$ to be conformal as well, given the fact that all data computed
in the next sections will turn out diagonal.

In Lorentzian signature, there is a subtlety: $h$ depends on operator ordering through
the branch choice $|p|\equiv \sqrt{p^2\pm i0}$ in eq.~\eqref{h in p}.
While eqs.~\eqref{hT} remain valid as long as the same branch is used for $h$ and $T$, this means
that taking discontinuities or commutators do not preserve $h$ eigenstates; one can explicitly see
in eqs.~\eqref{VVstcor}-\eqref{VVstcor odd} that even and odd structures acquire different phases.
This will be important below in our discussion of Lorentzian inversion.


\subsection{Simple operations: three-point pairings and shadow map}
\label{ssec:simple ops}

Since $h$ is a conformal operation, three-point structures with definite helicity will be orthogonal under all natural operations.  Here we review two simple  operations, which will form useful building blocks later.

The simplest may be the conformal pairing between three-point structures and shadow structures:
\bea
P^{\pa,\pb}_{123} &=& \left(T^{\pa}_{123},T_{\tilde{1}\tilde{2}\tilde{3}}^{\pb}\right) \equiv
\int \frac{d^d x_1 d^d x_2 d^d x_3}{{\rm vol(SO}(d+1,1))} \l\OO_1\OO_2\OO_3\r^a \l\tilde\OO_1\tilde\OO_2\tilde\OO_3\r^b
\\ &=& \fft{1}{2^d{\rm vol(SO}(d-1))}\sum_{\eps_1,\eps_2,\eps_3}
T^{\pa}_{123}(\eps_1,\eps_2,\eps_3)(1)\ T^{\pb}_{\tilde{1}\tilde{2}\tilde{3}}(\eps_1^*,\eps_2^*,\eps_3^*)(1)\,,
\label{pairing def}
\eea
where we have used the symmetry to put $x=1$. The denominator is the volume form of the ``little group'' that keeps the frame $(0,1,\infty)$ fixed \cite{Karateev:2018oml}.

A good way to compute index contraction is to use the differential operator (\ref{Tod})
\be
 \sum_\epsilon f(\eps^*) g(\eps) = \frac{1}{\j! (\tfrac{d-2}{2})_\j} f(\mathcal{D}_\eps) g(\eps) . \label{Tod pairing}
\ee
For example, for vector-vector-general $\langle V_1 V_2 \mathcal{O}\rangle$ case, the pairings between
Even or Odd structures (\ref{VVstcor}) and \eqref{VVstcor odd} is readily evaluated:
\be
P_{11\mathcal{O}}^{E}=16P_s N^{E}_{11\mathcal{O}}\left(
\begin{array}{cc}
 \frac{(J_3+1) (J_3+2)}{(J_3-1) J_3} & 0  \\
 0 & 1
\end{array}
\right)\,, \qquad
P_{11\mathcal{O}}^{O}=16P_s N^{O}_{11\mathcal{O}}\left(
\begin{array}{cc}
 \frac{(J_3+1) (J_3+2)}{(J_3-1) J_3} & 0  \\
 0 & 1
\end{array}
\right)\,, \label{P11}
\ee
where $P_s$ is just the pairing of two scalars and one spinning operator \cite{Karateev:2018oml}\footnote{%
Our scalar structures are larger by a factor $2^{\j_3/2}$ than those of \cite{Karateev:2018oml}: $P_s^{\rm here} = 2^{\j_3} P^{\rm there}_s$.}
\bea
P_s=\frac{1}{2^d {\rm vol}({\rm SO}(d-1))} \frac{(d-2)_{\j_3}}{\big(\tfrac{d-2}{2}\big)_{\j_3}}
\eea
and for latter convenience we introduce the $N$ factor, which is precisely the product of the gamma-functions in
eq.~\eqref{even odd} and its shadow:
\bea
&& N^E_{\j_1\j_2\mathcal{O}}= \big(\tfrac{\tau_1+\tau_2-\tau_3}{2}\big)_{\j_1+\j_2} \big(\tfrac{3+\beta_3-\beta_1-\beta_2}{2}\big)_{\j_1+\j_2}\,,
\cr &&
\cr &&
N^O_{\j_1\j_2\mathcal{O}} = \big(\tfrac{1+\tau_1+\tau_2-\tau_3}{2}\big)_{\j_1+\j_2-1} \big(\tfrac{4+\beta_3-\beta_1-\beta_2}{2}\big)_{\j_1+\j_2-1}\,.\label{NPEuc}
\eea
Many other examples can be straightforwardly worked out and it turns out that the three-point pairing is always orthogonal.
In fact there is a rather mechanical explanation: the $x$-space pairing is also proportional to the momentum-space one \cite{Karateev:2018oml}\footnote{This can be proven formally by moving gauge-fixing factors in the frame $(0,x,\infty)$:
\be
\int \frac{d^dx}{{\rm vol}({\rm SO}(d)\times {\rm SO(1,1)})} T(x)\tilde{T}(x)
= \int d^dx \int \frac{d^dp\ d^dp'\ e^{ix{\cdot}(p+p')}}{(2\pi)^{2d}{\rm vol}({\rm SO}(d)\times {\rm SO(1,1)})}
 T(p)\tilde{T}(p')\,.
\ee
The $x$ integral simply gives a delta-function setting $p'=-p$.
}:
\be
 P^{\pa,\pb}_{123} = \fft{1/(2\pi)^d}{2^d{\rm vol(SO)}(d-1)} \sum_{\eps_1,\eps_2,\eps_3}
T^{\pa}_{123}(\eps_1,\eps_2,\eps_3)(p)\ T^{\pb}_{\tilde{1}\tilde{1}\tilde{1}}(\eps_1^*,\eps_2^*,\eps_3^*)(-p)\,.
\ee
Due to this, the diagonal pairing would be rather trivially diagonal
in any $d$, using the momentum space basis discussed above eq.~\eqref{VVmom}.
Without derivation, we thus quote the diagonal $4\times4$ matrix of pairings in the $d=3$ helicity basis \eqref{Tpm}:
\be
 P^{(h_1,h_2),(\bar{h}_1',\bar{h}_2')}_{123}
  = \delta_{h_1}^{h_1'}\delta_{h_2}^{h_2'}  \times P_s \times 4^{|h_1|+|h_2|}(-1)^{|h_1-h_2|}\frac{(\j_3+1)_{|h_1-h_2|}}{(-\j_3)_{|h_1-h_2|}}\,.
\label{pairing general}
\ee
Taking Even/Odd combinations \eqref{even odd} simply adds the $N^{E/O}$ factors,
reproducing the $\j_1=\j_2=1$ example quoted in eq.~\eqref{P11}.
The fact that the pairing is diagonal (with $\bar{h}=-h$) is a first hint that the structures are well chosen.

A second natural and useful operation is the shadow transform
\be
\mathbf{S}[\mathcal{O}_1(x)]\equiv \int d^dy\langle \tilde{\mathcal{O}}_1(x)\tilde{\mathcal{O}}_1(y)\rangle \mathcal{O}_1(y)\,,
\label{shadowtrans}
\ee
which maps operators to their shadow operators nonlocally. Operating on three-point structures this
generally produces a shadow matrix $S([\mathcal{O}_1]\mathcal{O}_2\mathcal{O}_3)^{\pa}\,_{\pb}$:
\be
\langle \mathbf{S}[\mathcal{O}_1]\mathcal{O}_2\mathcal{O}_3\rangle^{\pa}=S([\mathcal{O}_1]\mathcal{O}_2\mathcal{O}_3)^{\pa}\,_{\pb} \langle \tilde{\mathcal{O}}_1\mathcal{O}_2\mathcal{O}_3\rangle^{\pb}\,.\label{shadowmatrix}
\ee
The shadow transform for conserved currents in $d=3$ is simple:
the two-point function in momentum space can be diagonalized by helicity, which is always maximal for conserved currents.
Using $2^{\Delta-\tilde{\Delta}}\mathcal{A}_{j,j}\big|_{\Delta=\j+1}$ from eq.~(E.11) of \cite{Karateev:2018oml}, we get simply
\be
 S([\tilde{\mathcal{O}}_1]\mathcal{O}_2\mathcal{O}_3)^{(h_1',h_2')}\,_{(h_1,h_2)} =
  \delta^{h_1'}_{h_1}\delta^{h_2'}_{h_2} (-4)^{\j_1}\pi^2 \times \mathcal{C}_{\j_1}, \qquad \mathcal{C}_{\j} \equiv \frac{1+\delta_{\j,0}}{2(2\j)!}\,.\label{Seasy}
\ee
This holds when acting on the shadow of a conserved current $\tilde{\mathcal{O}}_1$,
or a scalar with the same twist $\Delta_1=1$.
(We note that $S$ is not invertible and $S[\OO_1]=0$ acting on a conserved current.)
The transform in the Even/Odd basis
is of course also diagonal, but displays additional scalar factors
due to the gamma-functions in \eqref{even odd}.

The shadow transform with respect to $\mathcal{O}_3$ will be technically more difficult to compute;
we will find below (see \eqref{hard shadow}) that it is also diagonal.

\subsection{Spinning conformal blocks}

A more interesting and nontrivial object is the correlator of four operators. The Operator Product Expansion distills those in terms of a given theory's spectrum and OPE coefficients. Using conformal symmetry we can assume the four points
are at $(0,x,y^{-1},\infty)$ (where $y^{-1}$ is the point $y^\mu/y^2$).
Factoring out a conventional prefactor to trivialize the $x\to 0$ and $y\to0$ limits
\be
 \langle \OO_1(0)\OO_2(x)\OO_3(y^{-1})\OO_4(\infty)\rangle = \frac{|y|^{\Delta_3+\Delta_4}}{|x|^{\Delta_1+\Delta_2}} G(z,\zb).
\ee
Our notation $\OO_3(y^{-1})$ implies that we apply inversion tensors
to the indices on the third (and fourth) operator.
The complex variable $z$ (which is complex conjugate to $\zb$ in Euclidean signature)
encodes the sizes and angles of the vectors $x^\mu$ and $y^\mu$:
\be
 z\zb = x^2y^2, \qquad z+\zb = 2x{\cdot}y\ .
\ee
Inserting a complete basis of states between $\OO_1,\OO_2$ gives the operator product expansion
\be
 G(z,\zb) = \sum_{\Delta,\j,\pa,\pb} \lambda_{12\OO\pa}\lambda_{43\OO\pb} G^{\pa,\pb}_{\Delta,\j} (z,\zb)\label{OPEexp}
\ee
where the sum runs over the spectrum of the theory, and the $\lambda$'s are OPE coefficients.
When the external operators have spin, there are generally multiple index contractions $\pa$, $\pb$ to sum over representing
the different three-point structures, each of which has an independent coefficient.
The special functions $G^{\pa,\pb}_{\Delta,\j}(z,\zb)$ are the so-called conformal blocks,
which we normalize
so they approach as $x\to 0$
a simple product of three-point structures (summing over the polarizations of the intermediate operator $\OO$):
\be
 \lim_{x\to 0} G^{\pa,\pb}_{\Delta,\j} (z,\zb) =
 \frac{x^{\Delta_1+\Delta_2}}{y^{\Delta_3+\Delta_4}}\sum_{\eps_\OO} T_{12\OO}^{\pa}(x) T_{43\OO}^{\pb}(y) \equiv
  \mathcal{P}_{\Delta,\j}^{\pa,\pb}(x,y) \label{limit of block}
\ee
For example,
for scalar external operators in our normalization \eqref{T00} one finds
\be
\mathcal{P}_{\Delta,\j}(\hat{x},\hat{y}) = (|x||y|)^\Delta \frac{(d-2)_\j}{\big(\tfrac{d-2}{2}\big)_\j} \tilde{C}_\j\big(\tfrac{x{\cdot}y}{|x||y|}\big) \to_{z\ll \zb\ll 1} (z\zb)^{\Delta/2} (z/\zb)^{\j/2}
\label{gegen}
\ee
where $\tilde{C}_j(\xi) =C_\j(\xi)/C_\j(1) ={}_2F_1(-\j,\j+d-2,\tfrac{d-1}{2},\tfrac{1-\xi}{2})$ is a Gegenbauer normalized to unity at
$\xi=1$.  In terms of cross-ratios, $\tfrac{x{\cdot}y}{|x||y|}=\tfrac{z+\zb}{2\sqrt{z\zb}}$.

The conformal block $G$ contains an infinite tower of terms suppressed by powers of x (or $z,\zb$),
arising from exchange of descendants $\partial^k\OO_\Delta$.  Series expansions for these terms are available from
refs~\cite{Hogervorst:2013sma,Costa:2016xah,Fortin:2019gck}, as well as an efficient Zamolodchikov recursion algorithm, see \cite{Erramilli:2019njx,Erramilli:2020rlr}.
In practice we will use the spinning up/spinning down method.  We write the spinning block as a derivative of a scalar one,
\be \label{spin up}
G^{a,b}_{\Delta,\j}=\mathbb{P}^{\pa}_{(\alpha)}\mathbb{P}^{\pb}_{(\beta)}\mathcal{D}_{\uparrow}^{(\alpha,\beta)}G^{(\alpha,\beta)}_{\Delta,\j}\,.
\ee
Let us explain our notation here. The indices $\alpha, \beta,\cdots$ span the space of spinning-up operators (see eq.~\eqref{spin-ups} below), so that the $\mathbb{P}^{\pb}_{(\beta)}$ are constant matrices, that depend only on $\Delta, \j$ but not on spacetime coordinates;
$\tilde{G}^{(\alpha,\beta)}$ is a scalar conformal blocks, where the superscripts denote the specific shift of conformal dimensions associated with the particular spinning-up operator $(\alpha, \beta)$.
Explicit operators will be written in section \ref{MFT from Lorentzian} below;
a simple recursion for scalar conformal blocks is reviewed in appendix \ref{series}.

\subsection{Euclidean inversion formula}

The OPE sum runs over the spectrum of the theory, which we generally don't know exactly.
For analytics it is often better to replace the sum by an integral, the ``harmonic analysis'':
\be
 G(z,\zb) = \sum_{\j,\pa,\pb}\frac12\int_{d/2-i\infty}^{d/2+i\infty} \frac{d\Delta}{2\pi i} c_{\pa,\pb}(\Delta,\j)
 \left(G^{\pa,\pb}_{\Delta,\j} (z,\zb) + \mbox{shadow}\right).\label{Gint}
\ee
The ``shadow term'' is the same block with $\Delta\mapsto \tilde{\Delta}=d-\Delta$ and with a specific coefficient, see \cite{SimmonsDuffin:2012uy,Costa:2012cb}. This shadow term ensures that the
parenthesis is Euclidean  single-valued (i.e. does not have a branch cut) in the limits $x\to y^{-1}$ and $x\to \infty$.
Explicitly, this term is
\be
S(\mathcal{O}_1\mathcal{O}_2[\mathcal{O}])^{\pa}\,_{\pp{c}}\ (S(\mathcal{O}_3\mathcal{O}_4[\tilde{\mathcal{O}}])^{-1})^{\pb}\,_{\pp{d}}\ G^{\pp{c},\pp{d}}
_{\tilde{\Delta},\j}\,.
\ee

To obtain the OPE (\ref{OPEexp}) from the integral (\ref{Gint}) one simply closes the contour to the right in the $G$ term, and the formulas will match
provided
\be
 {-}\!\mathop{\rm Res}\limits_{\Delta'\to \Delta} c_{\pa,\pb}(\Delta,\j) = \lambda_{12\OO\pa}\lambda_{43\OO\pb}.
\label{residue}
\ee
The function $c_{\pa,\pb}(\Delta,\j) $ will be useful below since it simultaneously encodes the spectrum (through the location of its poles) and OPE coefficients (through the residues); this enables one to speak about OPE coefficients without having to first know the spectrum.

As single-valued eigenfunctions of a Casimir differential operator, the harmonic functions satisfy an orthogonality relation
\be
 \int \frac{d^dx_1\cdots d^d x_4}{{\rm vol(SO(}d+1,1))} \langle 1234\rangle_{\Delta,\j}^{\pa,\pb}
\langle \tilde1\tilde2\tilde3\tilde4\rangle_{\tilde{\Delta},\j}^{\pp{c},\pp{d}}
 = (\mathcal{N}(\Delta,\j)_{(a,b),(c,d)})^{-1}
\ [2\pi\delta(\nu-\nu')+\mbox{shadow}],\label{orthpartial}
\ee
where $\Delta=\tfrac{d}{2}+i\nu$ and the tildes denote shadow operators; tensor indices are meant to be contracted between each operator and its shadow. Note that
we abbreviate $(G_{\Delta,\j}+{\rm shadow})$ as $\langle1234\rangle_{\Delta,\j}$.
The symmetry can be used to fix the points to $(0,x,1,\infty)$ so the integral is really just over $x$.
The normalization $\mathcal{N}(\Delta,\j)$ can be expressed in terms of the
pairing $P^{(a,b)}$ of eq.~(\ref{pairing general}), since the $\delta$-function originates from the $x\to 0$ limit, where the blocks can be approximated by their limit (\ref{limit of block}).
Explicitly, the normalization reads \cite{Karateev:2018oml}
\bea \label{curly N}
\mathcal{N}(\Delta,\j)_{(a,b),(c,d)}=\mu(\Delta,\j)\
(P^{\pa,\pp{c}}_{12\OO})^{-1}\
 (P^{\pp{b'},\pp{d'}}_{34\tilde\OO})^{-1}\ S(34[\tilde\OO])^{b'}{}_{b}\ S(\tilde{3}\tilde{4}[\OO])^{d'}{}_d\,,
\eea
where the ``Plancherel measure'' is
\begin{small}
\be
\mu(\j,\Delta)=\frac{(d+2 J-2) \Gamma (d+J-2)\Gamma(\Delta-1) \Gamma(d-\Delta-1)(\Delta +J-1)(d-\Delta +J-1)}{2^d \pi^{d}{\rm vol}({\rm SO}(d))\Gamma (d-1) \Gamma (J+1) \Gamma(\frac{d}{2}-\Delta) \Gamma(\Delta -\frac{d}{2})}\,.
\ee
\end{small}
Evaluated in terms of cross-ratios, this gives an integral over the complex-$z$ plane\footnote{
We used eq.~\eqref{pairing def} and the relation
$\frac{{\rm vol(SO}(d-1))}{{\rm vol(SO}(d-2))}={\rm vol} S_{d-2}$ to write, for any conformal function
$(\cdots)$:
\be \int \frac{d^dx_1\cdots d^d x_4}{{\rm vol(SO(}d+1,1))} \frac{(\cdots)}{x_{12}^{2d} x_{34}^{2d}}
 = \frac{1}{2^{2d-1}{\rm vol(SO(}d-2))} \int \frac{d^2z}{z^2\zb^2} \left|\frac{z-\zb}{z\zb}\right|^{d-2}(\cdots)\,.
\ee}
\bea
 c_{\pa,\pb}(\Delta,\j) &=&
\frac{\mathcal{N}(\Delta,\j)_{(a,b),(c,d)}}{2^{2d-1}{\rm vol(SO(}d-2))}
\int \frac{d^2z}{z^2\zb^2} \left|\frac{z-\zb}{z\zb}\right|^{d-2}
\left(\tilde{G}^{\pp{c},\pp{d}}_{d-\Delta,\j} (z,\zb) + \mbox{non-shadow}\right)  G(z,\zb)
\cr &&\label{Eucross}
\eea 
where index contractions with $G(z,\zb)$ is again implied.
To extract the spectrum using this formula one would have to know the exact correlator $G(z,\zb)$, which of course is impractical unless one already has solved the theory. The usefulness of this formula is that it provides analytic estimates for the OPE data in certain limits.
Specifically,  following \cite{Karateev:2018oml}
we will use this formula to extract OPE data in mean field theory in section \ref{GFF}.

\subsection{Spinning Lorentzian inversion formula}

An effective method to go beyond MFT is to analytically continue the Euclidean inversion formula to Lorentzian signature, which
gives the Lorentzian inversion formula \cite{Caron-Huot:2017vep,Simmons-Duffin:2017nub,Kravchuk:2018htv}.
It expresses OPE data as a sum of so-called $t$- and $u$-channel double-discontinuities.

A practical advantage relevant for the present paper is that at tree-level
in theories with a large-$N$ expansion, the double-discontinuity is saturated by single-trace exchanges
\cite{Alday:2017vkk,Caron-Huot:2018kta}, effectively giving AdS cutting rules (see also \cite{Meltzer:2019nbs,Ponomarev:2019ofr}).

The formula was generalized to the spinning case in ref.~\cite{Kravchuk:2018htv}.
The $t$-channel contribution is given as:
\be
c^t_{\pa,\pb}(\Delta,\j)=\mathcal{N}_{(a,b),(c,d)}^L \int_0^1 \frac{dzd\zb}{z^2\zb^2} \left|\frac{z-\zb}{z\zb}\right|^{d-2}\tilde{G}^{\pp{c},\pp{d}}_{J+d-1,\Delta-d+1}(z,\bar{z}) {\rm dDisc}[G(z,\bar{z})]\,,\label{Lorentzian}
\ee
where the tilde denotes that the external operators are shadow operators, and the tensor indexes are contracted between $\tilde{G}$ and ${\rm dDisc}[G]$.
A key result of ref.~\cite{Kravchuk:2018htv} is an elegant way to calculate the normalization factor $\mathcal{N}^L$, which is generally a matrix, in terms of ``light-transforms''.
The light-transform of a spinning operator is defined as
\be
{\bf{L}}[\mathcal{O}](x,\epsilon)=\int_{-\infty}^{\infty}d\alpha (-\alpha)^{-\Delta-J}\mathcal{O}(x-\fft{\epsilon}{\alpha},\epsilon)\,.
\ee
(Despite appearance, the integral has no branch point at $\alpha=0$ due to the behavior of $\mathcal{O}$.
We refer to \cite{Kravchuk:2018htv} for further details on the precise branch choices, which we will ignore in this presentation.)
When the light-transform acts on the third operator of a three-point function, it simply induces a Weyl reflection for that operator $(\Delta\mapsto 1{-}J, J\mapsto 1{-}\Delta)$ with an overall light-transform matrix, i.e.
\be
\langle\mathcal{O}_1\mathcal{O}_2{\bf{L}}[\mathcal{O}_{\Delta,\j}]\rangle^{\pa}=L^{\pa}_{\pb}(\mathcal{O}_1\mathcal{O}_2[\mathcal{O}])
\langle\mathcal{O}_1\mathcal{O}_2\mathcal{O}_{1-\Delta,1-J}\rangle^{\pb}\,.\label{lighttrans}
\ee
We found that the integrand can be computed directly in our frame $(0,x,\infty)$,
using a special conformal transformation along the direction $\eps_3$ to keep $x_3=\infty$ and move $x_2$ instead.
This reduces to a simple substitution:
\be
x\mapsto x+\fft{x^2}{\alpha}\epsilon_3\,,\quad \epsilon_2\mapsto \epsilon_2+\Big(x\cdot\epsilon_2-\fft{x^2\epsilon_2
\cdot\epsilon_3}{\alpha+2x\cdot\epsilon_3}\Big)\fft{2\epsilon_3}{\alpha}-\fft{2\epsilon_2\cdot\epsilon_3}{\alpha+2x\cdot\epsilon_3}x
\ee
and we have to multiply three-point functions by $\left(1+2x\.\eps_3/\alpha\right)^{\Delta_1-\Delta_3}$.
Using this rule, and integrating over $\alpha$ following ref.~\cite{Kravchuk:2018htv}, we find that
the light-transform matrix in the Even/Odd basis is actually independent of $J_1$ and $J_2$:
\be
L^{E}(J_1J_2[\mathcal{O}_{\Delta,\j}])=L_s^{(0)}
\left(\begin{array}{c@{\hspace{2mm}}c} 0&1\\1&0\end{array}\right)
\,,\quad L^{O}(J_1J_2[\mathcal{O}_{\Delta,\j}])=L_s^{(1)}\left(\begin{array}{c@{\hspace{2mm}}c} 0&1\\1&0\end{array}\right)\,,
\label{light transform matrix}
\ee
where $L_s^{(\Delta_{12})}$ denotes the scalar light transform \cite{Kravchuk:2018htv}
 \be
L_s^{(\Delta_{12})}= -i\frac{2^{\fft{3-\beta}{2}}\pi\Gamma(\beta-1)}{\Gamma\big(\fft{\beta-\Delta_{12}}{2}\big)\Gamma\big(\fft{\beta+\Delta_{12}}{2}\big)}\,.
\ee
We note that the light transform is not diagonal.
Attempting to transform from the Even/Odd basis to the helicity basis (via eq.~\eqref{even odd})
would produce a matrix that is not only non-diagonal, but also dense.
The reason the light transform does not commute with helicity
is that its calculation requires taking a discontinuity, which does not commute with $h$, as
found at the end of subsection \ref{ssec:helicity is conf}.  We will therefore work in the Even/Odd basis,
where the simple form of eq.~\eqref{light transform matrix} will enable us to
write the Lorentzian inversion formula very explicitly below.

The remaining ingredient is the
inverse of a ``Lorentzian'' pairing between three-point structures \cite{Kravchuk:2018htv},
which reads in the $x=1$ gauge:
\bea
P_{12[\OO],L}^{\pa,\pb}= 
\fft{(-2\epsilon_3\cdot 1)^{d-2}}{2^{2d-2}{\rm vol}({\rm SO}(d-2))}\sum_{\epsilon_1,\epsilon_2}
T^{\pa}_{123}(\epsilon_1,\epsilon_2,\epsilon_3)
T^{\pb}_{\tilde{1}\tilde{2}3^{\bf S}}(\epsilon_1^\ast,\epsilon_2^\ast,\epsilon_3)(1)\,.
\eea
The tilde denotes the shadow, and the superscript ${\bf S}$ denotes the full shadow where both the scaling dimension and the spin are reflected $(\Delta\mapsto d{-}\Delta, J\mapsto 2{-}d{-}J)$. Similarly to the Euclidean pairing discussed above,
we find that it is nicely diagonal in the even/odd basis:
\bea
P_{J_1J_2[\mathcal{O}_{\bf L}],L}^{E}= (-4)^{J_1+J_2}P_{s,L}N^E_{J_1J_2\mathcal{O}_{\bf L}}\times\mathbb{I} \,,\quad P_{J_1J_2\mathcal{O}_{\bf L},L}^{O}=-(-4)^{J_1+J_2}P_{s,L} N^O_{J_1J_2\mathcal{O}_{\bf L}}\times\mathbb{I}\,,\qquad
\eea
where the factor $N^{E/O}$ is defined in eq.~\eqref{NPEuc} and the subscript ${\bf L}$ denotes Weyl reflection associated with the light-transform $(\Delta{\mapsto} 1{-}\j,\ \j{\mapsto} 1{-}\Delta)$. $P_{s,L}$ is simply the Lorentzian pairing of two scalars and one spinning operator \cite{Kravchuk:2018htv}
\be
P_{s,L}=\fft{(-1)^d 2^{1-\fft{3d}{2}}}{{\rm vol}({\rm SO}(d-2))}\,.
\ee
The normalization $\mathcal{N}^L_{(a,b),(c,d)}$ in the Lorentzian inversion formula \eqref{Lorentzian} is then given as
\be
\mathcal{N}^L_{(a,b),(c,d)}=\fft{1}{2^{\Delta+\j}(\Delta+\j-1)}
\hat{L}_{\pa,\pp{c}}(\mathcal{O}_1\mathcal{O}_2[\mathcal{O}_{J,\Delta}])
\hat{L}_{\pb,\pp{d}}(\mathcal{O}_3\mathcal{O}_4[\mathcal{O}_{J,\Delta}])\,,\label{NormLoren}
\ee
where $\hat{L}$ is a sort of inverse of the light transform with respect to the pairing:
\be
\hat{L}_{\pa,\pp{c}}(\mathcal{O}_1\mathcal{O}_2[\mathcal{O}_{J,\Delta}])
L^{\pp{d}}_{\pp{e}}(\mathcal{O}_1\mathcal{O}_2[\mathcal{O}_{J,\Delta}])
P_L^{\pp{c},\pp{e}}(\mathcal{O}_1\mathcal{O}_2[\mathcal{O}_{1-\Delta,1-J}])=-i\delta_{\pa}^{\pp{d}} P_{s,L}\,.\label{Lhat}
\ee
For scalars, it is straightforward to verify that the above expression reduces to
\be
\mathcal{N}^L_s = \frac14\kappa_{\beta}^{(\Delta_{12},\Delta_{34})},
\qquad
\kappa_{\beta}^{(\Delta_{12},\Delta_{34})}=\fft{\Gamma(\fft{\beta-\Delta_{12}}{2})\Gamma(\fft{\beta+\Delta_{12}}{2})\Gamma(\fft{\beta-\Delta_{34}}{2})\Gamma(\fft{\beta+\Delta_{34}}{2})}{2\pi^2 \Gamma(\beta-1)
\Gamma(\beta)}\,.
\ee
More generally, we can write explicitly the normalization factor in the spinning
Lorentzian inversion formula \eqref{Lorentzian} in the Even/Odd basis:
\be \label{Lorentzian norm}
 \mathcal{N}^L_{(a,b),(c,d)}= \frac14\frac{\kappa_\beta^{(\Delta_{12},\Delta_{34})}(-4)^{-\sum_{i=1}^4 J_i}}
 {N^{E/O}_{J_1J_2\mathcal{O}_{\bf L}} N^{E/O}_{J_3J_4\mathcal{O}_{\bf L}} }
 \left(\begin{array}{c@{\hspace{2mm}}c} 0&1\\1&0\end{array}\right)_{a,c}
 \left(\begin{array}{c@{\hspace{2mm}}c} 0&1\\1&0\end{array}\right)_{b,d}
\ee
where $a$ and $c$ must have the same parity, as well as $b$ and $d$.
We set $\Delta_{12}=0$ if $a,b$ are even and $\Delta_{12}=1$ if they are both odd, and similarly
for $\Delta_{34}$.  (If operator 1 or 2 is a scalar, there is only one structure $a$ and we drop the corresponding matrix.)

Performing (\ref{Lorentzian}) is a bit challenging because generally evaluating the spinning conformal blocks is a hard task. A nice idea, following ref.~\cite{Karateev:2017jgd},
is to ``integrate-by-parts'' the spin-up from eq.~\eqref{spin up} acting on the block
to get instead spinning-down operators acting on the correlator:
\be
c^t_{a,b}(\Delta,\j)=\mathcal{N}_{(a,b),(c,d)}^L \int \frac{dzd\zb}{z^2\zb^2} \left|\frac{z-\zb}{z\zb}\right|^{d-2}\tilde{G}^{(\alpha,\beta)}_{J+d-1,\Delta
-d+1}(z,\bar{z}) {\rm dDisc}[\mathbb{P}^{c}_{\alpha}\mathbb{P}^{d}_{\beta}\mathcal{D}^{(\alpha,\beta)}_{\downarrow}G(z,\bar{z})]\,,\label{refLoren}
\ee
which will effectively reduce us to the scalar Lorentzian inversion formula.
Eq.~\eqref{refLoren 1} below gives a concrete expression in a specific basis of spin-down operators.

\section{OPE data for spinning Generalized Free Fields}
\label{GFF}

\subsection{From Euclidean inversion and shadow representation}

Using the shadow transform, the OPE data in MFT can be efficiently evaluated by the Euclidean inversion formula \cite{Karateev:2018oml}. It is especially effective for three-point functions in momentum-space. To use this, it is best to write the Euclidean inversion formula \eqref{Eucross} in a covariant way
\be
c_{\pa,\pb}(\Delta,\j)= \hat{\mathcal{N}}(\Delta,\j)_{(a,b),(c,d)}
\int \fft{d^dx_1\cdots d^dx_4}{{\rm vol}({\rm SO}(d+1),1)}\langle 1234\rangle
(\Psi^{\tilde{\Delta}_i}_{\tilde{\Delta},\j})^{\pp{c},\pp{d}}\,,
\label{EucinvP}
\ee
where the tildes denote shadow operators.
The factor $\hat{\mathcal{N}}$ is the same as $\mathcal{N}$ in eq.~\eqref{curly N} but with
the factor $S(\tilde{3}\tilde{4}[O])$ dropped ({\it ie.} replaced by identity).
The harmonic function $\Psi^{\Delta_i}_{\Delta,\j}$ is a combination of block and shadow,
which, importantly, can be written as integral of two three-point functions (this is called the shadow representation):
\bea
(\Psi^{\Delta_i}_{\Delta,J})^{\pa,\pb}&=&S(34[\tilde{\mathcal{O}}])^{\pb}\,_{\pp{c}}\ (G_{J,\Delta}^{\pa,\pp{c}}+{\rm shadow})
\\ &=& \int d^dx\ \langle 12\mathcal{O}(x)\rangle^{\pa}\langle \tilde{\mathcal{O}}(x)34\rangle^{\pb}\,.
\eea
We now consider a Mean Field Theory four-point function:
\be
\langle 1234\rangle=\langle12\rangle\langle34\rangle+\langle13\rangle\langle24\rangle+\langle14\rangle\langle23\rangle\,.\label{meanfield}
\ee
We focus on the $t$-channel contribution $\langle23\rangle\langle14\rangle$ to illustrate the algorithm for computing the OPE data of MFT. The $u$-channel contributions $\langle13\rangle\langle24\rangle$ can be evaluated in the same way, while the $s$-channel $\langle12\rangle\langle34\rangle$ is trivial to evaluate: it contributes to the identity exchange. Considering the term $\langle23\rangle\langle14\rangle$, the integrals over $x_3, x_4$ in (\ref{EucinvP}) boil down to the shadow-transform for $\tilde{3}$ and $\tilde{4}$, and the remaining integrals are all removed by the gauge-fixing, leaving a simple pairing \cite{Karateev:2018oml}:
\be \label{cMFT}
 c_{\pa,\pb}^{t,\rm MFT}(\Delta,J) = \mu(\Delta,J) (P_{34\tilde{\mathcal{O}}}^{\pa,\pp{c}})^{-1}
 S([\tilde{1}]\tilde{2}\tilde{\mathcal{O}})^{\pp{c}}{}_{\pp{d}}
 S(1[\tilde{2}]\tilde{\mathcal{O}})^{\pp{d}}{}_{\pp{e}}
 S(12[\tilde{\mathcal{O}}])^{\pp{e}}{}_{\pb}\,.
\ee
This formula breaks the calculation of MFT coefficients into simple algebraic operations: three shadows and a pairing.

The pairing and first two shadows were presented earlier in subsection \ref{ssec:simple ops}.
Before we calculate the third shadow, let us revisit the shadow transform, defined in eq.~\eqref{shadowtrans}. It can be computed algebraically as multiplication in momentum space:
\be \label{shadow as mul}
\sum_{\epsilon_1'}K_{\tilde{1}\tilde{1}'}(p)T_{1'23}(p)^{\pa}=S([\mathcal{O}_1]\mathcal{O}_2\mathcal{O}_3)^{\pa}\,_{\pb} T_{\tilde{1}23}(p)^{\pb}\,,
\ee
where $K_{11'}$ is the Fourier transform of the two-point function of $\mathcal{O}_1$ \cite{Isono:2018rrb}:
\bea
&& K_{11'}(p)=\sum_{k=0}^J \mathcal{K}_{k}(\Delta,J)(\epsilon_1\cdot p)^k (p\cdot\epsilon_1')^k (\epsilon_1\cdot\epsilon_1')^{J-k}|p|^{2\Delta-d-2k}\,,
\cr &&
 \cr && \mathcal{K}_k(\Delta,J)=\frac{\pi ^{d/2} \Gamma (J+1) 2^{d-2 \Delta +k} \Gamma \left(\frac{d}{2}+k-\Delta \right) \Gamma (J-k+\Delta -1)}{\Gamma (\Delta -1) \Gamma (k+1) \Gamma (J+\Delta ) \Gamma (J-k+1)}\,.\label{twoptgene}
\eea
Applying this map to the helicity structures normalized as in eq.~\eqref{Tpm},
we find the simple result
\be \label{Seasy 1}
K_{11'}(p)T_{\tilde{1}'2\mathcal{O}}^{\pm,\pm}(p) = \frac{\pi^{\frac32}\Gamma(\tfrac32-\Delta_1)}{(\Delta_1+\j_1-1)\Gamma(\Delta_1-1)}\times T_{12\mathcal{O}}^{\pm,\pm}(p)\,,
\ee
which reproduces as $\Delta_1\to\j_1+1$  the formula for conserved current in eq.~(\ref{Seasy}).

The third shadow transform is technically more difficult to evaluate since we defined our structures in a frame where $x_3=\infty$.  The trick is to use the rule eq.~(\ref{exinforder}) to interchange
$x_1$ and $x_3$ in position space, then Fourier transform back to the momentum space, where we can apply eq.~\eqref{shadow as mul}.
These steps are somewhat lengthy (we found the Fourier transform \eqref{Fourier spinors} helpful),
but thankfully the last step turns out to simply multiply each structure by an overall factor.
This had to be the case since the shadow transform commutes with $h_1$ and $h_2$.
Trying a few cases we observe a simple pattern:
\be
S(12[\mathcal{O}])^{h_1,h_2}{}_{h_1',h_2'}=\delta^{h_1}_{h_1'}\delta^{h_2}_{h_2'}\
\fft{4^{2\Delta-3}\pi^{\fft{3}{2}}\Gamma(2+\j-\Delta)\Gamma(\Delta-\fft{3}{2})}{\Gamma(\Delta-1)\Gamma(\Delta+\j)}\times \fft{(2-\Delta)_{|h_1+h_2|}}{(\Delta-1)_{|h_1+h_2|}}\,.
\label{shadow T}
\ee
Combining the shadows \eqref{Seasy 1} and \eqref{shadow T} with the pairing \eqref{pairing general}
thus gives MFT coefficients \eqref{cMFT}:
\be\label{cMFT T}\begin{aligned}
c^{t,\rm MFT}_{h_1,h_2,\bar{h}_3,\bar{h}_4}(\Delta,\j)
&= \delta_{h_1}^{h_4}\delta_{h_2}^{h_3}\
2^{5-4\Delta}\pi \frac{\Gamma(\Delta-1)}{\Gamma(\Delta-\tfrac32)}
\frac{\Gamma(\Delta+\j)}{\Gamma(\j-\Delta+2)}\frac{\Gamma(\j+\tfrac32)}{\Gamma(\j+1)}\mathcal{C}_{\j_1}\mathcal{C}_{\j_2}
\\ &\quad\times
\frac{(-\j)_{|h_1-h_2|}}{(\j+1)_{|h_1-h_2|}}\frac{(\Delta-1)_{|h_1+h_2|}}{(2-\Delta)_{|h_1+h_2|}}\,,
\end{aligned}\ee
where the constant $\mathcal{C}_{\j}$ is defined in eq.~\eqref{Seasy}.
The u-channel identity (if operators $1$ and $3$ are identical) gives the same result times $(-1)^\j$
and with $h_3$ and $h_4$ swapped.

Eq.~\eqref{cMFT T} can be used in the harmonic decomposition \eqref{Gint}.  Where are the poles and corresponding OPE data?
To read off the local OPE data, we have to keep in mind that tensor structures in the helicity basis have poles at double-twist locations. To find OPE data from residues, it is best to convert to the Even/Odd basis defined in eq.~\eqref{even odd}, in which the position-space structures do not have poles.
Performing the rotation, we get extra gamma-functions which nicely combine to give scalar MFT coefficients, times the same matrix in the even and odd cases:
\be
c^{t,{\rm MFT},\rm E/O}(\Delta,J)=c^{\rm E/O}(\Delta,J)^s\left(
\begin{array}{cc}
\frac{(-J)_{J_1+J_2} (\Delta -1)_{|J_1-J_2|}}{(J+1)_{J_1+J_2} (2-\Delta )_{|J_1-J_2|}}  & 0 \\
 0 & \frac{(-J)_{|J_1-J_2|} (\Delta -1)_{J_1+J_2}}{(J+1)_{|J_1-J_2|} (2-\Delta )_{J_1+J_2}} \\
\end{array}
\right)\mathcal{C}_{\j_1}\mathcal{C}_{\j_2}\,,\label{free data}
\ee
where we normalized it by the OPE data for scalars of twist 1 or 2
in the even and odd cases, more precisely:
\be
c^{\rm E}(\Delta,\j)^s=c(1,1;\Delta,\j)^s\,,\quad c_J^{\rm O}(\Delta)^s=\tfrac{1}{2}
c(1,2;\Delta,\j)^s\,.
\ee
The scalar MFT data  $c(\Delta_1,\Delta_2;\Delta,\j)^s$ can be found from earlier literature \cite{Fitzpatrick:2011dm} and is recorded in
eq.~\eqref{GFF integral 0} (with $p=\Delta_1+\Delta_2,\ a=b=\tfrac{\Delta_2-\Delta_1}{2}$).

For future reference, let us summarize all the ingredients in the Even/Odd basis.
The products of ``easy'' shadows,
$S([\tilde{1}]\tilde{2}\tilde{\mathcal{O}})S(1[\tilde{2}]\tilde{\mathcal{O}})$, are given as
\bea
\mathbb{S}^{E}=\mathbb{S}_s^E N^E_{\j_1\j_2\tilde{\mathcal{O}}}(-4)^{\j_1+\j_2} \times \mathcal{C}_{\j_1}\mathcal{C}_{\j_2}\mathbb{I}\,,
\quad
\mathbb{S}^{O}=\ft{1}{2}\mathbb{S}^{O}_s N'_{\j_1\j_2\tilde{\mathcal{O}}}(-4)^{\j_1+\j_2+1} \times \mathcal{C}_{\j_1}\mathcal{C}_{\j_2}\mathbb{I}\,,
\eea
where $\mathbb{S}^{E/O}_s$ are just the scalar factor for $(\Delta_1,\Delta_2)=(1,1)$ and $(1,2)$ respectively \cite{Kravchuk:2018htv}
\be
\mathbb{S}^{E}=\fft{4\pi^4}{(1-\beta)(\tau-2)}\,,\quad \mathbb{S}^{O}=-2\pi^4\,.
\ee
The third shadow \eqref{shadow T} yields
\bea
S^{E/O}(12[\mathcal{O}])=S_s(12[\mathcal{O}])\left(
\begin{array}{cc}
\frac{(2-\Delta )_{|J_1-J_2|}}{(\Delta -1)_{|J_1-J_2|}} & 0 \\
 0 & \frac{(2-\Delta )_{J_1+J_2}}{(\Delta -1)_{J_1+J_2}}\\
\end{array}
\right)\,,\label{hard shadow}
\eea
with the same matrix for both even and odd, and
where $S_s$ is just the shadow coefficients of scalars \cite{Liu:2018jhs,Karateev:2018oml}
\be
S_s(12[\mathcal{O}])=\frac{\pi ^{d/2} \Gamma(\Delta -\frac{d}{2}) \Gamma (\j+\Delta -1) \Gamma\big(\ft{1}{2}(\j+\tilde{\Delta}+\Delta_{12})\big) \Gamma\big(\ft{1}{2}(\j+\tilde{\Delta} -\Delta_{12}\big)}{\Gamma(\Delta -1) \Gamma\big(\ft{1}{2}(\j+\Delta +\Delta_{12})\big)\Gamma\big(\ft{1}{2} (\j+\Delta -\Delta_{12})\big)\Gamma(\j+\tilde{\Delta})}
\ee
with $\Delta_{12}=0$ for parity-even and $\Delta_{12}=1$ for parity-odd cases.
Finally, the pairing \eqref{pairing general}:
\be
 P^{E/O}_{12\OO}
  = \delta_{h_1}^{h_1'}\delta_{h_2}^{h_2'}  \times P_s \times N_{12\OO}^{E/O}
  4^{|h_1|+|h_2|}(-1)^{|h_1-h_2|}\frac{(\j_3+1)_{|h_1-h_2|}}{(-\j_3)_{|h_1-h_2|}}\,.
\label{pairing general E/O}
\ee
Multiplying these ingredients again according to \eqref{cMFT} gives eq.~\eqref{free data}.

\subsection{OPE data and remarks on the leading trajectory}

Let us now describe the OPE data which stems from eq.~\eqref{free data}.
When computing the integral \eqref{Gint} as a sum of poles, one finds two sorts of terms:
double-twist poles at $\Delta-J=2+2n$ from the gamma-function in eq.~\eqref{free data}, and
spurious poles from the block, at $\Delta-J=3,4,\ldots$.
The position of the latter is set by their kinematical origin as zero-norm descendants (``null states'') of the exchanged primary.

We are in the unfortunate situation that the physical and spurious poles overlap.
In principle, we should subtract the spurious poles using
the results from ref.~\cite{Erramilli:2019njx} for the poles of spinning 3d blocks.
We pursue a simpler, heuristic method, to be justified in the next subsection.
For scalar mean-field-theory with $\Delta_1=\Delta_2=1$,
the poles are simpler and have been discussed in ref.~\cite{Caron-Huot:2017vep}.
Using eq.~(3.9) there, we find that the spurious poles effectively \emph{double}
the OPE coefficient.
On the other hand, the leading trajectory $n=0$ has no corresponding spurious pole and so does not double.

Such a relative factor $\tfrac12$ was also found in the spinning case \cite{Karateev:2018oml}, and so
our tentative guess is that the same happens in our basis and the spurious poles just double the non-leading trajectories, that is:
\begin{subequations}
\bea\label{evenOPEcoeff}
 \lambda_{12\OO}^{E}\lambda_{43\OO}^{E}\big|_{n,\j}&=& -2\!\!\mathop{\rm Res}\limits_{\Delta= 2+2n+\j} c^{E,{\rm MFT}}(\Delta,\j)\nonumber\\
  &=&
\frac{2\mathcal{C}_{\j_1}\mathcal{C}_{\j_2}}{2^{4n+2\j}}
 \frac{(\j+1)_{\frac12}(2n+\j+\tfrac12)_{\frac12}}{(n+\tfrac12)_{\frac12}(n+\j+1)_{\frac12}}
 M(2+2n+\j,\j) \hspace{5mm} (n=1,2,3, \ldots),\qquad
\\
  \lambda_{12\OO}^{O}\lambda_{43\OO}^{O}\big|_{n,\j}&=& -2\!\!\mathop{\rm Res}\limits_{\Delta= 2+2n+\j} c^{O,{\rm MFT}}(\Delta,\j) \nonumber\\
  &=&
\frac{2\mathcal{C}_{\j_1}\mathcal{C}_{\j_2}}{2^{4n+2\j}}
 \frac{(\j+1)_{\frac12}(2n+\j+\tfrac12)_{\frac12}}{(n+1)_{-\frac12}(n+\j+\frac32)_{-\frac12}}
M(2+2n+\j,\j) \hspace{5mm} (n=\tfrac12,\tfrac32,\tfrac52, \ldots)\,,\quad
\cr &&
\eea
\end{subequations}
where $M(\Delta,\j)$ is the $2\times 2$ matrix
\be
M(\Delta,\j)=\left(\begin{array}{cc}
\frac{(-\j)_{\j_1+\j_2}(\Delta-1)_{|\j_1-\j_2|}}{(\j+1)_{\j_1+\j_2}(2-\Delta)_{|\j_1-\j_2|}}
&0\\
0&\frac{(-\j)_{|\j_1-\j_2|}(\Delta-1)_{\j_1+\j_2}}{(\j+1)_{|\j_1-\j_2|}(2-\Delta)_{\j_1+\j_2}}\end{array}
\right).
\ee
Some comments are in order.  We recall that the first structure (opposite-helicity) exists only
for $\j\geq \j_1+\j_2$. This is reflected in an overall zero from $(-\j)_{\j_1+\j_2}$ in the first entry.
Even below this range, the denominator always have fewer zeros than the numerator,
so the vanishing is never ambiguous.  The range of the $\j$-sums is built-in!

The second structure (same-helicity) is more subtle.
It generically exists only for $\j\geq |\j_1-\j_2|$. But since $2-\Delta=-2n-\j$, it may look like the second
entry of the matrix $M$ diverges for the lowest few trajectories.  However, inspection of the
structures $T^{E}_{12\OO}$ reveals that these have corresponding zero for precisely those cases
(a special case  is visible in eq.~\eqref{VVstcor odd} with $n=\frac12,\j=1$).
The conformal blocks thus have a double zero, which shields the singularity from the denominator.
This means that mean-field-theory doesn't have operators at these places.
For $n=0$, we will find below that there is a single leading trajectory.

\begin{figure}
\centering
\begin{subfigure}{0.45\textwidth}
\centering \includegraphics[width=\textwidth]{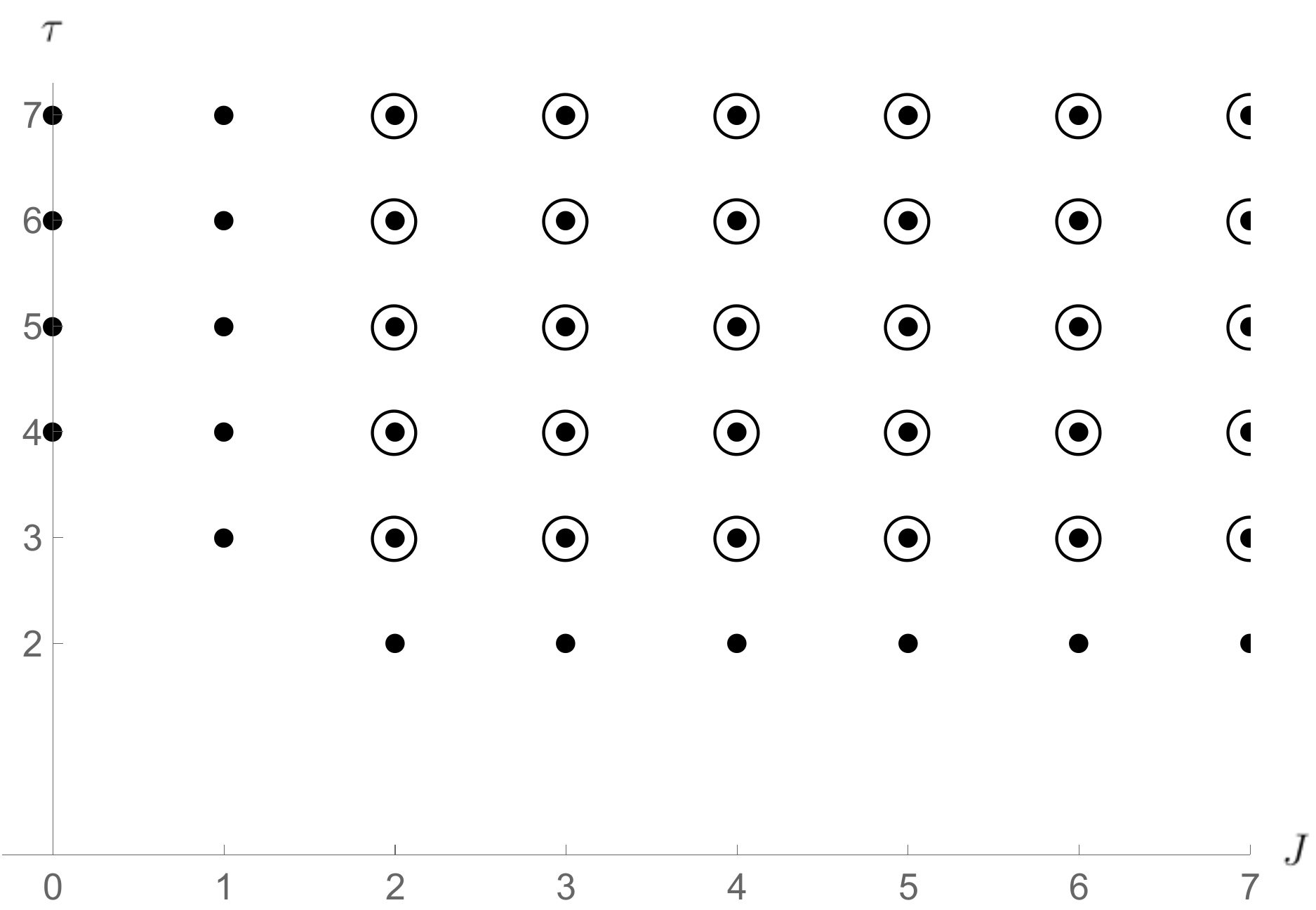}
 \caption{two currents ($\j_1=\j_2=1$)}
\end{subfigure}
\quad
\begin{subfigure}{0.45\textwidth}
\centering \includegraphics[width=\textwidth]{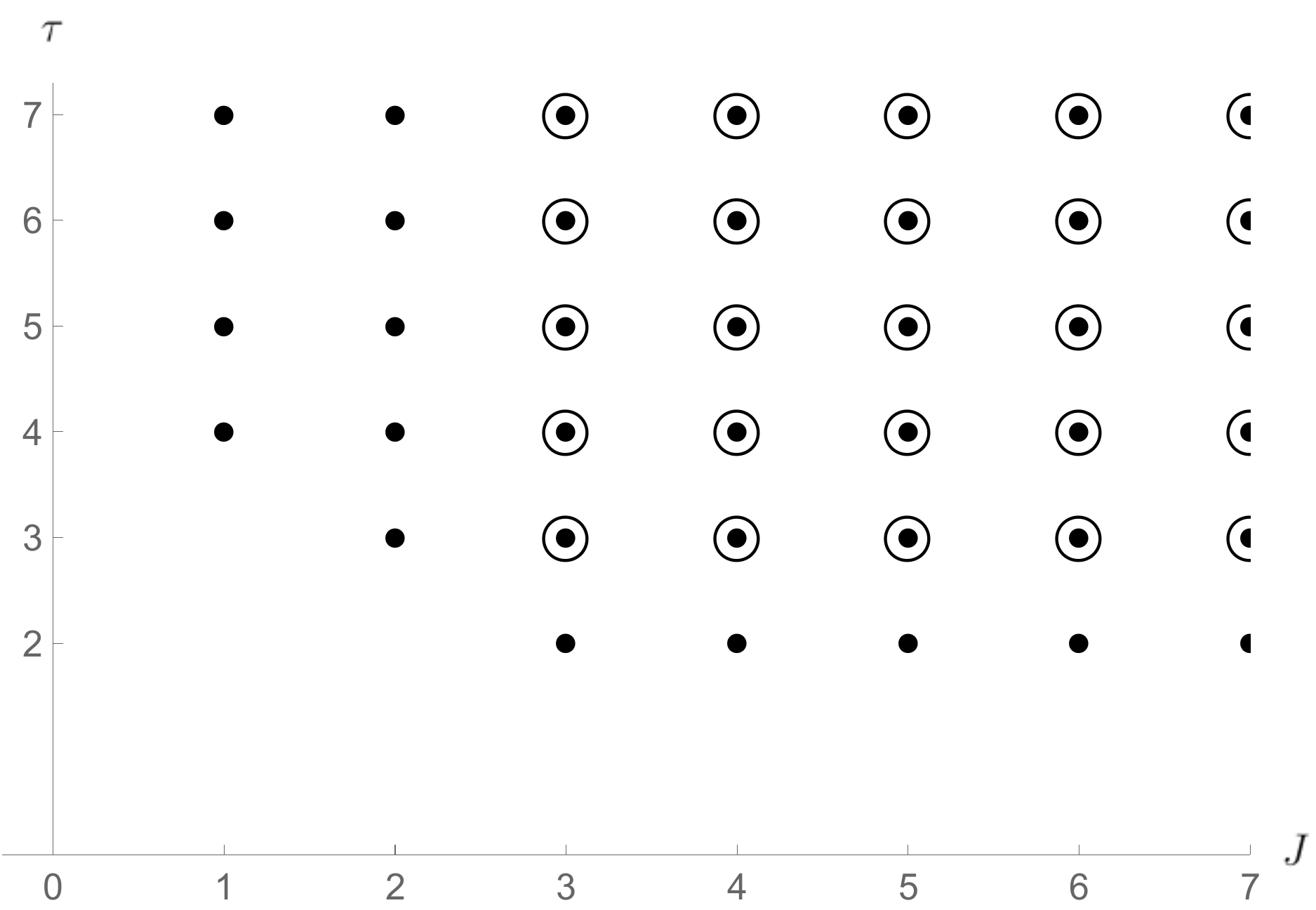}
 \caption{one current and one stress tensor}
\end{subfigure}
\caption{Spectrum of double-twist operators of the form
$[JJ]_{n,J}$ and $[JT]_{n,J}$. Double circles indicate multiplicity:
there is a single trajectory for $n=0$ and two for each $n\geq 1$.
\label{fig:doubletwists}}
\end{figure}

The set of operators appearing in MFT can thus be characterized as:
\begin{itemize}
\item Opposite-helicity: One operator for each $n\geq 0$ and $\j\geq \j_1+\j_2$
\item Same-helicity: One operator for each $n\geq 1$ and $\j\geq \max(|\j_1-\j_2|,\j_1+\j_2-n)$
\end{itemize}
This spectrum is depicted in fig.~\ref{fig:doubletwists}.
(The helicity of the $n=0$ double-twists is really undefined.)

Let us discuss more the leading trajectory, $n=0$. Since there are no spurious poles,
one might think that we should take half the above formula.
This is correct but misleading.
The reason is that when $n=0$ the same- and opposite- helicity structures become
degenerate, as visible from eq.~\eqref{VVstcor}.  Helicity is simply not defined for $n=0$.
One can verify that this happens whenever $\OO_1, \OO_2$ are spinning operators, of any spin.
The resolution is to rotate to a new basis near $n=0$:
\be
\left(\begin{array}{c} T^{E, \rm reg}_{123} \\ T^{E, \rm sing}_{123}\end{array}\right)
=
\left(\begin{array}{c@{\hspace{3mm}}c} 1 & 0\\
 -\frac{(-\j)_{\j_1+\j_2}(\j+1)_{|\j_1-\j_2|}}{(\j+1)_{\j_1+\j_2}(-\j)_{|\j_1-\j_2|}}\frac1n & \frac1n
\end{array} \right)
\left(\begin{array}{c} T^{E, \rm opp}_{123} \\ T^{E, \rm same}_{123}\end{array}\right)\,.
\label{rotation}
\ee
As the two structures degenerate, both combinations are smooth
around $n=0$.  Since the second structure $T^{E, \rm sing}_{123}$ has a non-vanishing double-discontinuity (in fact it has poles $1/x_{12}^2$), its coefficient is guaranteed to vanish in MFT.
The fact that the two structures become $T^{\rm reg}$ effectively doubles the \emph{real}
$n=0$ coefficient. In the rotated basis $(T^{E, \rm reg}_{123}, T^{E, \rm sing}_{123})$,
the leading-trajectory data is thus given by
\be\label{leading trajectory}
 \lambda_{12\OO}^{E,\rm rotated}\lambda_{43\OO}^{E,\rm rotated}\big|_{0,\j}
= \frac{2\Gamma(\j+1)^2}{\Gamma(2\j+1)}\times \mathcal{C}_{\j_1}\mathcal{C}_{\j_2} \frac{(-\j)_{\j_1+\j_2}(\j+1)_{|\j_1-\j_2|}}{(\j+1)_{\j_1+\j_2}(-\j)_{|\j_1-\j_2|}}\left(\begin{array}{cc} 1&0\\0&0\end{array}\right)\,.
\ee
The above fully describes the OPE decomposition of $t$-channel exchange.
To be fully explicit, let us write out the $s$-channel OPE decomposition
of the full MFT correlator including identity in all three-channels,
without any matrix, and including color indices in the case we have several currents:
\bea
\mathcal{G}^{abcd,\rm MFT} &=& \phantom{+}\delta^{ab}\delta^{cd} +
\nonumber \\
&&+ \sum_{n\geq 0}
 \sum_{\j\geq \j_1+\j_2}  \lambda_{12\OO}^{E,\rm same}\lambda_{43\OO}^{E,\rm same}\big|_{n,\j}
\left(\delta^{bc}\delta^{ad}+(-1)^\j \delta^{ac}\delta^{bd}\right)\ G^{E,{\rm same},\ E, \rm same}_{\Delta,\j}
 \nonumber\\ &&
 +\sum_{n\geq 1} \sum_{\j\geq \j_0(n)}
 \lambda_{12\OO}^{E,\rm opp}\lambda_{43\OO}^{E,\rm opp}\big|_{n,\j}
\left(\delta^{bc}\delta^{ad}+(-1)^\j \delta^{ac}\delta^{bd}\right)\ G^{E,{\rm opp},\ E, \rm opp}_{\Delta,\j}
\nonumber\\ &&
+  \sum_{n\geq \frac12}
 \sum_{\j\geq \j_1+\j_2}  \lambda_{12\OO}^{O,\rm same}\lambda_{43\OO}^{O,\rm same}\big|_{n,\j}
\left(\delta^{bc}\delta^{ad}+(-1)^\j \delta^{ac}\delta^{bd}\right)\ G^{O,{\rm same},\ O,\rm same}_{\Delta,\j}
 \nonumber\\
 &&
 +\sum_{n\geq \frac12} \sum_{\j\geq \j_0(n)}
 \lambda_{12\OO}^{O,\rm opp}\lambda_{43\OO}^{O,\rm opp}\big|_{n,\j}
\left(\delta^{bc}\delta^{ad}-(-1)^\j \delta^{ac}\delta^{bd}\right)\ G^{O,{\rm opp},\ O,\rm opp}_{\Delta,\j}\,,
\label{GFF OPE}
\eea
where $\j_0(n)=\max(|\j_1-\j_2|,\j_1+\j_2-n)$ and the $\lambda$'s refer to elements
of \eqref{evenOPEcoeff}.  The last two sums run over half-integer $n$.

Typically, one would further decompose the global symmetry indices into $s$-channel
irreps, and symmetrical versus antisymmetrical combinations. The $t$ and $u$ channels
contributions then effectively remove half the spins (the double-twist operators with the wrong symmetry),
and otherwise effectively double the coefficient.

\label{leadtwist}

Let us cross-check the above MFT spectrum. MFT operators
can be written as products of two operators and their derivatives:
$\partial^\# \OO_1\partial^\#\OO_2$; the game is to enumerate linear combinations that are primaries.
An equivalent exercise is to enumerate three-point structures of the form eq.~\eqref{Tpm} whose Fourier transform are polynomials in $x$. Although finding such explicit polynomials is somewhat cumbersome,
it is straightforward to count them by making a generating function.  We now summarize this exercise.

We make a generating function where a power $q^\Delta z^J$
represents an SO(3) multiplet of dimension $\Delta$ and spin $J$ (that is, $2J+1$ states).
Starting from a scalar operator $\phi$ of dimension $\Delta$, we could characterize
its descendants in terms of symmetric-traceless tensors, times Laplacian:
$(\partial^{\mu_1}\cdots \partial^{\mu_J}-{\rm traces})(\partial^2)^n\phi$,
which contributes a term $q^{\Delta+2n+\j}z^{\j}$.  Summing over $n$ and $\j$
gives a generating function $\frac{q^{\Delta}}{(1-q^2)(1-zq)}$ which enumerates descendants of a scalar.
Omitting steps, we find similar generating functions for the descendants of conserved currents and generic primaries:
\be
 Z_{\j}^{\rm conserved} = \frac{q^{\j+1}z^\j}{(1-q)(1-q z)}\,,\quad
  Z_{\Delta,J}^{\rm generic} = q^{\Delta}\frac{z^{\j} + q (1 + z) \frac{q^{\j} - z^{\j}}{q-z}}{(1 - q^2) (1 - q z)} \,.
\ee
For conserved currents, the dimension-one generator responsible for $\frac{1}{1-q}$ is simply the curl $\vec{\nabla}\times\bullet$, that is, the numerator of eq.~\eqref{higher spin}.
To find the primaries that enter the OPE product of two conserved currents, we have to match the generating functions:
\be
 Z_{\j_1}^{\rm conserved} \times Z_{\j_2}^{\rm conserved} = \sum_{n,\j} c_{n,\j} Z_{2+n+\j,\j}^{\rm generic}
\ee
where the $c$'s are multiplicities of the various representations appearing.
Putting in the multiplicities from fig.~\ref{fig:doubletwists} and comparing the series for various values of $\j_1, \j_2$, we find perfect agreement.

\subsection{From Lorentzian inversion formula}
\label{MFT from Lorentzian}

Beyond MFT, the Euclidean inversion formula is less efficient as
double-twist operators contaminate the cross-channel OPE.
We should thus seek another way to extract the relevant OPE data: using the Lorentzian inversion formula.
As a warm-up, we demonstrate that we can reproduce the above OPE data from the Lorentzian inversion formula, using spinning-down technology.  As we will explain, within this framework it is straightforward to disentangle physical and spurious poles, so this calculation will also confirm the decomposition \eqref{GFF OPE}. In this subsection, we restrict attention to parity-even four currents (``VVVV'') as a concrete example.

In $d=3$, all bosonic conformal blocks can be written as spin-ups of scalar conformal blocks.
In embedding space, a convenient set of spinning-up differential operators is \cite{Costa:2011dw}
\bea \label{spin-ups}
&& D_{ii}^{ij}=Z_i^A\left(
(X_i\cdot X_j)\fft{\partial}{\partial X_j^A}+(X_i\cdot Z_j)\fft{\partial}{\partial Z_j^A}
-X_j^A (X_i\cdot \fft{\partial}{\partial X_j})-Z_j^A (X_i\cdot \fft{\partial}{\partial Z_j})\right)\,,
\cr &&
\cr && D_{ij}^{ij}=Z_i^A\left(
(X_i\cdot X_j) \fft{\partial}{\partial X_i^A}+X_j^A (Z_i\cdot \fft{\partial}{\partial Z_i})
-X_j^A (X_i\cdot \fft{\partial}{\partial X_i})\right)\,,
\cr && D_{iO}^{ij} = \epsilon_{ABCDE}\
Z_i^A X_i^B \frac{\partial}{\partial X_{iC}}
\left( X_j^D \frac{\partial}{\partial X_{jE}}+Z_j^D \frac{\partial}{\partial Z_{jE}}\right)\,.
\eea
$D_{ii}^{ij}$ increases the spin and decreases the conformal dimension of $i$th operator by one unit simultaneously.
On the other hand, $D_{ij}^{ij}$ increases the spin of $i$th operator by one unit and decreases the conformal dimension of $j$th operator by one unit simultaneously, while the odd operator
$D_{iO}$ only changes the first spin but not the dimensions.
Using these operators, (for example) our two parity-even
three-point structures $\langle V_1V_2O\rangle$ can be constructed by acting
on scalar three-point functions $\langle \mathcal{O}_1\mathcal{O}_2\mathcal{O}\rangle$ with five spin-up operators
\bea
&& \langle V_1V_2O\rangle^{\pa}=\mathbb{P}^{\pa}_{(\alpha)}\mathcal{D}^{(\alpha)}_{\uparrow}\langle \mathcal{O}_1\mathcal{O}_2\mathcal{O}\rangle^{(\alpha)}\,,\quad \mathcal{D}^{(\alpha)}_{\uparrow}=\Big(D_{11}^{12}D_{22}^{21}, H_{12},
D_{12}^{12}D_{22}^{21}, D_{21}^{21}D_{11}^{12},D_{12}^{12}D_{21}^{21}\Big)\,,
\cr &&\label{spinupJJO}
\eea
where $H_{12}$ is
\be
H_{12}=2\big((X_1\cdot Z_2)(Z_1 \cdot X_2)-(X_1\cdot X_2)(Z_1 \cdot Z_2)\big)\,.\label{Hij}
\ee
As mentioned previously, it is important to note that the operators act on different three-point functions
$(\alpha)$ as the dimensions $\Delta_1$ and $\Delta_2$ are shifted differently for different operators.
For example, the first and the third structures are actually $(D_{11}^{12}D_{22}^{21},D_{12}^{12}D_{21}^{21})\langle \mathcal{O}_{\Delta_1+1}\mathcal{O}_{\Delta_2+1}\mathcal{O}_{\Delta,J}\rangle$, and the fourth structure is $D_{12}^{12}D_{22}^{21}\langle \mathcal{O}_{\Delta_1}\mathcal{O}_{\Delta_2+2}\mathcal{O}_{\Delta,J}\rangle$.
Each of these can be written as a combination of the five basis monomials in eq.~\eqref{xspace structure}
and ultimately we are interested only in the linear combinations which produce the two conserved
structures in our basis \eqref{VVstcor}.
We find that these combinations, when acting on the ``funny block''
$\tilde{G}^{(c,d)}_{\j+d-1,\Delta-d+1}$, \emph{with external shadow operators} are:
\begin{small}
\bea \label{Pmatrix}
&&\!\!\!\!\! \mathbb{P}^a_\alpha\!= \!\!\left(
\begin{array}{ccccc}
{-}\ft{\sqrt{2}(\beta +1) (4-\tau )}{(\Delta -3) (\Delta -2)} & \ft{(\beta +1) (\Delta -1) (4-\tau )}{\sqrt{2}(\Delta -3)} & \ft{\sqrt{2}(J+3) (\beta +1) (4-\tau )}{(J+1) (\Delta -3) (\Delta -2)} & \ft{\sqrt{2}(J+3) (\beta +1) (4-\tau )}{(J+1) (\Delta -3) (\Delta -2)} &  \ft{\sqrt{2}(\Delta  \tilde{\Delta} (J+5)-(J+1)^2 (J+4))}{(J+1) (\Delta -3) (\Delta -2)} \\
\ft{\sqrt{2} (\beta +1) (4-\tau)}{(J+1) (J+2)} & {-}\ft{J (\beta +1) (4-\tau)}{\sqrt{2}(J+2)} & \ft{\sqrt{2} (\beta +1) (4-\tau )}{(J+1) (J+2)} & \ft{\sqrt{2} (\beta +1) (4-\tau)}{(J+1) (J+2)} &  \ft{\sqrt{2}\left(\Delta  \tilde{\Delta}-J(J+1)\right)}{(J+1) (J+2)}
 \\
\end{array}
\right),
\cr &&
\eea
\end{small}
where $\beta=\Delta+J$ and $\tau=\Delta-J$.  (The coefficients are different if we want to get the currents instead of their shadows.)

After integrating by parts, the spinning-up operators $\mathcal{D}^{(\alpha)}_{\uparrow}$ become spinning-down operators,
in our case $\mathcal{D}^{(\alpha)}_{\downarrow}=\Big(\bar{D}_{22}^{21}\bar{D}_{11}^{12}, \bar{D}_{H_{12}}, \bar{D}_{22}^{21}\bar{D}_{12}^{12}, \bar{D}_{11}^{12}\bar{D}_{21}^{21},
\bar{D}_{21}^{21}\bar{D}_{12}^{12}\Big)$. The spinning-down operators can be constructed from weight-shifting operators in \cite{Karateev:2017jgd}, and we find convenient to define them so they are
adjoints to the above. This is readily done using
the operator $\mathcal{D}_Z$ from eq.~\eqref{Tod}\footnote{
While $\mathcal{D}_Z$ now acts on an embedding-space $5$-vector $Z$,
the dimension-dependent factor $\frac{d-2}{2}$ remains the \emph{same} as in eq.~\eqref{Tod}. See ref.~\cite{Costa:2011mg}.}:
\bea
&& \bar{D}_{ii}^{ij}= -\mathcal{D}_{Z_i}^A\left(
(X_i\cdot X_j)\fft{\partial}{\partial X_j^A}+(X_i\cdot Z_j)\fft{\partial}{\partial Z_j^A}
-X_j^A (X_i\cdot \fft{\partial}{\partial X_j})-Z_j^A (X_i\cdot \fft{\partial}{\partial Z_j})\right)\,,
\cr &&
\bar{D}_{ij}^{ij}=
-\mathcal{D}_{Z_i}^A\left(
(X_i\cdot X_j)\fft{\partial}{\partial X_i^A}-
X_{jA}\left(d-1+  (X_i\cdot \fft{\partial}{\partial X_i})+ (Z_i\cdot \fft{\partial}{\partial Z_i})\right)\right)\,,
\cr &&
\bar{D}_{H_{ij}}=2\big((X_i\cdot \mathcal{D}_{Z_j})(\mathcal{D}_{Z_i} \cdot X_j)-(X_i\cdot X_j)(\mathcal{D}_{Z_i} \cdot \mathcal{D}_{Z_j})\big)\,,
\cr &&
\bar{D}^{ij}_{iO} = -\epsilon_{ABCDE}\
\mathcal{D}_{Z_i}^A X_i^B \frac{\partial}{\partial X_{iC}}
\left( X_j^D \frac{\partial}{\partial X_{jE}}+Z_j^D \frac{\partial}{\partial Z_{jE}}\right)
\eea
These are adjoint to the $D$'s up to a spin-dependent factor which can be traced
to eq.~\eqref{Tod pairing}, namely:
\be
 \left( D_{11}^{12} T_{\j_1\j_2\ldots},T_{\j_1+1,\j_2\ldots}\right)
 = \frac{1}{(\j_1+\frac{d-2}{2})(\j_1+1)}\left( T_{\j_1\j_2\ldots},\bar{D}_{11}^{12} T_{\j_1+1,\j_2\ldots}\right)\,.
\ee
This identity makes it trivial to integrate-by-parts.\footnote{For the odd operators, we only verified that $D_{iO}$ is the adjoint of $\bar{D}_{iO}$ when acting on scalar operators, sufficient for our purposes.}
For $\bar{D}_{H_{ij}}$ there is an extra $\frac{1}{(\j_2+\frac{d-2}{2})(\j_2+1)}$ since both spins change.
Boundary terms cannot arise in the above pairing, because the integration variables are ultimately all
gauge-fixed to a point.

Interestingly, we find that $\bar{D}_{ij}^{ij}$ vanishes identically on conserved currents,
so the last three spin-down operators in our list vanish identically, reducing us to a two-dimensional basis.
It would be interesting to understand these simplifications from the perspective
of the bispinor formalism for AdS${}_4$/CFT${}_3$ \cite{Binder:2020raz}.

To find the spinned-down Lorentzian inversion formula, we now have two options.
The first, as described so far,
is to insert the matrix in eq.~\eqref{Pmatrix} inside eq.~\eqref{Lorentzian norm}
and integrate-by-parts. Since the last three spin-down operators vanish,
we can write eq.~\eqref{refLoren} in terms of two-by-two matrices.
Generally, we have\footnote{There are no possible boundary terms
because the potential limits $z,\zb=0,1$ are not really ``boundaries''. The limit $z\to 0$ is regulated, on the Euclidean and Regge sheets, by the fact that $\Delta$ is continuous and $\j>\j_*$, respectively.
Furthermore, as discussed in \cite{Caron-Huot:2017vep}, the integral over dDisc near $z\to 1$
is defined most precisely as a boundary-free ``keyhole'' type contour integral.
}
\be
c^t_{a,b}(\Delta,\j)=\sum_{\alpha,\beta}\frac{\kappa^{(\alpha,\beta)}_{\Delta+\j}}{4}
\int \frac{dzd\zb}{z^2\zb^2} \left|\frac{z-\zb}{z\zb}\right|^{d-2}\tilde{G}^{(\alpha,\beta)}_{\j+d-1,\Delta
-d+1}(z,\bar{z}) {\rm dDisc}[\hat{\mathbb{P}}_{a,\alpha}\hat{\mathbb{P}}_{b,\beta}\mathcal{D}^{(\alpha,\beta)}_{\downarrow}G(z,\bar{z})]\,,\label{refLoren 1}
\ee
where, from eq.~\eqref{Lorentzian norm},
\be
 \hat{\mathbb{P}}_{a,\alpha} =
\frac{(-4)^{\j_1+\j_2}}{\j_1!\j_2!(\tfrac12)_{\j_1}(\tfrac12)_{\j_2}}
\frac{1}{N^{E/O}_{\j_1\j_2\OO_{\bf L}}} \left(\begin{array}{c@{\,\,}c}0&1\\1&0\end{array}\right)_{ac}\mathbb{P}^c_\alpha\,.
\ee
Explicitly, for $\j_1=\j_2=1$, the parity-even matrix evaluates to:
\bea
&& \hat{\mathbb{P}}_{a,\alpha}^E=\fft{2\sqrt{2}}{(\beta -1)
 (\tau -2)} \times \left(
\begin{array}{cc}
\ft{-2}{(J+1) (J+2)}  & \ft{J}{(J+2)}
  \\ \frac{2}{(\Delta -3) (\Delta -2)} & -\ft{(\Delta -1)}{(\Delta -3)}
 \end{array}
\right)\,, \label{normVVVV}
\eea
where only $\Delta_{12}=0$ appears in $\kappa$ and the block.
For odd structures, in the spin-down basis
$\mathcal{D}_{\downarrow}^O=(\bar{D}_{1O}^{12}\bar{D}_{22}^{21},\bar{D}_{2O}^{21}\bar{D}_{11}^{12})$,
\be
 \hat{\mathbb{P}}_{a,\alpha}^O=\fft{-\sqrt{2}}{(\j+1)(\Delta-2)} \times \left(
\begin{array}{cc}
\ft{1}{(\j+2)(\Delta-1)}  & \ft{-1}{(\j+2)(\Delta-1)}
  \\ \ft{1}{\j(\Delta-3)}  & \ft{1}{\j(\Delta-3)}
 \end{array}
\right)\,. \label{normVVVV odd}
\ee
These matrices tell
us how to convert the scalar inversion of the spinned-down correlators (given below in eq.~\eqref{effGVV})
to OPE data in opposite/same-helicity structures.

There is a simple check: acting with the spin-down operators $\hat{\mathbb{P}}_{a,\alpha}\mathcal{D}_{\downarrow}^\alpha$ on the three-point spinning structure $T_{11\OO}^b$, we must get $\delta_a^b$ times a canonically normalized scalar three-point structure $T_{00\OO}$.
In fact this gives a second method to directly find the matrix $\hat{\mathbb{P}}_{a,\alpha}$, by-passing
the spinning Lorentzian inversion formula.  We find precise agreement between the two methods.
(The second one being admittedly more straightforward.)

These operators can be applied to any correlator.
We now consider $t$-channel identity exchange:
\be
G=\fft{H_{23}H_{14}}{(-2X_2\cdot X_3)^{\Delta_2+1}(-2X_1\cdot X_4)^{\Delta_1+1}}\,,
\ee
which gives for example the even spinned-down correlator $\mathcal{D}_{\downarrow}G$
\bea
&& \mathcal{D}^{(1,1)}_{\downarrow}G=-\frac{3}{2} y (\bar{y}+1)(24 y^4+3 y^3 (5-4 \bar{y})+3 y^2 (\bar{y} (4 \bar{y}+3)+1)-y (\bar{y}+1) (3 \bar{y} (4 \bar{y}+3)+1)
\cr && +3 (\bar{y}+1)^2 (\bar{y} (8 \bar{y}+7)+1))\,,
\cr &&
\cr && \mathcal{D}^{(2,2)}_{\downarrow}G=-y (\bar{y}+1) \left(y^2-y (\bar{y}+1)+(\bar{y}+1)^2\right)\,,
\cr &&
\cr && \mathcal{D}^{(1,2)}_{\downarrow}G=\mathcal{D}^{(2,1)}_{\downarrow}G=-\frac{1}{2} y (\bar{y}+1) \left(9 y^3+y^2 (1-5 \bar{y})+y (\bar{y}+1)
(5 \bar{y}+1)-3 (\bar{y}+1)^2 (3 \bar{y}+1)\right)\,,\label{effGVV}
\cr &&
\eea
where we reparameterized the cross-ratios by $(z=\frac{y}{1+y}, \bar{z}=\frac{1}{1+\bar{y}})$.

Inserting in eq.~\eqref{normVVVV} it remains to do the scalar inversion integrals of eqs.~\eqref{effGVV}.
A good strategy is to expand in $y\rightarrow 0$ to work out the integral over $z$ twist-by-twist.
This also requires the lightcone expansion $z\rightarrow0$ for
$\tilde{G}_{J+d-1,\Delta-d+1}(z,\bar{z})$ in the inversion formula (\ref{refLoren}),
which can be done by noting (see, eq.~(A.24) in \cite{Caron-Huot:2017vep})
\be
\fft{\kappa(\beta)}
{\kappa(\beta+2p)}(1-z)^{a+b}(1-\fft{z}{\bar{z}})^{d-2}G_{J+d-1,\Delta-d+1}\big|_{q,p}\sim B_{q,p}\,z^{\fft{J-\Delta}{2}+n+d-1}k_{\beta+2m}(\bar{z})\,,\label{expandfun}
\ee
where $B_{q,p}$ can be recursively solved by the quadratic Casimir equation \cite{Caron-Huot:2017vep}. Moreover, we can take use of the following integral formula to do the integral over $\bar{z}$ \cite{Caron-Huot:2017vep}
\bea
&& I_{\hat{\tau}}(\beta)=\int_0^1 \fft{d\bar{z}}{\bar{z}^2}(1-\bar{z})^{a+b}\kappa^{a,b}_\beta k^{a,b}_\beta(\bar{z})\,{\rm dDisc}[
\big(\fft{1-\bar{z}}{\bar{z}}\big)^{\fft{\hat{\tau}}{2}-b}(\bar{z})^{-b}]
\cr &&
\cr && =\fft{\Gamma(\fft{\beta}{2}-a)\Gamma(\fft{\beta}{2}+b)\Gamma(\fft{\beta}{2}-\fft{\hat{\tau}}{2})}
{\Gamma(-\fft{\hat{\tau}}{2}-a)\Gamma(-\fft{\hat{\tau}}{2}+b)
\Gamma(\beta-1)\Gamma(\fft{\beta}{2}+\fft{\hat{\tau}}{2}+1)}\,.\label{intresult}
\eea
With this strategy we can calculate the result analytically for any $n>0$, and find a simple common formula  given below.

The case $n=0$ is subtle as we discussed previously in subsection \ref{leadtwist}:
the structures become degenerate. In fact the whole matrix (\ref{normVVVV}) blows up as $\tau\to 2$.
The solution, as above, is to apply a further rotation to the basis in eq.~\eqref{rotation}.
In the $(T^{E, \rm reg}_{123},\ T^{E, \rm sing}_{123})$ basis,
the matrix \eqref{normVVVV} becomes:
\be
\hat{\mathbb{P}}_{a,\alpha}^{E,\rm rotated} = \sqrt{2}\left(
\begin{array}{cc}
 -\frac{2 J-1}{(J-1) J (J+1) (J+2) (2 J+1)} & \frac{J}{(J-1) (J+1) (J+2) (2 J+1)} \\
 \frac{1}{2 (J-1) J (2 J+1)} & -\frac{J+1}{4 (J-1) (2 J+1)} \\
\end{array}
\right),
\ee
which is now nicely finite.  The same rotation
will also work in the computation of anomalous dimensions in the next section.

For MFT correlators discussed here where $\mathcal{D}_{\downarrow}G$ is actually a finite sum of powers of cross-ratios times Gegenbauer polynomials, a more compact and comprehensive trick is available to extract the OPE data, see appendix \ref{GFFmore}.  Our result, for $n\geq 1$, the coefficients
of even (opposite/same) helicity structures are then:
\bea
 \lambda_{12\OO}^{E}\lambda_{43\OO}^{E}\big|_{n,\j}&=&
\frac{(\j+1)_{\frac12}(2n+\j+\tfrac12)_{\frac12}}{2^{4n+2\j+3}(n+\tfrac12)_{\frac12}(n+\j+1)_{\frac12}}
\left(
\begin{array}{cc}
\fft{\j(\j-1)}{(\j+2)(\j+1)}  & 0
  \\ 0 & \fft{(2n+\j+1)(2n+\j+2)}{(2n+\j)(2n+\j-1)}
 \end{array}
\right),\
\eea
which is precisely eq.~\eqref{evenOPEcoeff} with $\j_1=\j_2=1$.
For the leading trajectory, in the rotated basis we find
\be
 \lambda_{12\OO}^{E,\rm rotated}\lambda_{43\OO}^{E,\rm rotated}\big|_{0,\j}
= \frac{2\Gamma(\j+1)^2}{\Gamma(2\j+1)}\times \fft{\j(\j-1)}{16(\j+2)(\j+1)}\left(\begin{array}{cc} 1&0\\0&0\end{array}\right)\,,
\ee
which again agrees with eq.~\eqref{leading trajectory} with $\j_1=\j_2=1$.  This confirms that spurious poles simply double the $n>0$ trajectories.

\section{Application to AdS${}_4$/CFT${}_3$}
\label{application}

The simplicity and diagonal nature of the mean field OPE encourages
us to look at the leading corrections.  In this section, we study
CFT${}_3$ current correlators that are dual to bulk YM${}_4$ gluon amplitudes at tree-level.
The Lorentzian inversion formula will give us the corresponding anomalous dimensions in terms
$t$- and $u$- channel exchanges of conserved currents.

These correlation functions have been previously discussed in momentum space.
Results are remarkably tractable thanks to the fact that YM${}_4$ is conformally invariant (at tree-level) and
AdS${}_4$ is conformally flat \cite{Maldacena:2011nz,Raju:2012zs,Armstrong:2020woi,Albayrak:2020fyp,Albayrak:2018tam,Albayrak:2019asr}.
Our goal is to obtain the corresponding OPE anomalous dimension, which we will then compare with the flat space limit in the next section.
The flat space limit of AdS/CFT \cite{Gary:2009ae,Heemskerk:2009pn,Okuda:2010ym}
($R_{\rm AdS}\to\infty$) has not been much studied for spinning operators
(with a notable exception \cite{Raju:2012zr})
and we feel it is important to clarify it.
Similarly to the scalar case, one may expect (massless) amplitudes to be encoded in the $z\to \bar{z}^\circlearrowleft$ ``bulk-point'' limit \cite{Fitzpatrick:2011hu,Maldacena:2015iua}, or equivalently the large-twist limit of OPE data.
This will be confirmed in the next section.

\subsection{Setup for current correlators}

\begin{figure}
\centering \hspace{0mm}\def\svgwidth{50mm}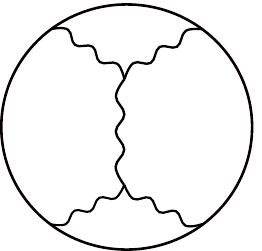
\caption{Witten diagram for $\langle VVVV\rangle$ with on-shell $t$-channel gluon exchange.
Two even and one odd coupling can be used in each vertex;
$u$-channel is similar with $1$ and $2$ swapped.}
\label{figure: Witten VVVV}
\end{figure}

Our strategy is to use spin-up/spin-down operators to reduce the calculation to scalar Lorentzian inversion formulas.
The spin-down operators were described and validated in section \ref{MFT from Lorentzian}, acting on identity exchange in the $t$- and $u$-channel.
The exchanged operator is now a current, as shown in fig.~\ref{figure: Witten VVVV}. (Double-trace exchanges do not contribute to tree-level accuracy, thanks to the double-discontinuity.)

From the CFT perspective, each current exchange involves two parity-even and one odd coupling, described below eq.~\eqref{bulk Yang-Mills}, which maps one-to-one with bulk on-shell three-gluon couplings.
These can be obtained from a bulk Lagrangian including higher-derivative corrections:
\be
\mathcal{L}= 
-\fft{1}{4g_{\rm YM}^2}F_{\mu\nu}^a F^{\mu\nu a}
+ \frac{\theta}{32\pi^2} F_{\mu\nu}^a \tilde{F}^{\mu\nu a}
-\frac{f^{abc}}{3g_{\rm YM}^3}\left(
g_{\rm H}\, F_\mu\,^{\nu a} F_\nu\,^{\rho b} F_\rho\,^{\mu c}
+
g'_{\rm H}\, \tilde{F}_\mu\,^{\nu a} \tilde{F}_\nu\,^{\rho b} \tilde{F}_\rho\,^{\mu c}\right)+\cdots\,,\label{bulk Yang-Mills}
\ee
where $\tilde{F}_{\mu\nu}=\frac12\epsilon_{\mu\nu\sigma\rho}F^{\sigma\rho}$.
We show that in appendix \ref{Witten} that the couplings satisfy:
\be
 \lambda^{(\rm e1)}_{VVV}  = \fft{g_{\rm YM}}{16\sqrt{2}},\qquad
 \lambda^{(\rm e2)}_{VVV}  = \fft{g_{\rm H}}{8\sqrt{2}},\qquad
 \lambda^{(\rm o2)}_{VVV}  = \fft{g_{\rm H}'}{4\sqrt{2}\pi} \label{dictionary}
\ee
where the structures refer to the even/odd basis in eq.~\eqref{even odd}.
(We recall that the first structure is the ``opposite helicity'' one which generically exists for spin $J\geq 2$.)
Having stated this dictionary, in this section we shall present results in terms of the CFT couplings $\lambda_{VVV}^{(i)}$.

We consider only the parity-even couplings.
There are then four ways to dress the
the graph in fig.~\ref{figure: Witten VVVV}:
\bea
&& G_{11}\,,\quad \text{Yang-Mills vertex to Yang-Mills vertex}\,,
\cr && G_{22}\,,\quad \text{higher-derivative vertex to higher-derivative vertex}\,,
\cr && G_{12}\,,\quad \text{Yang-Mills vertex to higher-derivative vertex}\,,
\cr && G_{21}\,,\quad \text{higher-derivative vertex to Yang-Mills vertex}\,.
\eea
In each case the $t$-channel block can be written as the spin-up of a scalar block, so after spinning down
$\mathcal{D}^{(c,d)}_{\downarrow}G(z,\bar{z})$ in eq.~\eqref{refLoren} gives a 8th order differential equation acting on scalar blocks.
The cross-channel
scalar blocks themselves are not known in closed form; in appendix \ref{tchannel} we provide the series expansion
of the $\log z$ term to any order in $z$, which is sufficient to calculate anomalous dimensions exactly,
in terms of $(y=z/(1-z),\bar{y}=(1-\bar{z})/\bar{z})$, i.e., eq.~(\ref{veczb11}).
For example, at the leading order in the lightcone expansion $y\rightarrow 0$, we find
\bea
\mathcal{D}_{\downarrow}G_{11}=\fft{\log y}{\pi}\left(
\begin{array}{cc}
 -\frac{9 y (\bar{y}+1) \left(\bar{y}^3+27 \bar{y}^2+675 \bar{y}+1225\right)}{32 \bar{y}^{9/2}} & \frac{3 y \left(3 \bar{y}^3-29 \bar{y}^2-123 \bar{y}-75\right)}{4 \bar{y}^{7/2}} \\
 \frac{3 y \left(3 \bar{y}^3-29 \bar{y}^2-123 \bar{y}-75\right)}{4 \bar{y}^{7/2}} & -\frac{2 y \left(9 \bar{y}^2+26 \bar{y}+9\right)}{\bar{y}^{5/2}} \\
\end{array}
\right)+\mathcal{O}(y^2)\,,\label{expT22}
\eea
where we parameterize $y=z/(1-z), \bar{y}=(1-\bar{z})/\bar{z}$. At the leading order, $\mathcal{D}_{\downarrow}G_{22}$ has the same expression as $\mathcal{D}_{\downarrow}G_{11}$, but differs at the second and higher orders.
Up to the leading order, $\mathcal{D}_{\downarrow}G_{12}=\mathcal{D}_{\downarrow}G_{21}$ is
\be
\mathcal{D}_{\downarrow}G_{12}=\fft{3\log y}{\pi}\left(
\begin{array}{cc}
 -\frac{3 y (\bar{y}+5) \left(\bar{y}^3-9 \bar{y}^2+171 \bar{y}+245\right)}{32 \bar{y}^{9/2}} & \frac{3 y \left(\bar{y}^3+\bar{y}^2-9 \bar{y}-25\right)}{4 \bar{y}^{7/2}} \\
 \frac{3 y \left(\bar{y}^3+\bar{y}^2-9 \bar{y}-25\right)}{4 \bar{y}^{7/2}} & -\frac{2 y \left(3 \bar{y}^2-2 \bar{y}+3\right)}{\bar{y}^{5/2}} \\
\end{array}
\right)+\mathcal{O}(y^2)\,.\label{expT12}
\ee
The above expansions eq.~(\ref{expT22}) and eq.~(\ref{expT12}) would then be used \emph{in principle} to obtain the leading-twist anomalous dimensions by simply integrating over $\bar{y}$ using the formula (\ref{intresult}). As discussed
in subsection \ref{leadtwist}, the leading-twist analysis is a bit subtle due to a degeneracy in three-point structures, and is discussed below.
As the rotation in eq.~\eqref{rotation} removes all divergences,
the anomalous dimension can be computed using just the logarithmic term in
eq.~\eqref{expT22}. Nontrivially, we find a result proportional to the leading order matrix
$\small\left(\begin{array}{cc}1&0\\0&0\end{array}\right)$, as required by the fact that there is a single leading-twist family (the number of operators can't change under small perturbations).
The anomalous dimension is then\footnote{Since the
second structure $T^{E, \rm sing}_{11\OO}$ has a nonvanishing discontinuity,
its $\sim (\lambda_{VVV}^{(\rm e1)})^2$ OPE coefficient will be required to
predict the one-loop dDisc, in addition to the given
anomalous dimension.}
\be
 \gamma_{11}^E\big|_{n=0} = -\Big(\ft{\beta ^4-4 \beta ^3+28 \beta ^2-48 \beta +32}{(\beta -4) (\beta -2) (\beta -1) \beta  (\beta +2)}
 \big((\lambda_{VVV}^{\rm e1})^2+(\lambda_{VVV}^{\rm e2})^2\big)+\ft{2}{1-\beta}\lambda_{VVV}^{\rm e1}\lambda_{VVV}^{\rm e2}\Big) (T+(-1)^\j U)\,.
\ee

At subleading twists, the calculation uses analogous expressions together with the
$s$-channel expansion (\ref{expandfun}) and (\ref{intresult}).

\subsection{Anomalous dimensions: Yang-Mills case}

This yields the anomalous dimensions as analytic functions of $\beta$ for fixed $n\geq 1$.
Including the $\hat{\mathbb{P}}$ matrix in eq.~(\ref{normVVVV}), we obtain $\langle c\gamma\rangle_{J,\Delta}$, which we then divide by the generalized free OPE data \eqref{free data} (with $\j_1=\j_2=1$), to arrive at anomalous dimensions.
It is important to include both $t$- and $u$-channel identity in the denominator, which effectively doubles it as discussed below \eqref{GFF OPE}.
In the pure Yang-Mills case we find:
\be
 \gamma_{11}^E=\ft{128 (\lambda_{VVV}^{(\rm e1)})^2}{\pi^2}\left( T+ (-1)^J U\right)
{\rm diag}
\left(\begin{array}{l}
 \psi_{\frac{\beta}2-n-2}\!-\psi_{\frac\beta2+n}-\ft{4}{(\beta-2 n)(\beta-2n+2)}+\ft{4}{-2 n+\beta-2}\\
 \psi_{\frac{\beta}2-n}\!-\psi_{\frac\beta2+n+2}+\ft{4}{(\beta+2 n-4)(\beta+2n-2)}+\ft{4}{2 n+\beta}
\end{array}\right)\,, \label{gamma11}
\ee
where ${\rm diag}$ represents the diagonal matrix, and we have factored out $T$- and $U$-channel color structures
\be
 T= f^{bce}f^{ade},\qquad U=f^{ace}f^{bde}\,. \label{T main text}
\ee
These should be viewed as operators acting on the initial pair, for example both have
the eigenvalue $T,U\mapsto C_A$ when acting on a color-singlet state $\delta^{ab}$.

Eq.~\eqref{gamma11} (for $n\geq 1$) gives the CFT${}_3$ analog of the four-point Parke-Taylor amplitude.
We note that to all orders in the $1/\beta$, the two entries are related by the reciprocity relation
$\beta\mapsto 2-\beta$, which could have been anticipated from the off-diagonal nature of the light transform in eq.~\eqref{light transform matrix}.  The fact that it is diagonal will match with
the vanishing of non-helicity-conserving flat space amplitudes at tree-level.

The Yang-Mills self-interaction also gives diagonal anomalous dimension the odd double-twists (which have half-integer $n$):
\be
 \gamma_{11}^O=\ft{128 (\lambda_{VVV}^{(\rm e1)})^2}{\pi^2}
{\rm diag}\left(\begin{array}{l}
 \left(\psi_{\frac{\beta}2-n}\!-\psi_{\frac\beta2+n}-\ft{8}{(\beta-2 n-2)(\beta-2n)}\right)\left(T-(-1)^J U\right) \\
 \left(\psi_{\frac{\beta}2-n}\!-\psi_{\frac\beta2+n}+\ft{8}{(\beta+2 n-2)(\beta+2n)}\right)\left(T+ (-1)^J U\right)
\end{array}\right).\label{gamma11 odd}
\ee

\subsection{Higher-derivative corrections}

Let us now record the pure higher-derivative corrections, which involve purely algebraic expressions:
\bea
\gamma_{22}^E&=&\ft{128 (\lambda_{VVV}^{(\rm e2)})^2}{\pi^2}
{\rm diag}\left(\begin{array}{l}
 \frac{(n (\beta -1)+2) \left(4 n^2+8 (\beta -1) n+(\beta -2) \beta +4\right)}{(2 n-\beta -2) (2 n-\beta ) (2 n-\beta +2) (2 n-\beta +4)}\left(-T- (-1)^J U\right) \\
  \frac{(n (\beta -1)-2) \left(4 (n+1)^2+\beta ^2-2 (4 n+1) \beta \right)}{(2 n+\beta -4) (2 n+\beta -2) (2 n+\beta ) (2 n+\beta +2)}
 \left(-T- (-1)^J U\right)
\end{array} \right)\,,
\\
\gamma_{22}^O&=&\ft{128 (\lambda_{VVV}^{(\rm e2)})^2}{\pi^2}
{\rm diag}\left(\begin{array}{l}
 \frac{(n (\beta -1)+2) \left(4 n^2+8 (\beta -1) n+(\beta -2) \beta +4\right)}{(2 n-\beta -2) (2 n-\beta ) (2 n-\beta +2) (2 n-\beta +4)}
 \left(T- (-1)^J U\right)\\
  \frac{(n (\beta -1)-2) \left(4 (n+1)^2+\beta ^2-2 (4 n+1) \beta \right)}{(2 n+\beta -4) (2 n+\beta -2) (2 n+\beta ) (2 n+\beta +2)}
 \left(T+ (-1)^J U\right)
\end{array}
\right)\,.\label{gamma22}
\eea
The even and odd matrices are identical up to some signs, and again reciprocity
$\beta\mapsto 2-\beta$ swaps the trajectories up to a minus sign.

The $G_{12}$ contributions (one Yang-Mills and one higher-derivative vertex)
violate helicity conservation and give purely off-diagonal anomalous dimensions.
Since the Lorentzian inversion formula gives us $\langle c\gamma\rangle_{J,\Delta}$,
we divide the off-diagonal terms by the geometric mean of MFT coefficients to define a symmetrical
anomalous dimension matrix $\gamma_{12}^{\rm even}=\gamma_{21}^{\rm even}$:
\bea
&& \gamma_{12}^E=\ft{128\lambda_{VVV}^{(\rm e1)}\lambda_{VVV}^{(\rm e2)}}{\pi^2}\ft{-4n (\beta -1)
\sqrt{(\beta -2 n)(\beta -2 n+2)} (\beta(\beta-2) +4 n^2-4)(T+(-1)^J U)}{(\beta-2 n+2) (\beta-2 n)
\sqrt{(\beta -2 n-4)(\beta -2 n-2)(\beta +2 n-4)(\beta +2 n-2)(\beta +2 n)(\beta +2 n+2)}} \left(
\begin{array}{cc}
 0 & 1 \\
 1 & 0 \\
\end{array}
\right)\,.\nonumber\\ &&\label{gamma12}
\eea
The odd $\gamma_{12}$ is the same and $\gamma_{21}$ is also identical up to
an overall minus sign (such that the sum vanishes: $\gamma_{12}^O+\gamma_{21}^O=0$,
which will be in agreement with symmetries of the scattering amplitude).

We end this section by giving the large-$n$ limit of above anomalous dimensions, which will be compared in the next section with flat-space $2$-to-$2$ gluon scattering amplitudes:
\bea
&& \gamma^E_{11}|_{n\rightarrow\infty}=\ft{128 (\lambda_{VVV}^{(\rm e1)})^2}{\pi^2}(T+(-1)^JU){\rm diag}\left(\begin{array}{l}
\psi_{J-1} -\log(2n)+\ft{2}{J}-\ft{1}{(J+1)(J+2)} \\
\psi_{J+1}-\log(2n) \end{array}\right)\,,
\cr &&
\cr && \gamma^O_{11}|_{n\rightarrow\infty}=\ft{128 (\lambda_{VVV}^{(\rm e1)})^2}{\pi^2}\left(\begin{array}{l}
\big(\psi_{J+1} -\log(2n)-\ft{2}{\j(\j+1)}\big)(T-(-1)^{\j}U) \\
\big(\psi_{J+1}-\log(2n)\big)(T+(-1)^{\j}U) \end{array}\right)\,,
\cr &&
\cr && \gamma^{E/O}_{22}|_{n\rightarrow\infty}=\ft{128 (\lambda_{VVV}^{(\rm e2)})^2}{\pi^2}\left(
\begin{array}{cc}
 \ft{12n^4 (\mp T+(-1)^JU)}{(J-1)J(J+1)(J+2)} & 0 \\
 0 & 0 \\
\end{array}
\right)\,,
\cr &&
\cr &&  \gamma^{E/O}_{12}|_{n\rightarrow\infty}=\ft{128\lambda_{VVV}^{(\rm e1)}\lambda_{VVV}^{(\rm e2)}}{\pi^2}
\ft{-n^2(T+(-1)^JU)}{\sqrt{(J-1)J(J+1)(J+2)}}
\left(\begin{array}{cc}
 0 & 1 \\
 1 & 0 \\
\end{array}
\right)\,,
\cr &&
\cr &&
\gamma^{E/O}_{21}|_{n\rightarrow\infty}=\ft{128\lambda_{VVV}^{(\rm e1)}\lambda_{VVV}^{(\rm e2)}}{\pi^2}
\ft{\mp n^2(T+(-1)^JU)}{\sqrt{(J-1)J(J+1)(J+2)}}
\left(\begin{array}{cc}
 0 & 1 \\
 1 & 0 \\
\end{array}
\right)\,.\label{largenVVVV}
\eea
We note that each higher-derivative correction $\lambda^{(e2)}_{VVV}$ comes accompanied
with a power of $n^2\sim s$, as expected from bulk dimensional analysis.
Furthermore, we see that the difference between even- and odd- same-helicity anomalous dimensions
vanishes at large-$n$:
\be
\gamma^{E,\rm same}_{11} - \gamma^{O,\rm same}_{11} =
\frac{128 (\lambda_{VVV}^{(\rm e1)})^2}{\pi^2} \frac{6(T+(-1)^JU)}{(\tfrac{\beta}{2}+n-2)_4} \sim \frac{1}{n^4}\,.
\label{++++}
\ee
This indicates that the same-helicity amplitude $\mathcal{M}_{++++}$
vanishes in the flat-space limit (as expected).
However, we find it remarkable that it is \emph{not} identically zero in AdS space.
This suggests that, in a more precise treatment where the flat-space limit is defined
as $R\to\infty$ as opposed to $s\to\infty$, a distributional term near $s=0$ may survive; such terms
could potentially give a new
perspective on four-dimensional unitarity and
the rational one-loop amplitude $\mathcal{M}_{++++}^{(1)}$. We leave this to future work.

\section{Large-$n$ limit from gluon scattering amplitudes}
\label{fromamp}

\begin{figure}
\centering \hspace{0mm}\def\svgwidth{50mm}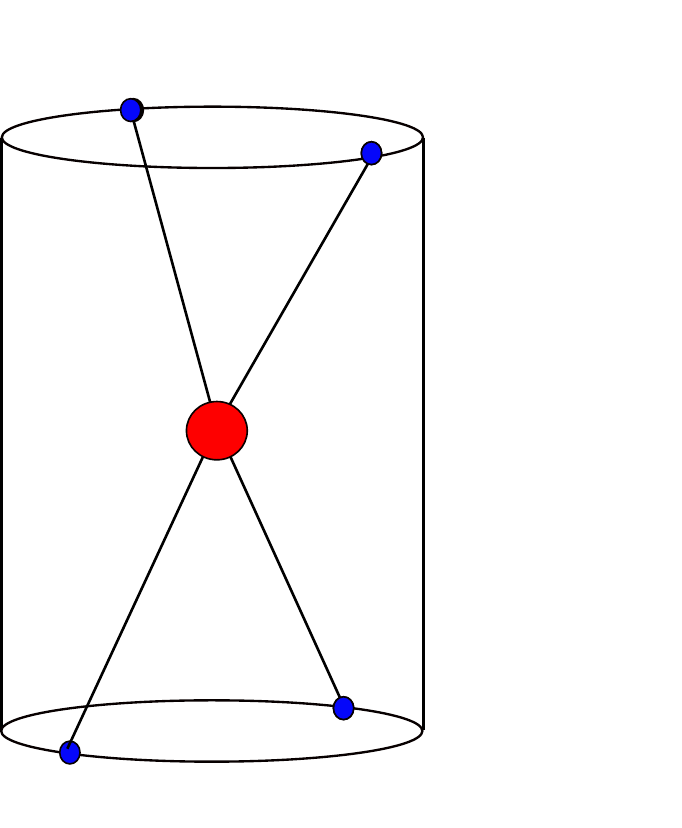
\caption{Bulk-point kinematics in Lorentzian cylinder of AdS. $X_1$ and $X_2$ are at Lorentzian time $-\pi/2$, $X_3$ and $X_4$ are at Lorentzian
time $\pi/2$, where particles are focused on the bulk-point $P$.}
\end{figure}

There is a close relation between the anomalous dimensions at large dimension in a CFT${}_d$
and the scattering amplitude of a dual QFT${}_{d+1}$ in the flat space limit of AdS.
This can be seen for example by considering kinematic configurations which focus particles --- such as the analytically continued $z\to\zb$ ``bulk-point'' limit, see for example \cite{Gary:2009ae,Heemskerk:2009pn,Okuda:2010ym,Penedones:2010ue,Fitzpatrick:2011hu,Maldacena:2015iua}\footnote{This kinematic configuration is, however, modified if external particles are massive \cite{Paulos:2016fap,Hijano:2019qmi,Komatsu:2020sag,Li:flatspace}.}.
For massless external particles (dual to our currents),
since the past and future states are connected by time $\pi$ on the cylinder, the scattering phase
is related to CFT anomalous dimensions by the simple dictionnary
\be
\gamma_{n,\j}|_{n\rightarrow \infty}\to -\frac{1}{\pi} a_{\j}\,,\quad sR^2=4n^2\,,\label{dictCFTQFT}
\ee
where $a_{\j}$ is the partial-wave amplitudes with angular momentum $\j$, $s$ is the  Mandelstam invariant of the bulk scattering process, and $R$ is the AdS radius. We often take $R=1$ below for simplicity and take $s\neq 0$, so the limit is equivalent to $n\to\infty$. (In general the amplitude maps to a weighted average of anomalous dimensions. A one-loop example is provided in \cite{Alday:2017vkk}.)
We expect this relation to work for spinning operators as well,
for suitably defined partial waves.

\subsection{Partial waves in massless QFT$_4$}
\label{setup}

Two-particle scattering states in QFT${}_4$ can be organized according to their SO(3) spin
in the rest frame of their total momentum, $P=p_1+p_2$.  Since rotations commute with helicity,
we can choose a basis of states with definite helicity.
For definiteness, we focus here on the case of two massless spin 1 particles.

We use the spinor-helicity formalism where each null momentum is factorized into a product of spinors,
$p\!\!\!/{}_i = |i]\langle i|$, see \cite{Elvang:2013cua}.
Under little-group rotations of spinors $|i]$ and $|i\rangle$ by opposite phases,
a state of helicity $h$ transforms like $|i]^{2h}$.
We treat two-particle states like a massive particle of momentum $P$ and spin $J$, which in
index-free notation is a polynomial $\sim |\eps\rangle^{2J}$ in a left-handed spinor $|\eps\rangle$.
(There is no need to use right-handed spinors, since $P$ can be used to convert one into the other, see \cite{Arkani-Hamed:2017jhn}.)
Lorentz and little-group symmetries then uniquely fix the matrix elements of two-particle states
$\Psi^J_{\pm}$:
\def\ab#1{{\langle #1\rangle}}
\be\begin{aligned}\label{formfact}
\langle 2^{-}1^{-}|\Psi_{--}^\j\rangle&= \fft{\ab{\eps1}^{\j}\ab{\eps 2}^{\j}}{\langle 12\rangle^{\j-1}[12]}\,,&\quad
\langle 2^{+}1^{-}|\Psi_{-+}^\j\rangle&= \fft{\ab{\eps1}^{\j+2}\ab{\eps2}^{\j-2}}{\langle 12\rangle^{\j}}\,,\\
\langle 2^{+}1^{+}|\Psi_{++}^\j\rangle&= \fft{\ab{\eps1}^{\j}\ab{\eps 2}^{\j}}{\langle 12\rangle^{\j+1}/[12]}\,,&\quad
\langle 2^{-}1^{+}|\Psi_{+-}^\j\rangle&= \fft{\ab{\eps1}^{\j-2}\ab{\eps2}^{\j+2}}{\langle 12\rangle^{\j}}\,.
\end{aligned}\ee
More precisely, symmetries fix the states up to a power of $s=-P^2$, which we chose so that
all states have the same dimension.
We further define the state $|\Psi_{h_1h_2}^\j\rangle$ to be orthogonal to gluons of other helicity.

\def\A{\mathcal{A}}

In the above kinematic factors we treat the two particles as distinguishable.
These are related to actual gluon states by adding color labels and accounting for Bose symmetry:
fully decorated states can be defined as
\be
\langle 3^{h_3 c}4^{h_4d}|\Psi_{h_1h_2}^{\j,ab}\rangle =
\delta^{ad}\delta^{bc} \delta^{h_4}_{h_1}\delta^{h_3}_{h_2} \langle 3^{h_3} 4^{h_4}|\Psi_{h_1h_2}^{\j}\rangle+
\delta^{ac}\delta^{bd} \delta^{h_3}_{h_1}\delta^{h_4}_{h_2} \langle 3^{h_3} 4^{h_4}|\Psi_{h_1h_2}^{\j}\rangle\,.
\label{color-decorated}
\ee
Since interactions can change helicities, the action of the S-matrix on these
states takes the form of a $4\times 4$ matrix:
\be
 \mathcal{S}|\Psi_{h_1a,h_2b}^J\rangle = \sum\limits_{h_3,h_4,c,d}  S^J_{h_1a,h_2b}{}^{h_4d,h_3c} |\Psi_{h_3c,h_4d}^J\rangle + \mbox{multi-particles}\,.
\ee
As is customary, we subtract the identity part: $\mathcal{S}=1+i\A$,
where $\A$ is the scattering amplitude.
In the $2\to2$ sector, $S^J_{12}{}^{43}=\frac12(\delta_{1}^4\delta_2^3+\delta_{1}^3\delta_2^4) +ia^J_{12}{}^{43}$, where we use collective indices in $\delta_1^4 = \delta_{h_1}^{h_4}\delta_a^d$.
The partial wave $a$ is then simply the amplitude in the $|\Psi\rangle$ basis:
\be
 a^J=\A\otimes |\Psi^J\rangle\,,
\ee
which can be computed as a phase-space integral.
To be fully explicit with indices (see also eq.~(2.16) of \cite{Caron-Huot:2016cwu}):
\bea
 a^{J}_{h_1a,h_2b}{}^{h_4 d,h_3c} &=& \frac12\sum_{h_1',h_2',a',b'} \int \frac{d\Omega}{64\pi^2}
  \frac{\langle 3^{h_3c} 4^{h_4d}|\A |1^{a'}_{h_1'}2^{b'}_{h_2'}\rangle\langle 1^{h_1'a'}2^{h_2'b'}|\Psi_{h_1a,h_2b}^{J}\rangle}
   {\langle 3^{h_3}4^{h_4}|\Psi_{h_3,h_4}^J\rangle}  \\
&=&  \frac{1}{16\pi}\int \frac{d\Omega}{4\pi}
  \langle 3^{h_3c} 4^{h_4d}|\A |1^{a}_{h_1}2^{b}_{h_2}\rangle
  \frac{\langle 1^{h_1}2^{h_2}|\Psi_{h_1,h_2}^{J}\rangle} {\langle 3^{h_3}4^{h_4}|\Psi_{h_3,h_4}^J\rangle}\,.
     \label{phase space}
\eea
The second form will be particularly useful for calculations.  Notice that the two terms in
eq.~\eqref{color-decorated} simply canceled the symmetry factor $\frac12$.
In this integral, $p_3$ and $p_4$ are held fixed and $d\Omega$ represents the solid angle of $\vec{p_1}$ in the rest frame
of $P$.

The angular integral can be conveniently parametrized in terms of spinors via \cite{Zwiebel:2011bx}
\be
|1\rangle=\cos\theta |4\rangle -\sin\theta e^{i\phi}|3\rangle\,,\quad
|2\rangle=\sin\theta e^{-i\phi} |4\rangle +\cos\theta |3\rangle\,, \label{spinorpara}
\ee
with analogous expressions for the conjugate spinors $|1]$ and $|2]$ with the phase reversed $\phi\mapsto -\phi$.
In the rest frame of $P$, the variables $\theta$ and $\phi$ represent physically (half) the azimuthal and polar angle with respect to $p_1$.
The measure is then
\be
\int \frac{d\Omega}{4\pi} = \int_0^{2\pi} \frac{d\phi}{2\pi} \int_0^{\frac{\pi}{2}} \sin(2\theta)d\theta\,.
\ee
It is important to note that both the numerator and denominator in eq.~\eqref{phase space} depend on $|\eps\rangle$, $p_3$ and $p_4$, in addition to the integration variables $\theta,\phi$.
However, since the result of the integral is determined by symmetry,
the ratio \emph{after} doing the integral is guaranteed to be a pure number independent of these variables.

This method allows us to define partial waves without having to worry about the normalization of the states.
The idea is that the eigenvalues of the matrix $S^J_{12}{}^{43}$ map to weighted averages
of CFT anomalous dimensions $e^{-i\pi \gamma}$. To leading order in perturbation theory, this relation gives simply, as quoted:
\be
 \gamma^J_{12}{}^{43} \approx -\frac{1}{\pi} a^J_{12}{}^{43}.
\ee
Surprisingly, the exact same relation has an interpretation purely in the context of
QFT: the phase of the S-matrix acting on form factors of local operators gives the dilatation operator of the QFT:
$\mathcal{S}\mathcal{F}^*=e^{-i\pi D}\mathcal{F}^*$ \cite{Caron-Huot:2016cwu}.
This was used there to compute anomalous dimensions of local operators
of a QFT${}_4$, as labelled by their two-particle form factors. (For example, the infrared-safe combination
$\gamma^0_{++}{}^{++}-\gamma^2_{+-}{}^{+-}$ acting on a color-singlet state computes the QCD $\beta$-function.)
Here $\gamma^J_{12}{}^{43}$ instead gives holographically
a CFT${}_3$ anomalous dimension $\gamma(n)$ where $4n^2=sR^2$ is large.
It is amusing that anomalous dimensions in the bulk QFT$_{d+1}$ and boundary CFT$_d$ are computed by literally the same formula.

\subsection{Anomalous dimensions in Yang-Mills theory}

On-shell amplitudes in YM$_4$ are recorded in appendix \ref{gluonampdo}. We use these on-shell amplitudes together with eq.~\eqref{phase space} to extract the corresponding partial-wave amplitudes, from which we will find perfect agreement with CFT eq.~\eqref{largenVVVV}.

We begin with the pure Yang-Mills theory, then add higher-derivative corrections.

\subsubsection{Pure Yang-Mills}


Using Yang-Mills amplitudes eq.~\eqref{pureYM}, we can readily evaluate (\ref{phase space}). For example, we obtain
\bea
&& (a^{{\rm YM}^2})_{-+}\,^{-+}=
\ft{g_{\rm YM}^2}{8\pi\langle\epsilon 3\rangle^{\j-2}\langle\epsilon 4\rangle^{\j+2}}\int_0^{2\pi}
 d\phi \int_0^{\fft{\pi}{2}}d\theta (\langle\epsilon 4\rangle\cos\theta-\langle\epsilon 3\rangle\sin\theta e^{i\phi})^{\j+2}
\times \cr &&
\qquad (\langle\epsilon 3\rangle\cos\theta+
\langle\epsilon 4\rangle\sin\theta e^{-i\phi})^{\j-2} \cos^4\theta\times (T \cot\theta+U \tan\theta)\,,
\cr &&
\cr && (a^{{\rm YM}^2})_{-+}\,^{+-}=
\ft{g_{\rm YM}^2}{8\pi\langle\epsilon 3\rangle^{\j-2}\langle\epsilon 4\rangle^{\j+2}}\int_0^{2\pi}
 d\phi \int_0^{\fft{\pi}{2}}d\theta  e^{4i\phi}(\langle\epsilon 4\rangle\cos\theta-\langle\epsilon 3\rangle\sin\theta e^{i\phi})^{\j-2}
\times \cr &&
\qquad (\langle\epsilon 3\rangle\cos\theta+
\langle\epsilon 4\rangle\sin\theta e^{-i\phi})^{\j+2}\sin^4\theta\times (T \cot\theta+U \tan\theta)\,,
\eea
and $(a^{{\rm YM}^2})_{+-}\,^{-+}=(a^{{\rm YM}^2})_{-+}\,^{+-}$ when the integral is evaluated. Same-helicity partial-wave amplitudes give
\bea
&& (a^{{\rm YM}^2})_{--}\,^{--}=(a^{{\rm YM}^2})_{++}\,^{++}=\ft{g_{\rm YM}^2}{8\pi\langle\epsilon 3\rangle^{\j}\langle\epsilon 4\rangle^{\j}}\int_0^{2\pi}d\phi\int_0^{\fft{\pi}{2}}d\theta
(\langle\epsilon 4\rangle\cos\theta-\langle\epsilon 3\rangle\sin\theta e^{i\phi})^{\j}\times
\cr && \qquad (\langle\epsilon 3\rangle\cos\theta+
\langle\epsilon 4\rangle\sin\theta e^{-i\phi})^{\j}\times \big(T\cot\theta+U \tan\theta\big)\,,
\eea
and other helicity-violation terms identically vanish, e.g., $a_{-+}\,^{--}=a_{--}\,^{-+}=0$. It is worth noting that above integrals fail to converge due to IR divergence. In the context of computing UV anomalous dimensions in QFT, these could be subtracted using that the stress-tensor is protected \cite{Caron-Huot:2016cwu}. However, in our context these reflect physical divergences of bulk anomalous dimensions as $R_{\rm AdS}\to\infty$. We thus regularize the above equations by introducing a small-angle cut-off $\epsilon<\theta<\frac{\pi}{2}-\epsilon$ which we will then compare with the bulk cutoff $n\rightarrow \infty$. The azimuthal integral can be readily evaluated, which gives
\bea
&& (a^{{\rm YM}^2})_{-+}\,^{-+}=
-\ft{g_{\rm YM}^2}{4\pi}\big(\gamma_{\rm E}+\log\epsilon+\psi_{\j-1}+\ft{2}{j}-\ft{1}{(j+2)(j+3)}+\ft{3}{(j-1)_4}\big)T+
\ft{3g_{\rm YM}^2}{4\pi (\j-1)_4}(-1)^{\j}U\,,
\cr &&
\cr && (a^{{\rm YM}^2})_{-+}\,^{+-}=
-\ft{g_{\rm YM}^2}{4\pi}\big(\gamma_{\rm E}+\log\epsilon+\psi_{\j-1}+\ft{2}{j}-\ft{1}{(j+2)(j+3)}+\ft{3}{(j-1)_4}\big)(-1)^{\j}U+\ft{3g_{\rm YM}^2}{2\pi (\j-1)_4}T\,,
\cr &&
\cr && (a^{{\rm YM}^2})_{--}\,^{--}=(a^{{\rm YM}^2})_{++}\,^{++}=-\fft{g_{\rm YM}^2}{4\pi}\big(\gamma_{\rm E}+\log\epsilon+\psi_{\j+1}\big)(T+(-1)^{\j}U)\,.\label{aYanghelicity}
\eea
As a simple check, acting on color-singlet states ($T,U\mapsto C_A$) and taking large spin,
we reproduce the famous logarithmic scaling of gauge theories,
$\gamma=\frac{-a}{\pi} \to +\frac{g_{\rm YM}^2}{2\pi^2}\log J$.

To compare with anomalous dimensions evaluated in CFT, we should rotate to parity basis
\bea
&& (a^{{\rm YM}^2})^{E}=\fft{1}{2}{\rm diag}\left(\begin{array}{l}
(a^{{\rm YM}^2})_{-+}\,^{-+}+(a^{{\rm YM}^2})_{-+}\,^{+-}+(+\leftrightarrow-) \\
(a^{{\rm YM}^2})_{--}\,^{--}+(a^{{\rm YM}^2})_{++}\,^{++} \end{array}\right)\,,
\cr && (a^{{\rm YM}^2})^{O}=\fft{1}{2}{\rm diag}\left(\begin{array}{l}
(a^{{\rm YM}^2})_{-+}\,^{-+}-(a^{{\rm YM}^2})_{-+}\,^{+-}+(+\leftrightarrow-) \\
(a^{{\rm YM}^2})_{--}\,^{--}+(a^{{\rm YM}^2})_{++}\,^{++} \end{array}\right)\,,
\eea
where $(+\leftrightarrow -)$ denotes flipping all helicity. Imposing following simple identification
\be
 \epsilon=\frac{e^{-\gamma_{\rm E}}}{2n}\,,
\ee
and using $ \lambda^{(\rm e1)}_{VVV}  = g_{\rm YM}/(16\sqrt{2})$ from eq.~\eqref{dictionary}, we then find a perfect match with the CFT anomalous dimension in eq.~\eqref{largenVVVV}:
\be
\gamma^{E/O}_{11}|_{n\rightarrow\infty}=-\fft{1}{\pi}(a^{{\rm YM}^2})^{E/O}\,.
\ee

\subsubsection{Higher-derivative corrections}

Let us start with the pure higher-derivative interaction (e.g. at both vertices).
Using the amplitudes recorded in eq.~\eqref{pureH}, we can immediately conclude that $(a^{{\rm H}^2})_{--}\,^{--}=(a^{{\rm H}^2})_{++}\,^{++}=0$, because $\mathcal{M}^{\rm H^2}_{1_{-}2_{-}3_{+}4_{+}}$ only have $s$-channel pole and thus is evaluated to be identically zero, which nicely agrees with predictions from CFT. On the other hand, $(a^{{\rm H}^2})_{-+}\,^{-+}$ and $(a^{{\rm H}^2})_{-+}\,^{+-}$
contributes with $T$ and $U$ factors separately, giving
\bea
&& (a^{{\rm H}^2})_{-+}\,^{-+}=
\ft{g_{\rm H}^2}{32\pi\langle\epsilon 3\rangle^{\j-2}\langle\epsilon 4\rangle^{\j+2}}\int_0^{2\pi}
 d\phi \int_0^{\fft{\pi}{2}}d\theta  (\langle\epsilon 4\rangle\cos\theta-\langle\epsilon 3\rangle\sin\theta e^{i\phi})^{\j+2}\times
 \cr &&
\qquad (\langle\epsilon 3\rangle\cos\theta+
\langle\epsilon 4\rangle\sin\theta e^{-i\phi})^{\j-2}(\cos\theta)^3\sin\theta(\cos(2\theta)-3)\times U\,,
\cr &&
\cr && (a^{{\rm H}^2})_{-+}\,^{+-}=
\ft{g_{\rm H}^2}{32\pi\langle\epsilon 3\rangle^{\j-2}\langle\epsilon 4\rangle^{\j+2}}\int_0^{2\pi}
 d\phi \int_0^{\fft{\pi}{2}}d\theta  e^{4i\phi}(\langle\epsilon 4\rangle\cos\theta-\langle\epsilon 3\rangle\sin\theta e^{i\phi})^{\j-2}\times
 \cr &&
\qquad(\langle\epsilon 3\rangle\cos\theta+
\langle\epsilon 4\rangle\sin\theta e^{-i\phi})^{\j+2}(\sin\theta)^3\cos\theta(\cos(2\theta)+3)\times T\,.
\eea
We can readily evaluate the integrals and find
\be
(a^{{\rm H}^2})_{-+}\,^{-+}=\fft{3g_{\rm H}^2 s^2}{4\pi^2(\j-1)_4}(-1)^{\j}U\,,
\quad (a^{{\rm H}^2})_{-+}\,^{+-}
=\fft{3g_{\rm H}^2 s^2}{4\pi^2(\j-1)_4}T\,,
\ee
and simultaneously flipping helicity $+\leftrightarrow-$ gives the same answer. Rotating to the
Even/Odd parity basis readily gives
\be
(a^{{\rm H}^2})^{E/O}=\fft{3g_{\rm H}^2 s^2}{4\pi^2(\j-1)_4}\left(
\begin{array}{cc}
 \big(\mp T+(-1)^{\j}U\big) & 0 \\
 0 & 0 \\
\end{array}
\right)\,.
\ee
Using $\lambda_{VVV}^{\rm e2}=g_H/(8\sqrt{2})$ from eq.~\eqref{dictionary} and $s=4n^2$ from eq.~\eqref{dictCFTQFT}, we achieve a perfect agreement with CFT anomalous dimensions
from eq.~\eqref{largenVVVV}.
\be
\gamma^{E/O}_{22}|_{n\rightarrow\infty}=-\fft{1}{\pi}(a^{{\rm H}^2})^{E/O}\,.
\ee

The contact ambiguity that has the same scaling
dimension as the $a_H^2$ interaction (see eq.~\eqref{pureH})
affects the $\j=2$ OPE data, making the preceding partial wave valid only for $\j>2$.
We believe that all other results are valid for $\j>1$
(with similar comments applying to the Lorentzian inversion formula results from the preceding section).

Finally, let us look at the product of Yang-Mills and higher-derivative couplings. Here, there are two kinds of amplitudes, for example $\mathcal{M}_{--+-}$ and $\mathcal{M}_{-+++}$, which is not symmetric and thus give slightly different partial-wave amplitudes that form a non-symmetric and anti-diagonal matrix; eigenvalues of the resulting matrix should agree with CFT eigenvalues
from eq.~\eqref{largenVVVV} (that is, we only compare up to similarity transformation).

For example, we find some of $(a^{\rm mix})=a\big|_{g_{\rm YM}g_{\rm H}}$
for $\mathcal{M}_{--+-}$ type mixing reads
\bea
&& (a^{{\rm mix}})_{-+}\,^{--}=-\ft{g_{\rm YM}g_{\rm H}s}{8\pi\langle\epsilon 3\rangle^{\j}\langle\epsilon 4\rangle^{\j}}\int_0^{2\pi}
 d\phi \int_0^{\fft{\pi}{2}}d\theta  e^{-2i\phi}(\langle\epsilon 4\rangle\cos\theta-\langle\epsilon 3\rangle\sin\theta e^{i\phi})^{\j+2}
\times \cr &&
\qquad(\langle\epsilon 3\rangle\cos\theta+
\langle\epsilon 4\rangle\sin\theta e^{-i\phi})^{\j-2}\sin(2\theta)\times (T \cot\theta+U \tan\theta)\,,
\cr &&
\cr && (a^{{\rm mix}})_{--}\,^{+-}=-\ft{g_{\rm YM}g_{\rm H}s}{8\pi\langle\epsilon 3\rangle^{\j-2}\langle\epsilon 4\rangle^{\j+2}}\int_0^{2\pi}
 d\phi \int_0^{\fft{\pi}{2}}d\theta  e^{2i\phi}(\langle\epsilon 4\rangle\cos\theta-\langle\epsilon 3\rangle\sin\theta e^{i\phi})^{\j}
\times \cr &&
\qquad(\langle\epsilon 3\rangle\cos\theta+
\langle\epsilon 4\rangle\sin\theta e^{-i\phi})^{\j}\sin^2\theta \cos^2\theta\times (T \cot\theta+U \tan\theta)\,.
\eea
$(a^{{\rm mix}})_{-+}\,^{++}$ gives the same as $(a^{{\rm mix}})_{-+}\,^{--}$, and $(a^{{\rm mix}})_{--}\,^{-+}$ is similar to $(a^{{\rm mix}})_{--}\,^{+-}$ but flipping $e^{2i\phi}\rightarrow e^{-2i\phi}$. Though the integrand looks a bit different when we flipping $+\leftrightarrow -$, we find they give the same result
\bea
&& (a^{{\rm mix}})_{-+}\,^{--}=(a^{{\rm mix}})_{-+}\,^{++}=-\fft{g_{\rm YM}g_{\rm H}s}{8\pi\j(\j-1)}\big(T+(-1)^{\j}U\big)\,,
\cr &&
\cr &&
(a^{{\rm mix}})_{--}\,^{-+}=(a^{{\rm mix}})_{--}\,^{+-}=-\fft{g_{\rm YM}g_{\rm H}s}{8\pi(\j+1)(\j+2)}\big(T+(-1)^{\j}U\big)\,,
\eea
and the same for flipping $\pm\rightarrow\mp$. Now we can rotate to the parity basis. To compare with CFT calculation where we record
$\gamma_{12}$ and $\gamma_{21}$ separately, we should be careful about clarifying $a^{\rm mix}_{12}$ and $a^{\rm mix}_{21}$:
$\gamma_{12}$ corresponds to $a^{\rm mix}$ with different helicity in $(h_2, h_3)$ , and $\gamma_{21}$ corresponds to $a^{\rm mix}$ with same helicity in $(h_2,h_3)$. We find
\bea
&& (a^{\rm mix}_{12})^{E/O}=\fft{1}{2}
\left(
\begin{array}{cc}
 0 & (a^{\rm mix})_{-+}\,^{--}+(a^{\rm mix})_{+-}\,^{++} \\
 (a^{\rm mix})_{--}\,^{+-}+(a^{\rm mix})_{++}\,^{-+} & 0 \\
\end{array}
\right)\,,
\cr &&
\cr && (a^{\rm mix}_{21})^{E/O}=\pm\fft{1}{2}
\left(
\begin{array}{cc}
 0 & (a^{\rm mix})_{-+}\,^{++}+(a^{\rm mix})_{+-}\,^{--} \\
 (a^{\rm mix})_{--}\,^{-+}+(a^{\rm mix})_{++}\,^{+-} & 0 \\
\end{array}
\right)\,.
\eea
The signs work out so that, when we add the contributions from the two vertices,
the parity-even part doubles and the odd part cancels out ($a_{12}^O+a_{21}^O=0$), as found in the preceding section.
Using the dictionary $\lambda_{VVV}^{\rm e2}=g_H/(8\sqrt{2})$ and $\lambda_{VVV}^{\rm e2}=g_H/(16\sqrt{2})$ from eq.~\eqref{dictionary} and $s=4n^2$, we find that the eigenvalues of $a^{\rm mix}$ precisely coincide with $\gamma_{12}$ and $\gamma_{21}$ in eq.~\eqref{largenVVVV} up to $-1/\pi$, i.e.,
\be
\gamma^{E/O}\big|_{n\rightarrow\infty}\sim -\fft{1}{\pi}(a^{\rm mix})^{E/O}\,,
\ee
and $\sim$ denotes the equivalence up to similarity transformation.

\section{Conclusion}
\label{conclusion}

In this paper, we introduced a helicity basis for conformal blocks
of conserved currents of any spins in three-dimensional CFTs.
We observed that the concept of helicity is conformally invariant (see subsection \ref{ssec:helicity is conf}) and can be defined without reference to any particular formalism such as momentum space.
This ensures that the helicity basis plays nicely with crossing symmetry.
We found evidence of this in the OPE decomposition of mean-field correlators, which turns out nicely diagonal (see eq.~\eqref{cMFT T}, and we further computed the CFT${}_3$ OPE
data dual to tree-level gluon scattering of Yang-Mills theory in AdS${}_4$, including higher-derivative corrections.

The YM${}_4$ calculation was done using the spinning Lorentzian inversion formula
(see eq.~\eqref{gamma11}, \eqref{gamma11 odd} and following), which gives the OPE data for
sufficiently large spin $J>J_*$, where we expect $J_*=1$ without including
higher-derivative corrections and $J_*=2$ with them.
The anomalous dimensions follow a simple diagonal / off-diagonal pattern and precisely match,
in the large-twist limit, with the partial waves in the flat space limit of the bulk theory, shown in eq.~\eqref{largenVVVV}.
We found a simple one-to-one dictionary between on-shell three-point interactions
in bulk AdS${}_4$ and three-point helicity structures (see eq.~\eqref{OPEg}).

We expect that a calculation of the $6j$ symbol (also known as crossing kernel) in the helicity basis
could thus greatly help bootstrap calculations involving conserved currents and stress tensors
in 3d CFTs.  We expect the $6j$ symbols to be diagonal in helicity basis.
It is also worth exploring if the helicity basis could also help numerical work by diagonalizing certain steps.

In higher spacetime dimensions, whether a basis exists which would diagonalize mean-field correlators
remains an open question. Better understanding the flat-space limit of massless-massless-massive three-point functions could shed light on this question.

In perturbation theory, our findings pave the way for a study of loop corrections in YM${}_4$
with a four-dimensional treatment of infrared effects.
Compared with flat space, AdS physics comes with a built-in infrared regulator,
and an interesting fact is that leading double-twist states (the $n=0$ trajectory)
do not have a definite helicity (see eq.~\eqref{leading trajectory}).
The notion that zero-energy gluons do not have helicity resonates with findings from the
asymptotic symmetry context (see for example \cite{He:2020ifr}), and it would be interesting
to make this connection closer.  Eq.~\eqref{++++} suggests that the tree-level amplitude for four
same-helicity gluons is not identically zero even in flat space, but retains a sort of distributional component around zero energy, which could be important for unitarity calculations in flat space.

Nonperturbatively, we expect the helicity basis to be particularly convenient for uncovering the implications
of crossing symmetry on stress tensor correlators in CFT${}_3$ and the dual gravitational physics.

\begin{acknowledgments}
We would like to thank for Nikhil Anand for useful conversations
and initial collaboration on AdS calculations, and Petr Kravchuk for discussions.
Work of S.C.-H. is supported by the National Science and Engineering Council of Canada, the Canada Research Chair program, the Fonds de Recherche du Qu\'ebec - Nature et Technologies,
the Simons Collaboration on the Nonperturbative Bootstrap,
and the Sloan Foundation.
Work of Y.-Z.L. is supported in parts by the Fonds de Recherche du Qu\'ebec - Nature et Technologies.
\end{acknowledgments}

\begin{appendix}

\section{$\langle VVV\rangle$ from Witten-diagram}
\label{Witten}

In this appendix, we start from AdS Lagrangian in $d=3$ to derive $\langle VVV\rangle$ three-point functions. From helicity basis we constructed in the main text, it follows that $\langle VVV\rangle$ has three independent structures, and it is expected the first structure corresponds to the Yang-Mills vertex and the higher-derivative coupling in AdS is captured by the second two (the odd and even ``same-helicity'' ones, which are analytic in spin for $J\geq 0$).
Our starting point is the following Lagrangian for Yang-Mills in AdS (omitting gravity):
\be
\mathcal{L}= 
-\fft{1}{4g_{\rm YM}^2}F_{\mu\nu}^a F^{\mu\nu a}
+ \frac{\theta}{32\pi^2} F_{\mu\nu}^a \tilde{F}^{\mu\nu a}
-\frac{f^{abc}}{3g_{\rm YM}^3}\left(
g_{\rm H}\, F_\mu\,^{\nu a} F_\nu\,^{\rho b} F_\rho\,^{\mu c}
+
g'_{\rm H}\, \tilde{F}_\mu\,^{\nu a} \tilde{F}_\nu\,^{\rho b} \tilde{F}_\rho\,^{\mu c}\right)+\cdots\,,\label{Lang}
\ee
where $a, b, c$ are ${\rm SU}(N)$ group indices, $f^{abc}$ is the structure constant, $\tilde{F}_{\mu\nu}=\frac12\epsilon_{\mu\nu\sigma\rho}F^{\sigma\rho}$ and $\cdots$ is other terms that are not relevant to our purpose.
After rescaling the fields by the coupling to make $A$ canonically normalized,
it follows that we have two three-point gluon vertices
\be\begin{aligned}
& \text{Yang-Mills:} &\qquad & -g_{\rm YM}\, f^{abc}\partial_\mu A_\nu^a A^{\mu b}A^{\nu c}\,,
\\
 & \text{Higher-derivative:} &\qquad &-\frac{g_{\rm H}}3 f^{abc} \, F_\mu\,^{\nu a} F_\nu\,^{\rho b} F_\rho\,^{\mu c}+ \mbox{odd part}\,,\label{threevert}
\end{aligned}\ee
where only the linearized part of $F_{\mu\nu}$ will contribute in the second case.

It is most convenient to work with the AdS embedding formalism \cite{Costa:2014kfa} where the bulk-to-boundary propagator with conformal dimension $\Delta$ and spin $\j$ is \cite{Costa:2014kfa}
\be
\Pi_{\Delta,\j}(Y,W;X,Z)=\mathcal{C}(\Delta,\j) \fft{\big((-2X\cdot Y)(W\cdot Z)+2(W\cdot X)(Z\cdot Y)\big)^\j}{(-2X\cdot Y)^{\Delta+\j}}\,,
\ee
where $X$ and $Z$ are embedding coordinate and auxiliary polarization respectively for boundary CFT,  similarly $Y$ and $W$ are $(d+2)$-dimensional embedding coordinate and polarization for the
bulk AdS${}_{d+1}$, which are constrained by
\bea
X^2=X{\cdot}Z=Z^2=0\,,\quad Y^2=-1\,,\quad Y{\cdot}W=W^2=0\,,
\eea
and have the further redundancy $Z\simeq Z+\alpha X$.
The normalization factor reads
\be
\mathcal{C}(\Delta,\j)=\fft{(\j+\Delta-1)\Gamma(\Delta)}{2\pi^{\fft{d}{2}}(\Delta-1)\Gamma(\Delta+1-\fft{d}{2})}\,.
\ee
Derivatives in AdS can be evaluated using the bulk covariant derivative operator \cite{Costa:2014kfa}
\bea
\nabla_A=\fft{\partial}{\partial Y^A}+Y_A (Y\cdot \fft{\partial}{\partial Y})+W_A (Y\cdot \fft{\partial}{\partial W})\,.
\eea
which commutes with the constraints.
It is also convenient to introduce the differential operator $K_A$ \cite{Costa:2014kfa}
\be
K_A^W =
\left(\fft{\partial}{\partial W^A}+Y_A (Y{\cdot}\fft{\partial}{\partial W})\right)\left(\fft{d-3}{2}+ W{\cdot}\fft{\partial}{\partial W}\right)
-\fft{1}{2}W_A\left(\fft{\partial^2}{\partial W{\cdot} \partial W}+(Y{\cdot}\fft{\partial}{\partial W})^2\right)\,,
\ee
which helps do index contractions in AdS:
\be
\sum_W f(W^\ast)g(W)=\fft{1}{\j! (\fft{d-1}{2})_\j}f(K^W)g(W)\,.
\ee
With these ingredients, we are ready to compute $\langle VVV\rangle$ by performing the following integrals over (Euclidean) AdS $Y^2=-1$:
\bea
\langle V(X_1)V(X_2)V(X_3)\rangle^{\rm YM}&=& -g_{\rm YM }f^{abc}\,\mathcal{C}_{d-1,1}^{\fft{3}{2}}\int_{\rm EAdS} dY \sum_{W_1,W_2}(W_1^\ast\cdot\nabla
\Pi_{d-1,\j}(Y,W_2^\ast;X_1,Z_1))
\cr && \times\Pi_{d-1,\j}(Y,W_1;X_2,Z_2)\Pi_{d-1,\j}(Y,W_2;X_3,Z_3) + \mbox{(5 permutations)}
\cr&&
\cr
\langle V(X_1)V(X_2)V(X_3)\rangle^{\rm H}&=&-2g_{\rm H}f^{abc}\,\mathcal{C}_{d-1,1}^{\fft{3}{2}} \int_{\rm EAdS} dY \sum_{W_1,W_2,W_3}
\cr &&\hspace{-12mm}\times
\big(W^\ast_1\cdot\nabla\Pi_{d-1,1}(Y,W_2;X_1,Z_1)-W_2\cdot \nabla \Pi_{d-1,1}(Y,W^\ast_1;X_1,Z_1)\big)
\cr &&\hspace{-12mm}\times
\big(W^\ast_2\cdot\nabla\Pi_{d-1,1}(Y,W_3;X_2,Z_2)-W_3\cdot \nabla \Pi_{d-1,1}(Y,W^\ast_2;X_2,Z_2)\big)
\cr && \hspace{-12mm}\times
\big(W^\ast_3\cdot\nabla\Pi_{d-1,1}(Y,W_1;X_3,Z_3)-W_1\cdot \nabla \Pi_{d-1,1}(Y,W^\ast_3;X_3,Z_3)\big)\,,
\eea
where the factor $\mathcal{C}_{d-1,1}^{\fft{3}{2}}$ ensures our $VV$ two-point function follows the CFT normalization.
The integrals can be done in elementary ways, for example using Feynman/Schwinger parameters.
We obtain (in $d=3$):
\bea
&& \langle VVV\rangle^{\rm YM}=\frac{3g_{\rm YM}}{16\sqrt{2}}f^{abc}\fft{H_{23}V_1+H_{13}V_2+H_{12}V_3+V_1V_2V_3}{(-2X_1\cdot X_2)^{\fft{3}{2}}
(-2X_1\cdot X_3)^{\fft{3}{2}}(-2X_2\cdot X_3)^{\fft{3}{2}}}\,,
\cr &&
\cr && \langle VVV\rangle^{\rm H}=\frac{-g_{\rm H}}{8\sqrt{2}} f^{abc}
\fft{H_{23}V_1+H_{13}V_2+H_{12}V_3+5V_1V_2V_3}{(-2X_1\cdot X_2)^{\fft{3}{2}}
(-2X_1\cdot X_3)^{\fft{3}{2}}(-2X_2\cdot X_3)^{\fft{3}{2}}}\,,
\eea
where $H_{ij}$ follows the definition in eq.~(\ref{Hij}) and $V_i$ is defined by (see \cite{Costa:2011mg} for more details)
\be
V_i:=V_{i,jk}=\fft{(X_i\cdot X_k)(Z_i\cdot X_j)-(X_i\cdot X_j)(Z_i\cdot X_k)}{X_j\cdot X_k}\,.
\ee
To project onto the conformal frame $(0,x,\infty)$, we parameterize $X_i, Z_i$ (in embedding lightcone coordinates) as
\bea
&& X_1=(1,0,0)\,,\quad Z_1=(0,0,\epsilon_1)\,,\quad X_2=(1,x^2,x)\,,\quad Z_2=(0,2\epsilon_2\cdot x,\epsilon_2)\,,
\cr &&  X_3=(0,1,0)\,,\quad Z_3=(0,0,\epsilon_3)\,.
\eea
We thus end up with
\bea
\langle VVV\rangle^{\rm YM}&=&\fft{3g_{\rm YM}f^{abc}}{16\sqrt{2}|x|^{3}}\left[
(x\cdot\epsilon_1)(\epsilon_2\cdot\epsilon_3)+(x\cdot \epsilon_2)(\epsilon_1\cdot\epsilon_3)
-(x\cdot\epsilon_3)(\epsilon_1\cdot\epsilon_2)+\frac{(x\cdot\epsilon_1)(x\cdot\epsilon_2)(x\cdot\epsilon_3)}{x^2}\right]\,,
\cr
\langle VVV\rangle^{\rm H}&=&\fft{-g_{\rm H}f^{abc}}{8\sqrt{2}|x|^{3}}\left[
(x\cdot\epsilon_1)(\epsilon_2\cdot\epsilon_3)+(x\cdot \epsilon_2)(\epsilon_1\cdot\epsilon_3)
-(x\cdot\epsilon_3)(\epsilon_1\cdot\epsilon_2)
-3\frac{(x\cdot\epsilon_1)(x\cdot\epsilon_2)(x\cdot\epsilon_3)}{x^2}\right]\,.
\cr &&
\eea
Comparing the above results with $M_V B_V$ (see eq.~(\ref{xspace structure}) and eq.~(\ref{VVstcor})) for conserved currents, the agreement can be easily observed and the OPE coefficients can be readily read off
\bea
\lambda_{VVV}^{(\rm e1)}=\fft{g_{\rm YM}}{16\sqrt{2}}\,,\qquad \lambda_{VVV}^{(\rm e2)}=\fft{g_{\rm H}}{8\sqrt{2}}\,,
\qquad \lambda_{VVV}^{(\rm o2)}=\fft{g'_{\rm H}}{4\sqrt{2}\pi}\,,
\label{OPEg}
\eea
where we strip off color factors by defining $\langle VVV\rangle$ three-point functions as
\be
\langle VVV\rangle^{\pa}=f^{abc}\times \lambda_{VVV}^{\pa}T^{\pa}_{111}\,,
\ee
in which $\pa$ runs through structures in eq.~(\ref{even odd}).

\section{Simplifying Fourier transforms using spinors}
\label{app:spinors}

We find that much of the calculations can be streamlined analytically by representing the polarization vectors as a product of two spinors (see also \cite{Karateev:2018oml}).

Given a two-component spinor $|\eps\rangle$, we define $\langle \eps| \equiv |\eps\rangle^T\.i\sigma_2$,
and parametrize the null polarizations as
\be
 \eps_i^\mu \equiv \frac12 \langle \eps_i| \sigma^\mu|\eps_i\rangle
\ee
where $\sigma^\mu$, $\mu=1,2,3$, are Pauli matrices. This vector is automatically null.
Other useful identities include:
\be\begin{aligned}
\langle a|\sigma^\mu|b\rangle\langle c|\sigma_\mu|d\rangle &= -\langle ac\rangle \langle bd\rangle-\langle ad\rangle \langle bc\rangle, \quad
(\eps_1,p,\eps_3) = \frac{i}{2}\langle \eps_1\eps_3\rangle \langle \eps_1|p|\eps_3\rangle\,.
\end{aligned}
\ee
The three-point helicity structures in eq.~\eqref{Tpm} are very simple in terms of spinors:
\be\begin{aligned}
 T^{\pm,\pm}_{123}(p) &= \frac{(4\pi)^{\fft{3}{2}}}{2^{\tau_1+\tau_2-\Delta_3}} \left(\frac{-i \langle \eps_3|p|\eps_3\rangle}{\sqrt{2}}\right)^{\j_3-\j_1-\j_2} \langle \eps_1\eps_3\rangle^{2\j_1}\langle \eps_2\eps_3\rangle^{2\j_2}|p|^{\beta_{12;3}-3}
\\
 &\quad \times \left(\frac{1-\xi_{1,p,3}}{2}\right)^{2\j_1}\left(\frac{1+\xi_{2,p,3}}{2}\right)^{2\j_2}
\label{Tpm spinor}
\end{aligned}\ee
where $\xi_{i,p,3} \equiv \frac{\langle \eps_i|p|\eps_3\rangle}{|p| \langle\eps_i\eps_3\rangle}$ is a
measure of spin along the $p$ axis.

When we go to Fourier space using eq.~\eqref{Fourier} and its derivatives, we find remarkable simplifications
thanks to the fact that the vector $\eps_3$ is orthogonal to all other vectors multiplying $p$.
In fact the Fourier-transform involves only similar-looking objects and
we were able to Fourier-transform the generic term analytically:
\be\begin{aligned}
 &\int \frac{d^3p}{(2\pi)^3} e^{ip{\cdot}x}p^{2k} \left(-i \langle \eps_3|p|\eps_3\rangle\right)^{\j} (\xi_{1,p,3})^a(\xi_{2,p,3})^b
 \\
 &\qquad\qquad= \frac{2^{2k+J}}{\pi^{\fft32}}\frac{\langle \eps_3|x|\eps_3\rangle^{\j}}{x^{2k+2\j+3}}\times
  \sum_{a',b'} f_{a',b'}^{a,b} \frac{\Gamma\big(\tfrac{a'+b'+3}{2}+k+\j\big)}{\Gamma\big(\tfrac{a+b}{2}-k\big)}
  (\xi_{1,x,3})^{a'}(\xi_{2,x,3})^{b'} \label{Fourier spinors}
\end{aligned}\ee
where the sum runs over $a'\leq a$, $b'\leq b$ such that $(a+b-a'-b')$ is even,
and $f$ is the following combinatorial factor
\be
 f_{a',b'}^{a,b} = \frac{(2i)^{a'+b'}}{2^{a+b}} \frac{a!}{a'(a-a')!}\frac{b!}{b'!(b-b')!}\frac{(a+b-a'-b')!}{\big(\tfrac{a+b-a'-b'}{2}\big)!}\,.
\ee
Using the integral \eqref{Fourier spinors} it is straightforward to convert the structures in eq.~\eqref{Tpm spinor} back and forth
between momentum and coordinate space.
The other operations also have simple forms:
\begin{itemize}
\item Conformal inversion: this takes $(\infty,x,0)\mapsto(0,x^\mu/x^2,\infty)$ and
$|\eps_2\rangle \mapsto i\frac{x|\eps_2\rangle}{|x|}$. The net effect is simply: $\xi_{2,x}\mapsto 1/\xi_{2,x}$
and $\langle \eps_2\eps_3\rangle \mapsto i\langle \eps_2\eps_3\rangle \xi_{2,x}$.
\item Shadow transform: two-point functions in position and Fourier space are simply:
\be\begin{aligned}
\langle \mathcal{O}_1(0)\mathcal{O}_2(x) \rangle &= \frac{\langle 1|x|2\rangle^{2J}}{(-2)^J|x|^{\Delta+J}},
\\
\langle \mathcal{O}_1(0)\mathcal{O}_2(p) \rangle &= \frac{(4\pi)^{\frac32}\Gamma(\tfrac{3}{2}-\Delta)(\tfrac12)_\j}
{(-2)^\j4^\Delta\Gamma(\Delta+\j)} |p|^{\Delta-\frac32} \langle \eps_1\eps_2\rangle^{2\j}\ \times {}_2F_1\big(-\j,\tfrac32-\Delta,\tfrac12,\xi_{1,p,2}^2\big)\,.
\end{aligned}\ee

\item Index contractions: the sum over a basis of spin-$\j$ states \eqref{Tod pairing} becomes:
\be
\sum_{\epsilon} f(\epsilon^*)g(\epsilon)=\frac{(-2)^{\j}}{(2\j)!} f(\partial_\epsilon)g(\epsilon)\,.
\ee

\end{itemize}

\section{More on conformal blocks}
\label{more conformal blocks}
\subsection{Series expansion of conformal blocks}
\label{series}

Here we review how to obtain a series expansion for conformal blocks
using the conformal Casimir operator, following the work of ref.~\cite{Hogervorst:2013sma} for scalar blocks.
The same recursion will come in handy for doing certain inversion integrals in the next subsection.
The conformal symmetry generators act on a spinning primary $\OO(x,\eps)$ of dimension $\Delta$ as
\be\begin{aligned}
 D &= x^\mu \partial_{x \mu}+\Delta, &\qquad
 J^{\mu\nu} &= x^\mu \partial_x^\nu - x^\nu \partial_x^\mu + \eps^\mu \partial_\eps^\nu-\eps^\nu \partial_\eps^\mu,\\
 P^\mu &= \partial_x^\mu, &\qquad
 K^\mu &= x^2\partial_x^\mu - 2x^\mu  D + 2(x{\cdot}\eps \partial_\eps^\mu - \eps^\mu x{\cdot}\partial_\eps),
\end{aligned}\ee
where $D$, $J$, $P$ and $K$
generate respectively dilations, rotations, translations and special conformal transformations.
The Casimir operator is then $\mathcal{C}_2= D^2-\tfrac12 J_{\mu\nu} J^{\mu\nu} +\tfrac12\{ P_\mu,K^\mu\}$,
which has eigenvalue $\mathcal{C}_{\Delta,\j} =\Delta(\Delta-d)+\j(\j+d-2)$ if $\OO$ is a rank-$\j$ tensor.

Four-point conformal blocks are (by definition) eigenfunctions of the Casimir acting
on the pair of operators $1,2$:
\be
\mathcal{C}= D_{12}^2 -\frac12 J_{12}^{\mu\nu}(J_{12})_{\mu\nu}  + \frac12\{P_{12}^\mu,(K_{12})_\mu \}
\ee
where the subscripts denote the fields on which the generators act: $D_{12}\equiv D_1+D_2$ etc.
This form of the Casimir operator however can't be used for the correlator in the frame $0,x,y^{-1},\infty$.
The problem is that $P_{12}$ does not preserve the condition $x_1=0$.
Fortunately, there is a simple solution:
we can use conformal invariance of the four-point correlator to rewrite $P_{12}\mapsto -P_{34}$.
Accounting for a commutator, the Casimir is
\be
\mathcal{C}= 
\left[D_{12}(D_{12}-d) -\frac12 J_{12}^{\mu\nu}(J_{12})_{\mu\nu}\right]+  K_{x}^\mu K_{y\mu} \equiv \mathcal{C}^{(0)} +\mathcal{C}^{(1)}.
\label{Casimir 0 infinity}
\ee
Notice that $\mathcal{C}^{(0)}$ is homogenous in $x$, while $\mathcal{C}^{(1)}$ increases the weight in $x$ and $y$ by one unit.
Furthermore, the former is diagonalized by the three-point structures $\mathcal{P}_{\Delta,\j}^{\pa\pb}$ in eq.~(\ref{gegen}).
This suggests writing the block as an infinite series in $\mathcal{P}_{\Delta,\j}^{\pa\pb}$:
\be
G^{(a,b)}_{\j,\Delta} (z,\zb) = \sum_{m=0}^{\infty} \sum_{k=-m}^m A^{(aa')(bb')}_{m,k} \mathcal{P}_{\Delta+m,\j+k}^{\pa,\pb}(\hat{x},\hat{y})\,,\label{seriesexp}
\ee
such that the Casimir (\ref{Casimir 0 infinity}) gives a recursion relation for the coefficients $A$.
For example, for scalar operators, applying the Casimir to the Gegenbauer polynomials (\ref{gegen}) gives
\be
 \mathcal{C}^{(0)} \mathcal{P}^{\pa,\pb}_{\Delta,\j} = \mathcal{C}_{\Delta,\j}\mathcal{P}^{\pa,\pb}_{\Delta,\j}, \qquad
 \mathcal{C}^{(1)} \mathcal{P}^{\pa,\pb}_{\Delta,\j} =
 \gamma^{a,b,-}_{\Delta,\j} \mathcal{P}^{\pa,\pb}_{\Delta+1,\j-1}+\gamma^{a,b,+}_{\Delta,\j} \mathcal{P}^{\pa,\pb}_{\Delta+1,\j+1},\label{Precur}
\ee
with
\bea
&& \gamma^{a,b,+}_{\Delta,J}=(\Delta+J+2a)(\Delta+J+2b)\,,
\cr && \gamma^{a,b,-}_{\Delta,J}=\frac{J (d+J-3) (-2 a+d-\Delta +J-2) (-2 b+d-\Delta +J-2)}{(d+2 J-4) (d+2 J-2)}\,,\label{gammarecur}
\eea
from which one deduces the recursion \cite{Hogervorst:2013sma}
\be
 \big(\mathcal{C}_{\Delta,\j} - \mathcal{C}_{\Delta+m,\j+k}\big) A_{m,k} = \gamma^-_{\Delta+m-1,\j+k-1} A_{m-1,k+1} + \gamma^+_{\Delta+m-1,\j+k+1} A_{m-1,k-1}.\label{recurs}
\ee
Note $a,b$ in $\gamma^{a,b,\pm}_{\Delta,\j}$ is not representing the structure index, they are simply $a=1/2(\Delta_2-\Delta_1), b=1/2(\Delta_3-\Delta_4)$. These coefficients eq.~(\ref{gammarecur}) will also play important role when we are dealing with MFT, see appendix \ref{GFFmore}.

This method allows to extend this result straightforwardly to spinning operators \cite{Kravchuk:2017dzd}.
We can use eq.~(\ref{limit of block}) to construct $\mathcal{P}^{\pa,\pb}_{\Delta,\j}$ from three-point functions, and in general $\mathcal{P}^{\pa,\pb}_{\Delta,\j}$ can be organized as Gegenbaur polynomials and their derivatives, which is consistent with group theoretical analysis for projectors \cite{Costa:2016hju}.

\subsection{Inverting powers of cross-ratios times Gegenbauers}
\label{GFFmore}

In this appendix, we present a more compact approach to deal with the spinning MFT. To be more precise, there is a surprisingly concise and powerful trick that can be used perform Lorentzian inversion formula for a scalar MFT correlator extended with Gegenbauer polynomial, namely
\be
\mathcal{G}=\fft{u^{\fft{p}{2}}}{v^{\fft{p}{2}+a}}\tilde{C}_{\j'}\big(\xi'\big)\,,\label{moreGFF}
\ee
where $u=z\bar{z}$, $v=(1-z)(1-\bar{z})$ and $\xi'=(1-u-v)/(2\sqrt{uv})$. The punchline is that we find a recursion relation for OPE data associated with above correlator, see eq.~(\ref{GFF integral}). This formula enjoys more general applications, since as just shown, conformal blocks admit series expansion of precisely this form (after interchanging operators $3$ and $4$ operators).
This was used in \cite{Caron-Huot:2020ouj} to estimate Lorentzian inversion integrals at large dimensions in the 3$d$-Ising model.
In this paper, we apply the formula to $\mathcal{G}=\mathcal{D}_{\downarrow}G$ for spinning MFT,
which is a finite sum of terms \eqref{moreGFF}.

The starting point of the recursion is the scalar case, $\j'=0$. The relevant OPE data can be found in literatures, at least for equal external operators $a=b$, e.g., \cite{Fitzpatrick:2011dm,Simmons-Duffin:2016wlq}. There is a trivial modification that also works for independent $a,b,p$:
\ba
 c^{a,b}_{0,p,\j}(\Delta)&=&
  \frac{\Gamma\big(\tfrac{d-p}{2}-a\big)\Gamma\big(\tfrac{d-p}{2}+b\big)}
 {2 \Gamma\big(\tfrac{p}{2}+a\big)\Gamma\big(\tfrac{p}{2}-b\big)}
 \frac{\Gamma\big(\tfrac{\Delta+\j}{2}+a\big)\Gamma\big(\tfrac{\Delta+\j}{2}-b\big)}
{\Gamma\big(\tfrac{d-\Delta+\j}{2}-a\big)\Gamma\big(\tfrac{d-\Delta+\j}{2}+b\big)}\nn
\\ && \times
\frac{\Gamma(\Delta-1)\Gamma\big(\j+\tfrac{d}{2}\big)\Gamma(d-\Delta+\j)}
 {\Gamma(\j+1)\Gamma\big(\Delta-\tfrac{d}{2}\big) \Gamma(\Delta-1+\j)}
\frac{
\Gamma\big(\tfrac{p-\Delta+\j}{2}\big)\Gamma\big(\tfrac{p-d+\Delta+\j}{2}\big)}
{\Gamma\big(\tfrac{-p+d+\Delta+\j}{2}\big)\Gamma\big(\tfrac{-p+2d-\Delta+\j}{2}\big)}.  \label{GFF integral 0}
\ea
This was tested by checking that the obtained OPE coefficients (obtained from the residues at $\Delta=p+\j+2m$) reproduce
the series expansion of the bracket in eq.~(\ref{moreGFF}) with $\j'=0$ to high order. To proceed on generalizing above OPE data to those with $\j'\neq0$, we shall slightly modify $\mathcal{P}_{\Delta,\j}$ in (\ref{seriesexp}) by interchanging operator $3$ and $4$, for which eq.~(\ref{Precur}) becomes
\bea
&& \mathcal{C}^{(0)} \fft{u^{\fft{p}{2}}}{v^{\fft{p}{2}+a}}\tilde{C}_{\j'}\big(\xi'\big) = \mathcal{C}_{p,\j'}\fft{u^{\fft{p}{2}}}{v^{\fft{p}{2}+a}}\tilde{C}_{\j'}\big(\xi'\big)\,,
\cr &&
\cr &&
 \mathcal{C}^{(1)} \fft{u^{\fft{p}{2}}}{v^{\fft{p}{2}+a}}\tilde{C}_{\j'}\big(\xi'\big) =
 \fft{u^{\fft{p+1}{2}}}{v^{\fft{p+1}{2}+a}}\big(\gamma^{a,-b,-}_{p,\j'} \tilde{C}_{\j'-1}\big(\xi'\big)+\gamma^{a,-b,+}_{p,\j'} \tilde{C}_{\j'+1}\big(\xi'\big)\big)\,,\label{Crecur}
\eea
with $\gamma^{a,b,\pm}_{\Delta,\j}$ already given in eq.~(\ref{gammarecur}). Since we can integrate-by-parts the Casimir operator in the inversion integral,
by eliminating $\tilde{C}_{\j'+1}$ from this equation, we get a recursion relation in $t$-channel spin $\j'$:
\ba
\gamma^{a,-b,+}_{p-1,\j'-1}\ c^{a,b}_{\j',p,\j}(\Delta) &=&
\Big(\mathcal{C}_{\Delta,\j}-\mathcal{C}_{p-1,\j'-1}\Big)
 c^{a,b}_{\j'-1,p-1,\j}(\Delta)-\gamma^{a,-b,-}_{p-1,\j'-1}\ c^{a,b}_{\j'-2,p,\j}(\Delta)\,.
\label{GFF integral}
\ea
Let's end by explaining how do we extract OPE data in spinning MFT by using above formula. We first decompose $\mathcal{D}_{\downarrow}G$, e.g., eq.~(\ref{effGVV}) into a finite sum of (\ref{moreGFF}), next we obtain OPE data for each term by using eq.~(\ref{GFF integral}) and in the end we can sum them over to get a final answer.

\subsection{Cross-channel expansion of blocks}
\label{tchannel}
In this appendix, we expand (scalar) conformal blocks as $\zb\to 1$ as an exact function of $z$.
To accomplish the computations of the anomalous dimensions in the main-text, we would need $t$-channel conformal blocks with scalar-exchange, conserved-current-exchange and stress-tensor-exchange. In particular, since we are only concerned about the anomalous dimensions, the logarithmic part of $t$-channel conformal blocks are enough for our purpose. Our formulae can be deduced from geodesic Witten-diagram \cite{Hijano:2015zsa} by doing a bit of guesswork as described in \cite{Li:2020dqm}, and is consistent with the most general $t$-channel conformal blocks in terms of $(u,v)$ rather than $(y,\bar{y})$ provided recently in \cite{Li:2019dix,Li:2019cwm} (see also \cite{Li:2020ijq}). Throughout this appendix, we use the variables:
\be
y=\frac{z}{1-z}, \qquad \bar{y}=\frac{1-\zb}{\zb}\,.
\ee
In the main text, we use these conformal blocks in the $t$-channel dDisc, where we take $z\mapsto 1-\zb$ (using $y$ variables, it is $y\rightarrow \bar{y}$).

\begin{itemize}
\item Scalar exchange

For scalar-exchange, we can provide a more general $t$-channel conformal blocks, beyond only picking up logarithmic part. The explicit series is given by
\begin{small}
\bea
&& G_{0,\Delta}(z,\bar{z})=y^{\fft{\Delta}{2}}(1+y)^b(1+\bar{y})^a\sum_k \Big(\fft{(-1)^k \bar{y}^{k+\fft{a+b}{2}} \Gamma(\Delta) \Gamma(-a-b-k)\big(a+\fft{\Delta}{2}\big)_k}{k! \Gamma(-b+\fft{\Delta}{2})
\Gamma(-a-k+\fft{\Delta}{2})}s_{a,b,k}(y)
\cr && + (a\rightarrow -a, b\rightarrow -b) \Big)\label{sctblokgene}
\eea
\end{small}
where
\be
s_{a,b,k}(y)=\,_3F_2\Big(\ft{\Delta -2 a}{2} ,\ft{2 b+\Delta}{2} ,\ft{-2 a-d+\Delta -2 k+2}{2} ;\ft{-d+2 \Delta +2}{2},\ft{-2 a+\Delta -2 k}{2} ,-y\Big)\,.
\ee

In practice, what we use in the main text is the logarithmic part $\log \bar{y}$ of above series from setting $a=b=0$. Note the first line of eq.~(\ref{sctblokgene}) does not have $\log \bar{y}$, and the second line gives us
\bea
G_{0,\Delta}(z,\bar{z})=-\sum_k \fft{\Gamma (\Delta )  \Gamma (k-\frac{\Delta }{2}+1) y^{\fft{\Delta}{2}}}{\Gamma(\frac{\Delta }{2})^2 \Gamma (k+1)^2 \Gamma (-k-\frac{\Delta }{2}+1)}\bar{y}^k \log\bar{y}\, s_{0,0,k}(y)\,.
\cr &&
\eea

\item Conserved-current exchange

The $\log\bar{y}$ part of $t$-channel conformal block with conserved-current-exchange is exhibited as follows:
\bea
&& G_{1,d-1}(z,\bar{z})=\sum_{k} \mathcal{N}^{(1)}_k\fft{\bar{y}^k\,y^{\fft{d-2}{2}}\log \bar{y} }{y+1}\big(
v_{\fft{d-2}{2},k,1}-\fft{2(d-2)ky}{(d-2k)(d-2+2k)}v_{\fft{d}{2},k,0}\big)\,,
\cr &&\label{veczb11}
\eea
where
\bea
\mathcal{N}^{(1)}_k=-\frac{2^{d-1} \Gamma \left(\frac{d+1}{2}\right) \Gamma \left(\frac{d}{2}+k\right)}{\sqrt{\pi } (k!)^2 \Gamma \left(\frac{d}{2}\right) \Gamma \left(\frac{d}{2}-k\right)}\,,\quad v_{p,k,m}=\,_2F_1(p,-k+m,p+1-k,-y)\,.
\eea

\item Stress-tensor exchange

The $\log\bar{y}$ part of $t$-channel conformal block with stress-tensor-exchange was also obtained in \cite{Li:2020dqm}, it is given by
\bea
&& G_{2,d}(z,\bar{z})=\sum_{k} \mathcal{N}^{(2)}_k\fft{\bar{y}^k\,y^{\fft{d-2}{2}}\log \bar{y} }{y+1}\big((d-2) (3 d (y+1)+2 (k y+k-2 y-1)) g_{\fft{d}{2},k}(y)
\cr && -2 \big(2 d^2 (y+1)+d (k (4 y+3)-6 y-5)+2 (k-1) (k y+k-2 y-1)\big) g_{\fft{d-2}{2},k}(y)\big)\,,
\cr &&\label{stresszb11}
\eea
where
\bea
&& \mathcal{N}^{(2)}_k= \frac{2^{d+1} \Gamma \left(\frac{d+3}{2}\right)  \Gamma \left(\frac{d}{2}+k+1\right)}{\sqrt{\pi }  (d+2 k-2) (d+2 k) \Gamma \left(\frac{d}{2}+1\right) \Gamma (k+1)^2 \Gamma \left(\frac{d}{2}-k+1\right)}\,,
\cr &&
\cr && g_{p,k}(y)=\,_2F_1\big(p,-k,\ft{1}{2}(d+2-2k),-y\big)\,.\label{stresszb12}
\eea
\end{itemize}

\section{Four-dimensional gluon amplitudes in flat space}
\label{gluonampdo}

Here we record the bulk
YM$_4$ tree-level gluon amplitudes corresponding to the Lagrangian in eq.~\eqref{Lang} used in the main text.
We start with the three-point ones, from which the four-point amplitudes are then determined
by factorization (see \cite{Elvang:2013cua}, whose conventions we follow, for a pedagogical introduction),
up to contact interactions with the mass dimension of $g_H^2$.
The form of on-shell three-point amplitudes are fixed by Lorentz and little-group symmetries, up to coupling-dependent prefactors, which we find to be
\bea
\mathcal{M}_{1_{-}2_{-}3_{+}}^{\rm YM}=i\sqrt{2}f^{abc}g_{\rm YM}\fft{\langle 12\rangle^3}{\langle 23\rangle\langle 31\rangle}\,,
\quad \mathcal{M}_{1_{-}2_{-}3_{-}}^{\rm H}=
i\sqrt{2}f^{abc}(g_{\rm H}-i g_{\rm H}')\langle12\rangle\langle23\rangle\langle31\rangle\,.
\eea
For $\mathcal{M}_{1_{+}2_{+}3_{-}}^{\rm YM}$ and $\mathcal{M}_{1_{+}2_{+}3_{+}}^{\rm H}$, we simply replace angle-bracket by square-bracket and reverse the odd coupling $g_{\rm H'}$.
Tree-level four-point amplitudes can be cut into a product of on-shell three-point amplitudes
\be
\mathcal{M}_{1234}\Big|_{p_I^2\rightarrow 0}=\fft{\mathcal{M}_{12I}\mathcal{M}_{I34}}{p_I^2}\,.\label{factorization}
\ee
We can use this factorization property to construct four-point amplitudes.

Let's first consider the pure Yang-Mills case.
One might try to directly use (\ref{factorization}) for all channels and sum them over, however, this overcounts the pole structures, since the $s$-channel residue has poles in $t$ or $u$ channel. The standard strategy (see \cite{Arkani-Hamed:2017jhn}) is to make an ansatz
which correctly counts helicity weight and number of derivatives without violating locality
\be
\mathcal{M}_{1_{-}2_{-}3_{+}4_{+}}=\langle12\rangle^2[34]^2(\fft{A}{st}+\fft{B}{su}+\fft{C}{tu})\,.
\ee
By demanding the factorization (\ref{factorization}), one can readily obtain the Parke-Taylor form
\be
\mathcal{M}^{\rm YM^2}= 2g_{\rm YM}^2\langle ij\rangle^4\Big(
\fft{T}{\langle12\rangle\langle23\rangle\langle34\rangle\langle41\rangle}
+
\fft{U}{\langle12\rangle\langle24\rangle\langle43\rangle\langle31\rangle}
\Big)\,,\label{pureYM}
\ee
where $i,j$ are gluons that have negative helicity, $T=f^{bce}f^{ade}$ is the $t$-channel color
factor previously defined in eq.~\eqref{T main text}, and $U$ is the same with $a$ and $b$ swapped.

Note that the first term above actually contains $s$ and $t$-channel poles, and the second term contains $s$ and $u$ poles.

For pure higher-derivative coupling, the nonvanishing amplitudes again all have two gluons of each helicity:
$\mathcal{M}^{\rm H^2}_{1_{-}2_{-}3_{+}4_{+}}$, $\mathcal{M}^{\rm H^2}_{1_{-}2_{+}3_{+}4_{-}}$ and $\mathcal{M}^{\rm H^2}_{1_{-}2_{+}3_{-}4_{+}}$ that arise from $s$-channel, $t$-channel and $u$-channel respectively.
Using the factorization (\ref{factorization}) and Bose symmetry, we obtain:
\bea
 \mathcal{M}^{\rm H^2}_{1_{-}2_{+}3_{+}4_{-}}&=&2(g_{\rm H}^2 +g_{\rm H}'^2)
 \langle14\rangle^2[23]^2T\ \frac{u-s}{2t}+ c \langle 14\rangle^2[23]^2\,,\label{pureH}
\eea
and permutations thereof.
The contact ambiguity $c$ depends on higher-derivative terms in the Lagrangian but doesn't contribute to the analysis
in the main text as it has finite support in spin.
(The tree-level all-$+$ amplitude, also a pure contact term but controlled by a different constant,
similarly does not contribute.)

Finally, the mixed $g_{\rm YM}g_{\rm H}$ amplitudes (including the higher-derivative correction on both vertices)
are quite similar to pure Yang-Mills amplitudes.
For example, for $\mathcal{M}^{\rm mix}_{1_{-}2_{+}3_{+}4_{+}}$ we consider an ansatz
suggested by its helicity scaling and derivative order: $\langle 12\rangle\langle14\rangle[23][34][24]^2$
times two-channel poles like $1/(st)$. We then obtain:
\bea
\mathcal{M}^{\rm YM-H}_{1_{-}2_{-}3_{-}4_{+}}&=&2g_{\rm YM}(g_{\rm H}-ig'_{\rm H})
T\fft{\langle12\rangle\langle23\rangle\langle13\rangle^2}{\langle 34\rangle \langle41\rangle}+
 (1\leftrightarrow2)\,,
\cr
\mathcal{M}^{\rm YM-H}_{1_{-}2_{+}3_{+}4_{+}}&=&2g_{\rm YM}(g_{\rm H}+ig'_{\rm H})T\fft{[23][34][24]^2}{[12][41]}+
 (3\leftrightarrow4)\,,\label{Mix}
\eea
and permutations thereof.
\end{appendix}

\bibliographystyle{JHEP}
\bibliography{refs}
\end{document}

%% file: threepoint_frame.pdf_tex
\begingroup%
  \makeatletter%
  \providecommand\color[2][]{%
    \errmessage{(Inkscape) Color is used for the text in Inkscape, but the package 'color.sty' is not loaded}%
    \renewcommand\color[2][]{}%
  }%
  \providecommand\transparent[1]{%
    \errmessage{(Inkscape) Transparency is used (non-zero) for the text in Inkscape, but the package 'transparent.sty' is not loaded}%
    \renewcommand\transparent[1]{}%
  }%
  \providecommand\rotatebox[2]{#2}%
  \newcommand*\fsize{\dimexpr\f@size pt\relax}%
  \newcommand*\lineheight[1]{\fontsize{\fsize}{#1\fsize}\selectfont}%
  \ifx\svgwidth\undefined%
    \setlength{\unitlength}{161.96423368bp}%
    \ifx\svgscale\undefined%
      \relax%
    \else%
      \setlength{\unitlength}{\unitlength * \real{\svgscale}}%
    \fi%
  \else%
    \setlength{\unitlength}{\svgwidth}%
  \fi%
  \global\let\svgwidth\undefined%
  \global\let\svgscale\undefined%
  \makeatother%
  \begin{picture}(1,0.5400697)%
    \lineheight{1}%
    \setlength\tabcolsep{0pt}%
    \put(0,0){\includegraphics[width=\unitlength,page=1]{threepoint_frame.pdf}}%
    \put(0.55080841,0.49911856){\color[rgb]{0,0,0}\makebox(0,0)[lt]{\lineheight{1.25}\smash{\begin{tabular}[t]{l}$x$\end{tabular}}}}%
    \put(0,0){\includegraphics[width=\unitlength,page=2]{threepoint_frame.pdf}}%
    \put(0.14949566,0.28299719){\color[rgb]{0,0,0}\makebox(0,0)[lt]{\lineheight{1.25}\smash{\begin{tabular}[t]{l}$\mathcal{O}_1(0)$\end{tabular}}}}%
    \put(0.29068905,0.3882557){\color[rgb]{0,0,0}\makebox(0,0)[lt]{\lineheight{1.25}\smash{\begin{tabular}[t]{l}$\mathcal{O}_2(x)$\end{tabular}}}}%
    \put(0.45769837,0.28359472){\color[rgb]{0,0,0}\makebox(0,0)[lt]{\lineheight{1.25}\smash{\begin{tabular}[t]{l}$\mathcal{O}_3(\infty)$\end{tabular}}}}%
    \put(0,0){\includegraphics[width=\unitlength,page=3]{threepoint_frame.pdf}}%
  \end{picture}%
\endgroup%

%% file: four_point.pdf_tex
\begingroup%
  \makeatletter%
  \providecommand\color[2][]{%
    \errmessage{(Inkscape) Color is used for the text in Inkscape, but the package 'color.sty' is not loaded}%
    \renewcommand\color[2][]{}%
  }%
  \providecommand\transparent[1]{%
    \errmessage{(Inkscape) Transparency is used (non-zero) for the text in Inkscape, but the package 'transparent.sty' is not loaded}%
    \renewcommand\transparent[1]{}%
  }%
  \providecommand\rotatebox[2]{#2}%
  \newcommand*\fsize{\dimexpr\f@size pt\relax}%
  \newcommand*\lineheight[1]{\fontsize{\fsize}{#1\fsize}\selectfont}%
  \ifx\svgwidth\undefined%
    \setlength{\unitlength}{501.90725706bp}%
    \ifx\svgscale\undefined%
      \relax%
    \else%
      \setlength{\unitlength}{\unitlength * \real{\svgscale}}%
    \fi%
  \else%
    \setlength{\unitlength}{\svgwidth}%
  \fi%
  \global\let\svgwidth\undefined%
  \global\let\svgscale\undefined%
  \makeatother%
  \begin{picture}(1,0.28194526)%
    \lineheight{1}%
    \setlength\tabcolsep{0pt}%
    \put(0,0){\includegraphics[width=\unitlength,page=1]{four_point.pdf}}%
    \put(-0.00120554,0.13978184){\color[rgb]{0,0,0}\makebox(0,0)[lt]{\lineheight{1.25}\smash{\begin{tabular}[t]{l}$\langle\mathcal{O}_1\mathcal{O}_2\mathcal{O}_3\mathcal{O}_4\rangle\,\,\,\,\,=\,\,\,\,\Sigma_{\mathcal{O}}$\end{tabular}}}}%
    \put(0.36416764,0.26776775){\color[rgb]{0,0,0}\makebox(0,0)[lt]{\lineheight{1.25}\smash{\begin{tabular}[t]{l}$\mathcal{O}_1$\end{tabular}}}}%
    \put(0.36124808,0.01117897){\color[rgb]{0,0,0}\makebox(0,0)[lt]{\lineheight{1.25}\smash{\begin{tabular}[t]{l}$\mathcal{O}_2$\end{tabular}}}}%
    \put(0.83987957,0.00352639){\color[rgb]{0,0,0}\makebox(0,0)[lt]{\lineheight{1.25}\smash{\begin{tabular}[t]{l}$\mathcal{O}_3$\end{tabular}}}}%
    \put(0.84266215,0.26579974){\color[rgb]{0,0,0}\makebox(0,0)[lt]{\lineheight{1.25}\smash{\begin{tabular}[t]{l}$\mathcal{O}_4$\end{tabular}}}}%
    \put(0.60447611,0.17188928){\color[rgb]{0,0,0}\makebox(0,0)[lt]{\lineheight{1.25}\smash{\begin{tabular}[t]{l}$\mathcal{O}$\end{tabular}}}}%
  \end{picture}%
\endgroup%

%% file: gluon_scattering.pdf_tex
\begingroup%
  \makeatletter%
  \providecommand\color[2][]{%
    \errmessage{(Inkscape) Color is used for the text in Inkscape, but the package 'color.sty' is not loaded}%
    \renewcommand\color[2][]{}%
  }%
  \providecommand\transparent[1]{%
    \errmessage{(Inkscape) Transparency is used (non-zero) for the text in Inkscape, but the package 'transparent.sty' is not loaded}%
    \renewcommand\transparent[1]{}%
  }%
  \providecommand\rotatebox[2]{#2}%
  \newcommand*\fsize{\dimexpr\f@size pt\relax}%
  \newcommand*\lineheight[1]{\fontsize{\fsize}{#1\fsize}\selectfont}%
  \ifx\svgwidth\undefined%
    \setlength{\unitlength}{75.44593514bp}%
    \ifx\svgscale\undefined%
      \relax%
    \else%
      \setlength{\unitlength}{\unitlength * \real{\svgscale}}%
    \fi%
  \else%
    \setlength{\unitlength}{\svgwidth}%
  \fi%
  \global\let\svgwidth\undefined%
  \global\let\svgscale\undefined%
  \makeatother%
  \begin{picture}(1,0.95765545)%
    \lineheight{1}%
    \setlength\tabcolsep{0pt}%
    \put(0,0){\includegraphics[width=\unitlength,page=1]{gluon_scattering.pdf}}%
    \put(0.16445252,0.87188071){\color[rgb]{0,0,0}\makebox(0,0)[lt]{\lineheight{1.25}\smash{\begin{tabular}[t]{l}$2$\end{tabular}}}}%
    \put(0.14491737,0.04176899){\color[rgb]{0,0,0}\makebox(0,0)[lt]{\lineheight{1.25}\smash{\begin{tabular}[t]{l}$1$\end{tabular}}}}%
    \put(0.76696766,0.86852546){\color[rgb]{0,0,0}\makebox(0,0)[lt]{\lineheight{1.25}\smash{\begin{tabular}[t]{l}$3$\end{tabular}}}}%
    \put(0.78703217,0.05450107){\color[rgb]{0,0,0}\makebox(0,0)[lt]{\lineheight{1.25}\smash{\begin{tabular}[t]{l}$4$\end{tabular}}}}%
  \end{picture}%
\endgroup%

%% file: bulk_point.pdf_tex
\begingroup%
  \makeatletter%
  \providecommand\color[2][]{%
    \errmessage{(Inkscape) Color is used for the text in Inkscape, but the package 'color.sty' is not loaded}%
    \renewcommand\color[2][]{}%
  }%
  \providecommand\transparent[1]{%
    \errmessage{(Inkscape) Transparency is used (non-zero) for the text in Inkscape, but the package 'transparent.sty' is not loaded}%
    \renewcommand\transparent[1]{}%
  }%
  \providecommand\rotatebox[2]{#2}%
  \newcommand*\fsize{\dimexpr\f@size pt\relax}%
  \newcommand*\lineheight[1]{\fontsize{\fsize}{#1\fsize}\selectfont}%
  \ifx\svgwidth\undefined%
    \setlength{\unitlength}{196.28793473bp}%
    \ifx\svgscale\undefined%
      \relax%
    \else%
      \setlength{\unitlength}{\unitlength * \real{\svgscale}}%
    \fi%
  \else%
    \setlength{\unitlength}{\svgwidth}%
  \fi%
  \global\let\svgwidth\undefined%
  \global\let\svgscale\undefined%
  \makeatother%
  \begin{picture}(1,1.21474999)%
    \lineheight{1}%
    \setlength\tabcolsep{0pt}%
    \put(0,0){\includegraphics[width=\unitlength,page=1]{bulk_point.pdf}}%
    \put(0.02071316,0.02694037){\color[rgb]{0,0,0}\makebox(0,0)[lt]{\lineheight{1.25}\smash{\begin{tabular}[t]{l}$X_1$\end{tabular}}}}%
    \put(0.40568401,0.12290527){\color[rgb]{0,0,0}\makebox(0,0)[lt]{\lineheight{1.25}\smash{\begin{tabular}[t]{l}$X_2$\end{tabular}}}}%
    \put(0.09705314,1.09713775){\color[rgb]{0,0,0}\makebox(0,0)[lt]{\lineheight{1.25}\smash{\begin{tabular}[t]{l}$X_3$\end{tabular}}}}%
    \put(0.51723757,0.90006021){\color[rgb]{0,0,0}\makebox(0,0)[lt]{\lineheight{1.25}\smash{\begin{tabular}[t]{l}$X_4$\end{tabular}}}}%
    \put(0.38337339,0.59142934){\color[rgb]{0,0,0}\makebox(0,0)[lt]{\lineheight{1.25}\smash{\begin{tabular}[t]{l}$P$\end{tabular}}}}%
  \end{picture}%
\endgroup%

%% file: spinning-OPE.bbl
\providecommand{\href}[2]{#2}\begingroup\raggedright\begin{thebibliography}{10}

\bibitem{Poland:2018epd}
D.~Poland, S.~Rychkov and A.~Vichi, \emph{{The Conformal Bootstrap: Theory,
  Numerical Techniques, and Applications}},
  \href{https://doi.org/10.1103/RevModPhys.91.015002}{\emph{Rev. Mod. Phys.}
  {\bfseries 91} (2019) 015002},
  [\href{https://arxiv.org/abs/1805.04405}{{\ttfamily 1805.04405}}].

\bibitem{Albayrak:2019gnz}
S.~Albayrak, D.~Meltzer and D.~Poland, \emph{{More Analytic Bootstrap:
  Nonperturbative Effects and Fermions}},
  \href{https://doi.org/10.1007/JHEP08(2019)040}{\emph{JHEP} {\bfseries 08}
  (2019) 040}, [\href{https://arxiv.org/abs/1904.00032}{{\ttfamily
  1904.00032}}].

\bibitem{Iliesiu:2017nrv}
L.~Iliesiu, F.~Kos, D.~Poland, S.~S. Pufu and D.~Simmons-Duffin,
  \emph{{Bootstrapping 3D Fermions with Global Symmetries}},
  \href{https://doi.org/10.1007/JHEP01(2018)036}{\emph{JHEP} {\bfseries 01}
  (2018) 036}, [\href{https://arxiv.org/abs/1705.03484}{{\ttfamily
  1705.03484}}].

\bibitem{Dymarsky:2017yzx}
A.~Dymarsky, F.~Kos, P.~Kravchuk, D.~Poland and D.~Simmons-Duffin, \emph{{The
  3d Stress-Tensor Bootstrap}},
  \href{https://doi.org/10.1007/JHEP02(2018)164}{\emph{JHEP} {\bfseries 02}
  (2018) 164}, [\href{https://arxiv.org/abs/1708.05718}{{\ttfamily
  1708.05718}}].

\bibitem{Karateev:2019pvw}
D.~Karateev, P.~Kravchuk, M.~Serone and A.~Vichi, \emph{{Fermion Conformal
  Bootstrap in 4d}}, \href{https://doi.org/10.1007/JHEP06(2019)088}{\emph{JHEP}
  {\bfseries 06} (2019) 088},
  [\href{https://arxiv.org/abs/1902.05969}{{\ttfamily 1902.05969}}].

\bibitem{Erramilli:2020rlr}
R.~S. Erramilli, L.~V. Iliesiu, P.~Kravchuk, W.~Landry, D.~Poland and
  D.~Simmons-Duffin, \emph{{blocks\_3d: Software for general 3d conformal
  blocks}},  \href{https://arxiv.org/abs/2011.01959}{{\ttfamily 2011.01959}}.

\bibitem{Costa:2011mg}
M.~S. Costa, J.~Penedones, D.~Poland and S.~Rychkov, \emph{{Spinning Conformal
  Correlators}}, \href{https://doi.org/10.1007/JHEP11(2011)071}{\emph{JHEP}
  {\bfseries 11} (2011) 071},
  [\href{https://arxiv.org/abs/1107.3554}{{\ttfamily 1107.3554}}].

\bibitem{Costa:2011dw}
M.~S. Costa, J.~Penedones, D.~Poland and S.~Rychkov, \emph{{Spinning Conformal
  Blocks}}, \href{https://doi.org/10.1007/JHEP11(2011)154}{\emph{JHEP}
  {\bfseries 11} (2011) 154},
  [\href{https://arxiv.org/abs/1109.6321}{{\ttfamily 1109.6321}}].

\bibitem{Karateev:2017jgd}
D.~Karateev, P.~Kravchuk and D.~Simmons-Duffin, \emph{{Weight Shifting
  Operators and Conformal Blocks}},
  \href{https://doi.org/10.1007/JHEP02(2018)081}{\emph{JHEP} {\bfseries 02}
  (2018) 081}, [\href{https://arxiv.org/abs/1706.07813}{{\ttfamily
  1706.07813}}].

\bibitem{Afkhami-Jeddi:2016ntf}
N.~Afkhami-Jeddi, T.~Hartman, S.~Kundu and A.~Tajdini, \emph{{Einstein gravity
  3-point functions from conformal field theory}},
  \href{https://doi.org/10.1007/JHEP12(2017)049}{\emph{JHEP} {\bfseries 12}
  (2017) 049}, [\href{https://arxiv.org/abs/1610.09378}{{\ttfamily
  1610.09378}}].

\bibitem{Costa:2017twz}
M.~S. Costa, T.~Hansen and J.~Penedones, \emph{{Bounds for OPE coefficients on
  the Regge trajectory}},
  \href{https://doi.org/10.1007/JHEP10(2017)197}{\emph{JHEP} {\bfseries 10}
  (2017) 197}, [\href{https://arxiv.org/abs/1707.07689}{{\ttfamily
  1707.07689}}].

\bibitem{Karateev:2018oml}
D.~Karateev, P.~Kravchuk and D.~Simmons-Duffin, \emph{{Harmonic Analysis and
  Mean Field Theory}},
  \href{https://doi.org/10.1007/JHEP10(2019)217}{\emph{JHEP} {\bfseries 10}
  (2019) 217}, [\href{https://arxiv.org/abs/1809.05111}{{\ttfamily
  1809.05111}}].

\bibitem{Sleight:2018epi}
C.~Sleight and M.~Taronna, \emph{{Spinning Mellin Bootstrap: Conformal Partial
  Waves, Crossing Kernels and Applications}},
  \href{https://doi.org/10.1002/prop.201800038}{\emph{Fortsch. Phys.}
  {\bfseries 66} (2018) 1800038},
  [\href{https://arxiv.org/abs/1804.09334}{{\ttfamily 1804.09334}}].

\bibitem{Kulaxizi:2017ixa}
M.~Kulaxizi, A.~Parnachev and A.~Zhiboedov, \emph{{Bulk Phase Shift, CFT Regge
  Limit and Einstein Gravity}},
  \href{https://doi.org/10.1007/JHEP06(2018)121}{\emph{JHEP} {\bfseries 06}
  (2018) 121}, [\href{https://arxiv.org/abs/1705.02934}{{\ttfamily
  1705.02934}}].

\bibitem{Kologlu:2019bco}
M.~Kologlu, P.~Kravchuk, D.~Simmons-Duffin and A.~Zhiboedov, \emph{{Shocks,
  Superconvergence, and a Stringy Equivalence Principle}},
  \href{https://doi.org/10.1007/JHEP11(2020)096}{\emph{JHEP} {\bfseries 11}
  (2020) 096}, [\href{https://arxiv.org/abs/1904.05905}{{\ttfamily
  1904.05905}}].

\bibitem{Kologlu:2019mfz}
M.~Kologlu, P.~Kravchuk, D.~Simmons-Duffin and A.~Zhiboedov, \emph{{The
  light-ray OPE and conformal colliders}},
  \href{https://doi.org/10.1007/JHEP01(2021)128}{\emph{JHEP} {\bfseries 01}
  (2021) 128}, [\href{https://arxiv.org/abs/1905.01311}{{\ttfamily
  1905.01311}}].

\bibitem{Weinberg:1965nx}
S.~Weinberg, \emph{{Infrared photons and gravitons}},
  \href{https://doi.org/10.1103/PhysRev.140.B516}{\emph{Phys. Rev.} {\bfseries
  140} (1965) B516--B524}.

\bibitem{Britto:2004ap}
R.~Britto, F.~Cachazo and B.~Feng, \emph{{New recursion relations for tree
  amplitudes of gluons}},
  \href{https://doi.org/10.1016/j.nuclphysb.2005.02.030}{\emph{Nucl. Phys. B}
  {\bfseries 715} (2005) 499--522},
  [\href{https://arxiv.org/abs/hep-th/0412308}{{\ttfamily hep-th/0412308}}].

\bibitem{Britto:2005fq}
R.~Britto, F.~Cachazo, B.~Feng and E.~Witten, \emph{{Direct proof of tree-level
  recursion relation in Yang-Mills theory}},
  \href{https://doi.org/10.1103/PhysRevLett.94.181602}{\emph{Phys. Rev. Lett.}
  {\bfseries 94} (2005) 181602},
  [\href{https://arxiv.org/abs/hep-th/0501052}{{\ttfamily hep-th/0501052}}].

\bibitem{Komargodski:2012ek}
Z.~Komargodski and A.~Zhiboedov, \emph{{Convexity and Liberation at Large
  Spin}}, \href{https://doi.org/10.1007/JHEP11(2013)140}{\emph{JHEP} {\bfseries
  11} (2013) 140}, [\href{https://arxiv.org/abs/1212.4103}{{\ttfamily
  1212.4103}}].

\bibitem{Fitzpatrick:2012yx}
A.~L. Fitzpatrick, J.~Kaplan, D.~Poland and D.~Simmons-Duffin, \emph{{The
  Analytic Bootstrap and AdS Superhorizon Locality}},
  \href{https://doi.org/10.1007/JHEP12(2013)004}{\emph{JHEP} {\bfseries 12}
  (2013) 004}, [\href{https://arxiv.org/abs/1212.3616}{{\ttfamily 1212.3616}}].

\bibitem{Kaviraj:2015cxa}
A.~Kaviraj, K.~Sen and A.~Sinha, \emph{{Analytic bootstrap at large spin}},
  \href{https://doi.org/10.1007/JHEP11(2015)083}{\emph{JHEP} {\bfseries 11}
  (2015) 083}, [\href{https://arxiv.org/abs/1502.01437}{{\ttfamily
  1502.01437}}].

\bibitem{Alday:2015eya}
L.~F. Alday, A.~Bissi and T.~Lukowski, \emph{{Large spin systematics in CFT}},
  \href{https://doi.org/10.1007/JHEP11(2015)101}{\emph{JHEP} {\bfseries 11}
  (2015) 101}, [\href{https://arxiv.org/abs/1502.07707}{{\ttfamily
  1502.07707}}].

\bibitem{Kaviraj:2015xsa}
A.~Kaviraj, K.~Sen and A.~Sinha, \emph{{Universal anomalous dimensions at large
  spin and large twist}},
  \href{https://doi.org/10.1007/JHEP07(2015)026}{\emph{JHEP} {\bfseries 07}
  (2015) 026}, [\href{https://arxiv.org/abs/1504.00772}{{\ttfamily
  1504.00772}}].

\bibitem{Alday:2015ewa}
L.~F. Alday and A.~Zhiboedov, \emph{{An Algebraic Approach to the Analytic
  Bootstrap}}, \href{https://doi.org/10.1007/JHEP04(2017)157}{\emph{JHEP}
  {\bfseries 04} (2017) 157},
  [\href{https://arxiv.org/abs/1510.08091}{{\ttfamily 1510.08091}}].

\bibitem{Alday:2016njk}
L.~F. Alday, \emph{{Large Spin Perturbation Theory for Conformal Field
  Theories}}, \href{https://doi.org/10.1103/PhysRevLett.119.111601}{\emph{Phys.
  Rev. Lett.} {\bfseries 119} (2017) 111601},
  [\href{https://arxiv.org/abs/1611.01500}{{\ttfamily 1611.01500}}].

\bibitem{Kravchuk:2016qvl}
P.~Kravchuk and D.~Simmons-Duffin, \emph{{Counting Conformal Correlators}},
  \href{https://doi.org/10.1007/JHEP02(2018)096}{\emph{JHEP} {\bfseries 02}
  (2018) 096}, [\href{https://arxiv.org/abs/1612.08987}{{\ttfamily
  1612.08987}}].

\bibitem{Simmons-Duffin:2016gjk}
D.~Simmons-Duffin, \emph{{The Conformal Bootstrap}},  in \emph{{Proceedings,
  Theoretical Advanced Study Institute in Elementary Particle Physics: New
  Frontiers in Fields and Strings (TASI 2015): Boulder, CO, USA, June 1-26,
  2015}}, pp.~1--74, 2017, \href{https://arxiv.org/abs/1602.07982}{{\ttfamily
  1602.07982}}, \href{https://doi.org/10.1142/9789813149441_0001}{DOI}.

\bibitem{Bzowski:2013sza}
A.~Bzowski, P.~McFadden and K.~Skenderis, \emph{{Implications of conformal
  invariance in momentum space}},
  \href{https://doi.org/10.1007/JHEP03(2014)111}{\emph{JHEP} {\bfseries 03}
  (2014) 111}, [\href{https://arxiv.org/abs/1304.7760}{{\ttfamily 1304.7760}}].

\bibitem{Costa:2014rya}
M.~S. Costa and T.~Hansen, \emph{{Conformal correlators of mixed-symmetry
  tensors}}, \href{https://doi.org/10.1007/JHEP02(2015)151}{\emph{JHEP}
  {\bfseries 02} (2015) 151},
  [\href{https://arxiv.org/abs/1411.7351}{{\ttfamily 1411.7351}}].

\bibitem{Dobrev:1975ru}
V.~K. Dobrev, V.~B. Petkova, S.~G. Petrova and I.~T. Todorov, \emph{{Dynamical
  Derivation of Vacuum Operator Product Expansion in Euclidean Conformal
  Quantum Field Theory}},
  \href{https://doi.org/10.1103/PhysRevD.13.887}{\emph{Phys. Rev. D} {\bfseries
  13} (1976) 887}.

\bibitem{Bzowski:2017poo}
A.~Bzowski, P.~McFadden and K.~Skenderis, \emph{{Renormalised 3-point functions
  of stress tensors and conserved currents in CFT}},
  \href{https://doi.org/10.1007/JHEP11(2018)153}{\emph{JHEP} {\bfseries 11}
  (2018) 153}, [\href{https://arxiv.org/abs/1711.09105}{{\ttfamily
  1711.09105}}].

\bibitem{Bzowski:2018fql}
A.~Bzowski, P.~McFadden and K.~Skenderis, \emph{{Renormalised CFT 3-point
  functions of scalars, currents and stress tensors}},
  \href{https://doi.org/10.1007/JHEP11(2018)159}{\emph{JHEP} {\bfseries 11}
  (2018) 159}, [\href{https://arxiv.org/abs/1805.12100}{{\ttfamily
  1805.12100}}].

\bibitem{Jain:2021wyn}
S.~Jain, R.~R. John, A.~Mehta, A.~A. Nizami and A.~Suresh, \emph{{Momentum
  space parity-odd CFT 3-point functions}},
  \href{https://arxiv.org/abs/2101.11635}{{\ttfamily 2101.11635}}.

\bibitem{Maldacena:2011nz}
J.~M. Maldacena and G.~L. Pimentel, \emph{{On graviton non-Gaussianities during
  inflation}}, \href{https://doi.org/10.1007/JHEP09(2011)045}{\emph{JHEP}
  {\bfseries 09} (2011) 045},
  [\href{https://arxiv.org/abs/1104.2846}{{\ttfamily 1104.2846}}].

\bibitem{Baumann:2020dch}
D.~Baumann, C.~Duaso~Pueyo, A.~Joyce, H.~Lee and G.~L. Pimentel, \emph{{The
  Cosmological Bootstrap: Spinning Correlators from Symmetries and
  Factorization}},  \href{https://arxiv.org/abs/2005.04234}{{\ttfamily
  2005.04234}}.

\bibitem{Kravchuk:2018htv}
P.~Kravchuk and D.~Simmons-Duffin, \emph{{Light-ray operators in conformal
  field theory}}, \href{https://doi.org/10.1007/JHEP11(2018)102}{\emph{JHEP}
  {\bfseries 11} (2018) 102},
  [\href{https://arxiv.org/abs/1805.00098}{{\ttfamily 1805.00098}}].

\bibitem{Raju:2012zs}
S.~Raju, \emph{{Four Point Functions of the Stress Tensor and Conserved
  Currents in AdS$_4$/CFT$_3$}},
  \href{https://doi.org/10.1103/PhysRevD.85.126008}{\emph{Phys. Rev. D}
  {\bfseries 85} (2012) 126008},
  [\href{https://arxiv.org/abs/1201.6452}{{\ttfamily 1201.6452}}].

\bibitem{Skvortsov:2018uru}
E.~Skvortsov, \emph{{Light-Front Bootstrap for Chern-Simons Matter Theories}},
  \href{https://doi.org/10.1007/JHEP06(2019)058}{\emph{JHEP} {\bfseries 06}
  (2019) 058}, [\href{https://arxiv.org/abs/1811.12333}{{\ttfamily
  1811.12333}}].

\bibitem{Nagaraj:2018nxq}
B.~Nagaraj and D.~Ponomarev, \emph{{Spinor-Helicity Formalism for Massless
  Fields in AdS$_4$}},
  \href{https://doi.org/10.1103/PhysRevLett.122.101602}{\emph{Phys. Rev. Lett.}
  {\bfseries 122} (2019) 101602},
  [\href{https://arxiv.org/abs/1811.08438}{{\ttfamily 1811.08438}}].

\bibitem{Closset:2012vp}
C.~Closset, T.~T. Dumitrescu, G.~Festuccia, Z.~Komargodski and N.~Seiberg,
  \emph{{Comments on Chern-Simons Contact Terms in Three Dimensions}},
  \href{https://doi.org/10.1007/JHEP09(2012)091}{\emph{JHEP} {\bfseries 09}
  (2012) 091}, [\href{https://arxiv.org/abs/1206.5218}{{\ttfamily 1206.5218}}].

\bibitem{Hogervorst:2013sma}
M.~Hogervorst and S.~Rychkov, \emph{{Radial Coordinates for Conformal Blocks}},
  \href{https://doi.org/10.1103/PhysRevD.87.106004}{\emph{Phys. Rev. D}
  {\bfseries 87} (2013) 106004},
  [\href{https://arxiv.org/abs/1303.1111}{{\ttfamily 1303.1111}}].

\bibitem{Costa:2016xah}
M.~S. Costa, T.~Hansen, J.~a. Penedones and E.~Trevisani, \emph{{Radial
  expansion for spinning conformal blocks}},
  \href{https://doi.org/10.1007/JHEP07(2016)057}{\emph{JHEP} {\bfseries 07}
  (2016) 057}, [\href{https://arxiv.org/abs/1603.05552}{{\ttfamily
  1603.05552}}].

\bibitem{Fortin:2019gck}
J.-F. Fortin, V.~Prilepina and W.~Skiba, \emph{{Conformal Four-Point
  Correlation Functions from the Operator Product Expansion}},
  \href{https://doi.org/10.1007/JHEP08(2020)115}{\emph{JHEP} {\bfseries 08}
  (2020) 115}, [\href{https://arxiv.org/abs/1907.10506}{{\ttfamily
  1907.10506}}].

\bibitem{Erramilli:2019njx}
R.~S. Erramilli, L.~V. Iliesiu and P.~Kravchuk, \emph{{Recursion relation for
  general 3d blocks}},
  \href{https://doi.org/10.1007/JHEP12(2019)116}{\emph{JHEP} {\bfseries 12}
  (2019) 116}, [\href{https://arxiv.org/abs/1907.11247}{{\ttfamily
  1907.11247}}].

\bibitem{SimmonsDuffin:2012uy}
D.~Simmons-Duffin, \emph{{Projectors, Shadows, and Conformal Blocks}},
  \href{https://doi.org/10.1007/JHEP04(2014)146}{\emph{JHEP} {\bfseries 04}
  (2014) 146}, [\href{https://arxiv.org/abs/1204.3894}{{\ttfamily 1204.3894}}].

\bibitem{Costa:2012cb}
M.~S. Costa, V.~Goncalves and J.~Penedones, \emph{{Conformal Regge theory}},
  \href{https://doi.org/10.1007/JHEP12(2012)091}{\emph{JHEP} {\bfseries 12}
  (2012) 091}, [\href{https://arxiv.org/abs/1209.4355}{{\ttfamily 1209.4355}}].

\bibitem{Caron-Huot:2017vep}
S.~Caron-Huot, \emph{{Analyticity in Spin in Conformal Theories}},
  \href{https://doi.org/10.1007/JHEP09(2017)078}{\emph{JHEP} {\bfseries 09}
  (2017) 078}, [\href{https://arxiv.org/abs/1703.00278}{{\ttfamily
  1703.00278}}].

\bibitem{Simmons-Duffin:2017nub}
D.~Simmons-Duffin, D.~Stanford and E.~Witten, \emph{{A spacetime derivation of
  the Lorentzian OPE inversion formula}},
  \href{https://doi.org/10.1007/JHEP07(2018)085}{\emph{JHEP} {\bfseries 07}
  (2018) 085}, [\href{https://arxiv.org/abs/1711.03816}{{\ttfamily
  1711.03816}}].

\bibitem{Alday:2017vkk}
L.~F. Alday and S.~Caron-Huot, \emph{{Gravitational S-matrix from CFT
  dispersion relations}},
  \href{https://doi.org/10.1007/JHEP12(2018)017}{\emph{JHEP} {\bfseries 12}
  (2018) 017}, [\href{https://arxiv.org/abs/1711.02031}{{\ttfamily
  1711.02031}}].

\bibitem{Caron-Huot:2018kta}
S.~Caron-Huot and A.-K. Trinh, \emph{{All tree-level correlators in
  AdS$_{5}$xS$_{5}$ supergravity: hidden ten-dimensional conformal symmetry}},
  \href{https://doi.org/10.1007/JHEP01(2019)196}{\emph{JHEP} {\bfseries 01}
  (2019) 196}, [\href{https://arxiv.org/abs/1809.09173}{{\ttfamily
  1809.09173}}].

\bibitem{Meltzer:2019nbs}
D.~Meltzer, E.~Perlmutter and A.~Sivaramakrishnan, \emph{{Unitarity Methods in
  AdS/CFT}}, \href{https://doi.org/10.1007/JHEP03(2020)061}{\emph{JHEP}
  {\bfseries 03} (2020) 061},
  [\href{https://arxiv.org/abs/1912.09521}{{\ttfamily 1912.09521}}].

\bibitem{Ponomarev:2019ofr}
D.~Ponomarev, \emph{{From bulk loops to boundary large-N expansion}},
  \href{https://doi.org/10.1007/JHEP01(2020)154}{\emph{JHEP} {\bfseries 01}
  (2020) 154}, [\href{https://arxiv.org/abs/1908.03974}{{\ttfamily
  1908.03974}}].

\bibitem{Isono:2018rrb}
H.~Isono, T.~Noumi and G.~Shiu, \emph{{Momentum space approach to crossing
  symmetric CFT correlators}},
  \href{https://doi.org/10.1007/JHEP07(2018)136}{\emph{JHEP} {\bfseries 07}
  (2018) 136}, [\href{https://arxiv.org/abs/1805.11107}{{\ttfamily
  1805.11107}}].

\bibitem{Fitzpatrick:2011dm}
A.~L. Fitzpatrick and J.~Kaplan, \emph{{Unitarity and the Holographic
  S-Matrix}}, \href{https://doi.org/10.1007/JHEP10(2012)032}{\emph{JHEP}
  {\bfseries 10} (2012) 032},
  [\href{https://arxiv.org/abs/1112.4845}{{\ttfamily 1112.4845}}].

\bibitem{Liu:2018jhs}
J.~Liu, E.~Perlmutter, V.~Rosenhaus and D.~Simmons-Duffin,
  \emph{{$d$-dimensional SYK, AdS Loops, and $6j$ Symbols}},
  \href{https://doi.org/10.1007/JHEP03(2019)052}{\emph{JHEP} {\bfseries 03}
  (2019) 052}, [\href{https://arxiv.org/abs/1808.00612}{{\ttfamily
  1808.00612}}].

\bibitem{Binder:2020raz}
D.~J. Binder, D.~Z. Freedman and S.~S. Pufu, \emph{{A Bispinor Formalism for
  Spinning Witten Diagrams}},
  \href{https://arxiv.org/abs/2003.07448}{{\ttfamily 2003.07448}}.

\bibitem{Armstrong:2020woi}
C.~Armstrong, A.~E. Lipstein and J.~Mei, \emph{{Color/Kinematics Duality in
  AdS$_4$}},  \href{https://arxiv.org/abs/2012.02059}{{\ttfamily 2012.02059}}.

\bibitem{Albayrak:2020fyp}
S.~Albayrak, S.~Kharel and D.~Meltzer, \emph{{On duality of color and
  kinematics in (A)dS momentum space}},
  \href{https://arxiv.org/abs/2012.10460}{{\ttfamily 2012.10460}}.

\bibitem{Albayrak:2018tam}
S.~Albayrak and S.~Kharel, \emph{{Towards the higher point holographic momentum
  space amplitudes}},
  \href{https://doi.org/10.1007/JHEP02(2019)040}{\emph{JHEP} {\bfseries 02}
  (2019) 040}, [\href{https://arxiv.org/abs/1810.12459}{{\ttfamily
  1810.12459}}].

\bibitem{Albayrak:2019asr}
S.~Albayrak, C.~Chowdhury and S.~Kharel, \emph{{New relation for Witten
  diagrams}}, \href{https://doi.org/10.1007/JHEP10(2019)274}{\emph{JHEP}
  {\bfseries 10} (2019) 274},
  [\href{https://arxiv.org/abs/1904.10043}{{\ttfamily 1904.10043}}].

\bibitem{Gary:2009ae}
M.~Gary, S.~B. Giddings and J.~Penedones, \emph{{Local bulk S-matrix elements
  and CFT singularities}},
  \href{https://doi.org/10.1103/PhysRevD.80.085005}{\emph{Phys. Rev. D}
  {\bfseries 80} (2009) 085005},
  [\href{https://arxiv.org/abs/0903.4437}{{\ttfamily 0903.4437}}].

\bibitem{Heemskerk:2009pn}
I.~Heemskerk, J.~Penedones, J.~Polchinski and J.~Sully, \emph{{Holography from
  Conformal Field Theory}},
  \href{https://doi.org/10.1088/1126-6708/2009/10/079}{\emph{JHEP} {\bfseries
  10} (2009) 079}, [\href{https://arxiv.org/abs/0907.0151}{{\ttfamily
  0907.0151}}].

\bibitem{Okuda:2010ym}
T.~Okuda and J.~Penedones, \emph{{String scattering in flat space and a scaling
  limit of Yang-Mills correlators}},
  \href{https://doi.org/10.1103/PhysRevD.83.086001}{\emph{Phys. Rev. D}
  {\bfseries 83} (2011) 086001},
  [\href{https://arxiv.org/abs/1002.2641}{{\ttfamily 1002.2641}}].

\bibitem{Raju:2012zr}
S.~Raju, \emph{{New Recursion Relations and a Flat Space Limit for AdS/CFT
  Correlators}}, \href{https://doi.org/10.1103/PhysRevD.85.126009}{\emph{Phys.
  Rev. D} {\bfseries 85} (2012) 126009},
  [\href{https://arxiv.org/abs/1201.6449}{{\ttfamily 1201.6449}}].

\bibitem{Fitzpatrick:2011hu}
A.~L. Fitzpatrick and J.~Kaplan, \emph{{Analyticity and the Holographic
  S-Matrix}}, \href{https://doi.org/10.1007/JHEP10(2012)127}{\emph{JHEP}
  {\bfseries 10} (2012) 127},
  [\href{https://arxiv.org/abs/1111.6972}{{\ttfamily 1111.6972}}].

\bibitem{Maldacena:2015iua}
J.~Maldacena, D.~Simmons-Duffin and A.~Zhiboedov, \emph{{Looking for a bulk
  point}}, \href{https://doi.org/10.1007/JHEP01(2017)013}{\emph{JHEP}
  {\bfseries 01} (2017) 013},
  [\href{https://arxiv.org/abs/1509.03612}{{\ttfamily 1509.03612}}].

\bibitem{Penedones:2010ue}
J.~Penedones, \emph{{Writing CFT correlation functions as AdS scattering
  amplitudes}}, \href{https://doi.org/10.1007/JHEP03(2011)025}{\emph{JHEP}
  {\bfseries 03} (2011) 025},
  [\href{https://arxiv.org/abs/1011.1485}{{\ttfamily 1011.1485}}].

\bibitem{Paulos:2016fap}
M.~F. Paulos, J.~Penedones, J.~Toledo, B.~C. van Rees and P.~Vieira, \emph{{The
  S-matrix bootstrap. Part I: QFT in AdS}},
  \href{https://doi.org/10.1007/JHEP11(2017)133}{\emph{JHEP} {\bfseries 11}
  (2017) 133}, [\href{https://arxiv.org/abs/1607.06109}{{\ttfamily
  1607.06109}}].

\bibitem{Hijano:2019qmi}
E.~Hijano, \emph{{Flat space physics from AdS/CFT}},
  \href{https://doi.org/10.1007/JHEP07(2019)132}{\emph{JHEP} {\bfseries 07}
  (2019) 132}, [\href{https://arxiv.org/abs/1905.02729}{{\ttfamily
  1905.02729}}].

\bibitem{Komatsu:2020sag}
S.~Komatsu, M.~F. Paulos, B.~C. Van~Rees and X.~Zhao, \emph{{Landau diagrams in
  AdS and S-matrices from conformal correlators}},
  \href{https://doi.org/10.1007/JHEP11(2020)046}{\emph{JHEP} {\bfseries 11}
  (2020) 046}, [\href{https://arxiv.org/abs/2007.13745}{{\ttfamily
  2007.13745}}].

\bibitem{Li:flatspace}
Y.-Z. Li, \emph{{Notes on flat-space limit of AdS/CFT}}, {\emph{work in
  progress} }.

\bibitem{Elvang:2013cua}
H.~Elvang and Y.-t. Huang, \emph{{Scattering Amplitudes}},
  \href{https://arxiv.org/abs/1308.1697}{{\ttfamily 1308.1697}}.

\bibitem{Arkani-Hamed:2017jhn}
N.~Arkani-Hamed, T.-C. Huang and Y.-t. Huang, \emph{{Scattering Amplitudes For
  All Masses and Spins}},  \href{https://arxiv.org/abs/1709.04891}{{\ttfamily
  1709.04891}}.

\bibitem{Caron-Huot:2016cwu}
S.~Caron-Huot and M.~Wilhelm, \emph{{Renormalization group coefficients and the
  S-matrix}}, \href{https://doi.org/10.1007/JHEP12(2016)010}{\emph{JHEP}
  {\bfseries 12} (2016) 010},
  [\href{https://arxiv.org/abs/1607.06448}{{\ttfamily 1607.06448}}].

\bibitem{Zwiebel:2011bx}
B.~I. Zwiebel, \emph{{From Scattering Amplitudes to the Dilatation Generator in
  N=4 SYM}}, \href{https://doi.org/10.1088/1751-8113/45/11/115401}{\emph{J.
  Phys. A} {\bfseries 45} (2012) 115401},
  [\href{https://arxiv.org/abs/1111.0083}{{\ttfamily 1111.0083}}].

\bibitem{He:2020ifr}
T.~He and P.~Mitra, \emph{{Covariant Phase Space and Soft Factorization in
  Non-Abelian Gauge Theories}},
  \href{https://arxiv.org/abs/2009.14334}{{\ttfamily 2009.14334}}.

\bibitem{Costa:2014kfa}
M.~S. Costa, V.~Goncalves and J.~a. Penedones, \emph{{Spinning AdS
  Propagators}}, \href{https://doi.org/10.1007/JHEP09(2014)064}{\emph{JHEP}
  {\bfseries 09} (2014) 064},
  [\href{https://arxiv.org/abs/1404.5625}{{\ttfamily 1404.5625}}].

\bibitem{Kravchuk:2017dzd}
P.~Kravchuk, \emph{{Casimir recursion relations for general conformal blocks}},
  \href{https://doi.org/10.1007/JHEP02(2018)011}{\emph{JHEP} {\bfseries 02}
  (2018) 011}, [\href{https://arxiv.org/abs/1709.05347}{{\ttfamily
  1709.05347}}].

\bibitem{Costa:2016hju}
M.~S. Costa, T.~Hansen, J.~a. Penedones and E.~Trevisani, \emph{{Projectors and
  seed conformal blocks for traceless mixed-symmetry tensors}},
  \href{https://doi.org/10.1007/JHEP07(2016)018}{\emph{JHEP} {\bfseries 07}
  (2016) 018}, [\href{https://arxiv.org/abs/1603.05551}{{\ttfamily
  1603.05551}}].

\bibitem{Caron-Huot:2020ouj}
S.~Caron-Huot, Y.~Gobeil and Z.~Zahraee, \emph{{The leading trajectory in the
  2+1D Ising CFT}},  \href{https://arxiv.org/abs/2007.11647}{{\ttfamily
  2007.11647}}.

\bibitem{Simmons-Duffin:2016wlq}
D.~Simmons-Duffin, \emph{{The Lightcone Bootstrap and the Spectrum of the 3d
  Ising CFT}}, \href{https://doi.org/10.1007/JHEP03(2017)086}{\emph{JHEP}
  {\bfseries 03} (2017) 086},
  [\href{https://arxiv.org/abs/1612.08471}{{\ttfamily 1612.08471}}].

\bibitem{Hijano:2015zsa}
E.~Hijano, P.~Kraus, E.~Perlmutter and R.~Snively, \emph{{Witten Diagrams
  Revisited: The AdS Geometry of Conformal Blocks}},
  \href{https://doi.org/10.1007/JHEP01(2016)146}{\emph{JHEP} {\bfseries 01}
  (2016) 146}, [\href{https://arxiv.org/abs/1508.00501}{{\ttfamily
  1508.00501}}].

\bibitem{Li:2020dqm}
Y.-Z. Li and H.-Y. Zhang, \emph{{More on heavy-light bootstrap up to
  double-stress-tensor}},
  \href{https://doi.org/10.1007/JHEP10(2020)055}{\emph{JHEP} {\bfseries 10}
  (2020) 055}, [\href{https://arxiv.org/abs/2004.04758}{{\ttfamily
  2004.04758}}].

\bibitem{Li:2019dix}
W.~Li, \emph{{Closed-form expression for cross-channel conformal blocks near
  the lightcone}}, \href{https://doi.org/10.1007/JHEP01(2020)055}{\emph{JHEP}
  {\bfseries 01} (2020) 055},
  [\href{https://arxiv.org/abs/1906.00707}{{\ttfamily 1906.00707}}].

\bibitem{Li:2019cwm}
W.~Li, \emph{{Lightcone expansions of conformal blocks in closed form}},
  \href{https://doi.org/10.1007/JHEP06(2020)105}{\emph{JHEP} {\bfseries 06}
  (2020) 105}, [\href{https://arxiv.org/abs/1912.01168}{{\ttfamily
  1912.01168}}].

\bibitem{Li:2020ijq}
W.~Li, \emph{{Factorized lightcone expansion of conformal blocks}},
  \href{https://arxiv.org/abs/2012.09710}{{\ttfamily 2012.09710}}.

\end{thebibliography}\endgroup
